\documentclass{aa}

\usepackage{natbib}
\bibpunct{(}{)}{;}{a}{}{,} 
\usepackage{array}

\usepackage{newtxtext,newtxmath}

\usepackage[T1]{fontenc}
\usepackage{ae,aecompl}

\usepackage{graphicx}	
\usepackage{amsmath}	
\usepackage{amssymb}	
\usepackage{lscape}
\usepackage{comment}
\usepackage{adjustbox}

\begin{document}


\title{The XXL Survey: XXXI. Classification and host galaxy properties of 2.1 GHz ATCA XXL-S radio sources}

\author{Andrew Butler\inst{\ref{inst1}}\thanks{E-mail: andrew.butler@icrar.org} \and Minh Huynh\inst{\ref{inst1},\ref{inst2}} \and Ivan Delvecchio\inst{\ref{inst4}} \and Anna Kapi\'{n}ska\inst{\ref{inst1},\ref{inst3}} \and Paolo Ciliegi\inst{\ref{inst5}} \and Nika Jurlin\inst{\ref{inst4}} \and Jacinta Delhaize\inst{\ref{inst4}} \and Vernesa Smol\v{c}i\'{c}\inst{\ref{inst4}} \and Shantanu Desai\inst{\ref{inst6}} \and Sotiria Fotopoulou\inst{\ref{inst7}} \and Chris Lidman\inst{\ref{inst8}} \and Marguerite Pierre\inst{\ref{inst9}} \and Manolis Plionis\inst{\ref{inst10},\ref{inst11}}}

\institute{International Centre for Radio Astronomy Research (ICRAR), University of Western Australia, 35 Stirling Hwy, Crawley WA 6009, Australia\label{inst1}
\and
CSIRO Astronomy and Space Science, 26 Dick Perry Ave, Kensington WA 6151, Australia\label{inst2}
\and
ARC Centre of Excellence for All-Sky Astrophysics (CAASTRO), University of Western Australia, 35 Stirling Hwy, Crawley WA 6009, Australia\label{inst3}
\and
Physics Department, University of Zagreb, Bijeni\v{c}ka cesta 32, 10002 Zagreb, Croatia\label{inst4}
\and
INAF - Osservatorio Astronomico di Bologna, via Gobetti 93/3, 40129 Bologna, Italy\label{inst5}
\and
Department of Physics, IIT Hyderabad, Kandi, Telangana-502285, India\label{inst6}
\and
Department of Astronomy, University of Geneva, Ch. d'\'{E}cogia 16, 1290 Versoix, Switzerland\label{inst7}
\and
Australian Astronomical Observatory, 105 Delhi Rd, North Ryde NSW 2113, Australia\label{inst8}
\and
Service d'Astrophysique AIM, CEA/DSM/IRFU/SAp, CEA Saclay, 91191 Gif-sur-Yvette, France\label{inst9}
\and
National Observatory of Athens, Lofos Nymfon, 11851 Athens, Greece\label{inst10}
\and
Physics Department, Aristotle University of Thessaloniki, 54124 Thessaloniki, Greece\label{inst11}
}

\date{Received date / Accepted date }

\abstract{The classification of the host galaxies of the radio sources in the 25 deg$^2$ ultimate XMM extragalactic survey south field (XXL-S) is presented.  XXL-S was surveyed at 2.1 GHz with the Australia Telescope Compact Array (ATCA) and is thus far the largest area radio survey conducted down to rms flux densities of $\sigma \sim 41$ $\mu$Jy beam$^{-1}$.  Of the 6287 radio sources in XXL-S, 4758 (75.7\%) were cross-matched to an optical counterpart using the likelihood ratio technique.  There are 1110 spectroscopic redshifts and 3648 photometric redshifts available for the counterparts, of which 99.4\% exist out to $z \sim 4$.  A number of multiwavelength diagnostics, including X-ray luminosities, mid-infrared colours, spectral energy distribution fits, radio luminosities, and optical emission lines and colours, were used to classify the sources into three types: low-excitation radio galaxies (LERGs), high-excitation radio galaxies (HERGs), and star-forming galaxies (SFGs). The final sample contains 1729 LERGs (36.3\%), 1159  radio-loud HERGs (24.4\%), 296  radio-quiet HERGs (6.2\%), 558  SFGs (11.7\%), and 1016  unclassified sources (21.4\%). The XXL-S sub-mJy radio source population is composed of $\sim$75\% active galactic nuclei and $\sim$20\% SFGs down to 0.2 mJy.  The host galaxy properties of the HERGs in XXL-S are independent of the HERG selection, but the XXL-S LERG and SFG selection is, due to the low spectral coverage, largely determined by the known properties of those populations.  Considering this caveat, the LERGs tend to exist in the most massive galaxies with low star formation rates and redder colours, whereas the HERGs and SFGs exist in galaxies of lower mass, higher star formation rates, and bluer colours.  The fraction of blue host galaxies is higher for radio-quiet HERGs than for radio-loud HERGs. LERGs and radio-loud  HERGs are found at all radio luminosities, but radio-loud HERGs tend to be more radio luminous than LERGs at a given redshift. These results are consistent with the emerging picture in which LERGs exist in the most massive quiescent galaxies typically found in clusters with hot X-ray halos and HERGs are associated with ongoing star formation in their host galaxies via the accretion of cold gas.}



\keywords{galaxies: general -- galaxies: evolution -- galaxies: active -- radio continuum: galaxies -- galaxies: statistics -- galaxies: stellar content}

\authorrunning{Butler et al.}
\maketitle



\section{Introduction}
\label{sec:intro}

Supermassive black holes (SMBHs) play an important role in galaxy evolution: the masses of SMBHs are correlated with the stellar velocity dispersions of the bulges of their host galaxies (e.g. \citealp{magorrian1998, gebhardt2000, graham2008}), and the growth curve of black holes and that of stellar mass in galaxies have the same shape \citep{shankar2009}.  This indicates that the evolution of galaxies and SMBHs are closely linked.  One probable reason for this link  is that SMBHs  exhibit energetic phenomena, called active galactic nuclei (AGN), which can affect the stellar and gas content of their host galaxies, fundamentally altering their properties. 
This occurs via `feedback', an outflow from the AGN that heats the interstellar medium, which would otherwise collapse to form stars \citep{bohringer1993,binney1995,forman2005,best2006,mcnamara2007,cattaneo2009,fabian2012}.  Galaxy formation models that include this extra AGN component are able to more accurately reproduce many observed galaxy properties for $z \leq 0.2$, including the optical luminosity function, colours, stellar ages, and morphologies (e.g. \citealp{croton2006,bower2006}). Therefore, AGN feedback is a crucial component to galaxy evolution models and fundamental to overall galaxy evolution.

Observations of AGN host galaxies have suggested that there are two distinct AGN populations.  One population is the classical AGN associated with quasars and radio galaxies, which are well described by the unified AGN model \citep{antonucci1985,barthel1989,urry1995}.  They exhibit a luminous accretion disk \citep{shakura1973} that produces ionising photons, which generate regions of broad and narrow optical emission lines outside the disk.  They also feature X-ray emission, which is generated by inverse Compton scattering, and a dusty torus, which is heated by UV and optical emission from the accretion disk.  As a result, the torus re-radiates in the mid-infrared (MIR).  If the torus is oriented edge-on relative to the line of sight, the broad-line region is obscured, giving rise to type II AGN \citep{antonucci1993}.  Because these AGN radiate across the electromagnetic spectrum \citep{elvis1994} and 
their optical spectra exhibit strong high-excitation emission lines (such as [\ion{O}{III}]), they are labelled 
high-excitation radio galaxies (HERGs). 

The second AGN population displays low levels of high-excitation emission lines in their spectra or none at all \citep{hine1979,laing1994,jackson1997,best2012}.  The X-ray emitting corona, optical continuum emission from an accretion disk, and MIR emission from an obscuring torus are absent, and their AGN signatures can only be detected at radio wavelengths (e.g. \citealp{hardcastle2007,hickox2009}).  Accordingly, these sources are called low-excitation radio galaxies (LERGs).  The typical host galaxies for LERGs are the most massive quiescent galaxies associated with clusters, while HERGs are hosted by galaxies that have lower mass and higher star formation rates (e.g. \citealp{tasse2008,smolcic2009b,best2012,best2014}).

It  has been hypothesised that the two AGN populations exhibit fundamentally different black hole accretion modes that result in two distinct forms of feedback (e.g. \citealp{hardcastle2007}).  LERGs accrete the hot X-ray emitting phase of the intergalactic medium (IGM) found in clusters, generating radiatively inefficient feedback, whereas HERGs accrete the cold phase of the IGM, allowing for radiatively efficient feedback and associated star formation \citep{heckman2014}.  However, the details of feedback as a function of host galaxy and cosmic time and the precise origin of HERG and LERG differences are poorly understood.

The  most direct way to measure the evolution of radio sources is via the construction of the radio luminosity function (RLF). Only a few studies have constructed individual RLFs for HERGs and LERGs \citep{best2012,best2014,pracy2016}.  The results indicate that for $z \lesssim 1$, HERGs evolve strongly and LERGs exhibit space densities that are consistent with weak or no evolution.  However, the radio data in these studies probed no deeper than $S_{\rm{1.4GHz}} \sim 3$ mJy (over very large areas of $\gtrsim$800 deg$^2$), and so the RLFs are not well constrained at the low-luminosity end ($L_{\rm{1.4GHz}} \lesssim 10^{22}$ W Hz$^{-1}$).  In addition, the HERG and star-forming galaxy (SFG) RLFs from these papers disagree with each other for $L_{\rm{1.4GHz}} \lesssim 10^{24}$ W Hz$^{-1}$, indicating that these two populations can be difficult to identify at low radio luminosities.  On the other hand, \cite{smolcic2009a} studied the evolution of the low-luminosity ($L_{\rm{1.4GHz}} \lesssim 5 \times 10^{25}$ W Hz$^{-1}$) and high-luminosity ($L_{\rm{1.4GHz}} \gtrsim 5 \times 10^{25}$ W Hz$^{-1}$) radio AGN down to $S_{\rm{1.4GHz}}~\gtrsim$~50~$\mu$Jy and out to $z = 1.3$ in the $\sim$2 deg$^2$ COSMOS field \citep{schinnerer2007}, but did not 
distinguish between HERGs and LERGs.

A full understanding of the HERG and LERG luminosity functions, host galaxies, and cosmic evolution is needed in order to understand the physical driver for their differences, and therefore the role of the two different feedback modes in galaxy evolution as a function of time (\citealp{best2012} and references therein).  Such a precise understanding can test and inform AGN synthesis models, such as that of \cite{merloni2008}.  Therefore, it is important to identify the HERG and LERG populations across different cosmic epochs so that their impact on galaxy evolution is not conflated.  This requires a deep radio survey conducted over a relatively wide area with a large amount of complementary multiwavelength coverage in order to separate between the LERG, HERG, and SFG populations, especially at low radio luminosities.  The 2.1 GHz radio survey of the 25 deg$^2$ ultimate XMM extragalactic survey (\citealp{pierre2016_xxl1}, hereafter XXL Paper I) south field (XXL-S; $\alpha=23^{\text{h}}30^{\text{m}}00^{\text{s}}$, $\delta=-55^{\circ}00'00''$), conducted with the Australia Telescope Compact Array (ATCA), is thus far the largest area radio survey conducted down to rms flux densities of $\sigma \sim 41 \mu$Jy beam$^{-1}$ (\citealp{butler2017_xxl18}, hereafter XXL Paper XVIII).  The excellent multiwavelength coverage of XXL-S allows for the construction of the HERG and LERG RLFs in multiple redshift bins.  When completed, they will be the most sophisticated RLFs to date.

First, however, the radio sources must be separated into the AGN and SFG populations.  This paper describes the process and results of the classification of the radio sources in XXL-S as LERGs, HERGs, and SFGs.  In Section~\ref{sec:data}, the data, cross-matching procedure, and final sample properties are described.  Section~\ref{sec:agn_sfg_class} details the multiwavelength classification scheme and decision tree used to classify each source.  In Section~\ref{sec:discussion}, AGN selection effects, possible misclassifications, the distribution of sub-mJy radio sources, and the properties of the host galaxies of LERGs, HERGs, and SFGs are discussed.  Section~\ref{sec:conclusions} contains the summary and conclusions.  Throughout this paper the following cosmology is adopted: $H_0 = 69.32$ km s$^{-1}$ Mpc$^{-1}$, $\Omega_{\rm{m}} = 0.287$, and $\Omega_{\Lambda} = 0.713$ \citep{hinshaw2013}.  The following notation for radio spectral index ($\alpha_R$) is used: $S_{\nu} \propto \nu^{\alpha_R}$.

\section{Data and sample}
\label{sec:data}

\subsection{Radio data}
\label{sec:radio_data}

The full 25 deg$^2$ of XXL-S was observed with the Australia Telescope Compact Array (ATCA) at 2.1 GHz, with a 2.0 GHz bandwidth, over $\sim$220 hours on source.  The radio data reached a median rms of $\sigma \sim 41$ $\mu$Jy beam$^{-1}$ and a resolution of $\sim$4.8$''$.  The number of radio sources extracted above 5$\sigma$ is 6287.  More details of the observations, data reduction, and source statistics can be found in XXL Paper XVIII and \cite{smolcic2016_xxl11} (XXL Paper XI).  The final source sample in this paper is selected from the radio sources that have reliable optical counterparts (see Section~\ref{sec:LR} for the cross-matching method).

\subsection{Multiwavelength catalogue}
\label{sec:mw_cat}

There is a wealth of complementary photometric and spectroscopic data available for XXL-S, from X-rays to the mid-infrared, which is contained in a single multiwavelength catalogue.  The details of the construction of this catalogue can be found in \cite{fotopoulou2016_xxl6} (hereafter XXL Paper VI) and Fotopoulou et al. (in prep.).  The following summarises the data available in the multiwavelength catalogue at each wavelength regime.

\subsubsection{X-ray data}
\label{sec:xray_data}

The \emph{XMM-Newton} X-ray telescope observed XXL-S as part of the \emph{XMM-Newton} Extragalactic Legacy survey.  The survey achieved a depth of $\sim$5 $\times$ 10$^{-15}$ ergs s$^{-1}$ cm$^{-2}$ in the 0.5$-$2 keV band and $\sim$2 $\times$ 10$^{-14}$ erg s$^{-1}$ cm$^{-2}$ in the 2$-$10 keV band. More details of the X-ray observations, data processing, and simulations can be found in XXL Paper I.

\subsubsection{Ultraviolet data}

The GALEX satellite surveyed the entire sky in two ultraviolet bands, 1344--1786 \AA~ (FUV) and 1771--2831 \AA~(NUV), reaching an all-sky depth of $m_{\rm{AB}}$ $\sim$ 20.  More details of the mission can be found in \cite{morrissey2005} and \cite{martin2005}.

\subsubsection{Optical photometry}

The Blanco Cosmology Survey (BCS) observed $\sim$50 deg$^2$ centred at $\alpha=23^{\text{h}}00^{\text{m}}00^{\text{s}}$ and  $\delta=-55^{\circ}00'00''$, giving full coverage of XXL-S.  The Mosaic2 imager was used on the Cerro Tololo Inter-American Observatory (CTIO) 4m Blanco telescope to generate data in the $griz$ bands, reaching 10$\sigma$ point-source depths of 23.9, 24.0, 23.6, and 22.1 mag (AB).  More details can be found in \cite{desai2012}.  In addition, the Dark Energy Camera (DECam; \citealp{flaugher2015}) was used with the Blanco telescope to observe XXL-S (PI: C. Lidman) in the $griz$ bands (4850--9000 $ $\AA).  The limiting magnitudes reached in each band (defined as the third quartile of the corresponding magnitude distribution) are 25.73, 25.78, 25.6, and 24.87 mag (AB).  More details of the observations can be found in XXL Paper VI and a description of the data reduction can be found in 
\cite{desai2012,desai2015}.

\subsubsection{Optical spectra}
\label{sec:spectra}

Spectra in the wavelength range of $\sim$370$-$900 nm were gathered for thousands of objects in XXL-S in two observing runs in 2013 and 2016 with the Australian Astronomical Telescope (AAT) using the two-degree field (2dF) fibre positioner \citep{lewis2002} with the AAOmega spectrograph \citep{smith2004}.  The first run obtained 3660 reliable redshifts for X-ray selected objects (\citealp{lidman2016}, XXL Paper XIV), 136 of which are also radio sources; the second run (PI: M. Plionis) obtained a total of 2813 reliable redshifts for a mix of X-ray, optical, and radio-selected objects (\citealp{chiappetti2017}, XXL Paper XXVII, submitted).  In the second run, 1148 radio sources whose optical counterparts are brighter than $r_{\rm{AB}} < 22.0$ (equivalent to $r_{\rm{Vega}} < 21.8$) were targeted, from which 986 reliable redshifts were obtained.  \citealp{adami2011}, Table 3 in XXL Paper I, \cite{adami2017} (XXL Paper XX, submitted), and \cite{guglielmo2017} (XXL Paper XXII) contain information on a few other spectroscopic surveys that covered XXL-S, but these surveys added a total of only 12 redshifts to the final sample of this paper.

\subsubsection{Near-infrared data}

The Vista Hemisphere Survey (VHS) is observing the entire southern sky in the near-infrared (NIR) using the Visible and Infrared Survey Telescope for Astronomy (VISTA), and it has also observed the $\sim$5000 deg$^2$ southern Galactic cap, which includes the whole of XXL-S.  In this region, it reached 5$\sigma$ point-source depths of $J=20.6$, $H = 19.8$ and $K_s=18.1$ mag (Vega).  More information can be found in \cite{mcmahon2013}.

\subsubsection{Mid-infrared data}
\label{sec:mir_data}

The Wide-field Infrared Survey Explorer (WISE) mission observed the whole sky in four MIR bands: $W1$ = 3.4 $\mu$m, $W2$ = 4.6 $\mu$m, $W3$ = 12 $\mu$m, and $W4$ = 22 $\mu$m \citep{wright2010}.  The survey reached 5$\sigma$ depths of 16.5, 15.5, 11.2, and 7.9 mag (Vega), respectively.  The WISE MIR data used in this paper was extracted with a fixed aperture 3$''$ wide at 1$\sigma$ depths of 17.62, 17.01, 13.50, and 10.64 mag (Vega), respectively.  These have been matched to $\sim$15 other filters in the multiwavelength catalogue for XXL-S (XXL Paper VI; Fotopoulou et al., in prep.), and are therefore unlikely to be spurious.  Furthermore, the \emph{Spitzer} satellite has observed the 94 deg$^2$ \emph{Spitzer} South Pole Telescope Deep Field (SSDF), centred at $\alpha=23^{\text{h}}00^{\text{m}}00^{\text{s}}$ and  $\delta=-55^{\circ}00'00''$, with its Infrared Array Camera (IRAC; \citealp{fazio2004}).
This field fully covers XXL-S.  IRAC channels 1 (3.6 $\mu$m) and 2 (4.5 $\mu$m) were used, reaching 5$\sigma$ (Vega) sensitivities of 19.0 and 18.2 mag (7.0 and 9.4 $\mu$Jy), respectively.  Additional details can be found in \cite{ashby2013}.

\subsection{Cross-matching procedure}
\label{sec:cross-matching}

This  section summarises the method used to cross-match the radio sources to the multiwavelength catalogue.

\subsubsection{Initial radial match}

The XXL-S radio source catalogue from XXL Paper XVIII was initially cross-matched to the sources in the multiwavelength catalogue from XXL Paper VI and Fotopoulou et al. (in prep.)  using a simple radial match with a radius of 3$''$; the position for each source in the multiwavelength catalogue is based on the position in the highest resolution band available for the source.  This produced a total of 6277 potential optical counterparts to all 6287 radio sources in XXL-S.

\begin{figure}
        \includegraphics[width=\columnwidth]{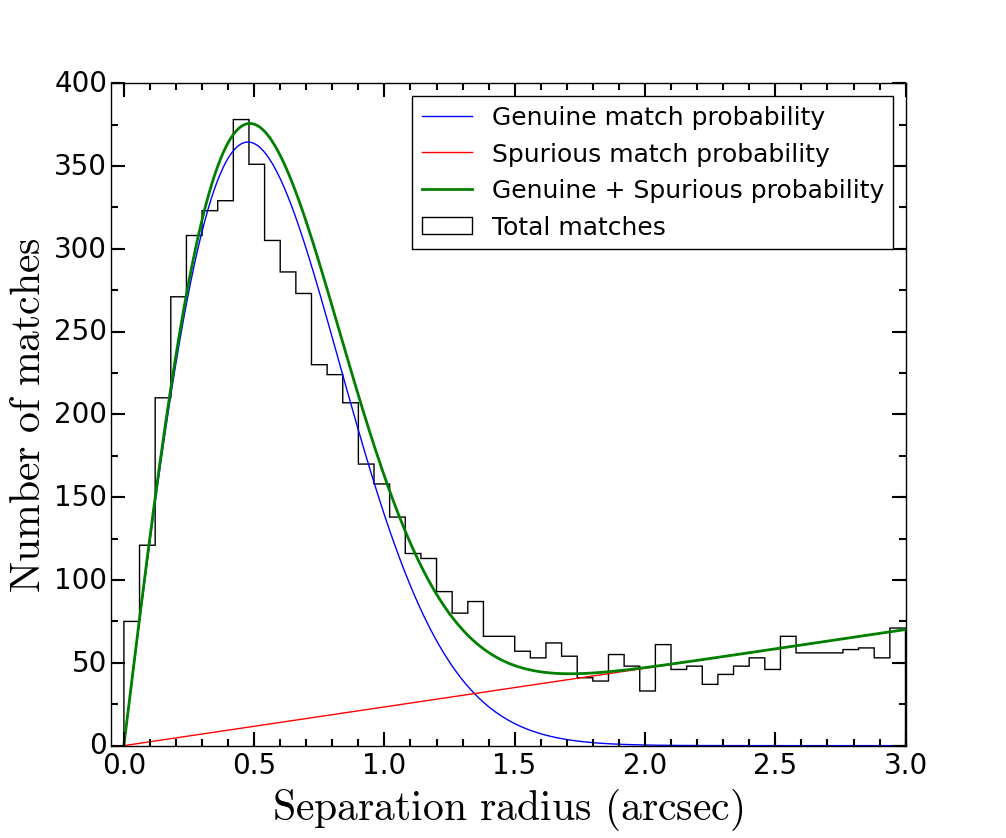}
    \caption{Histogram of the total number of optical matches vs. separation radius for the 6287 XXL-S radio sources (a total of 6277 optical matches are within 3$''$).  The blue curve represents the theoretical number of genuine matches as a function of radius (Eq. \ref{eq:gen_matches}), the red curve the theoretical number of spurious matches as a function of radius (Eq. \ref{eq:spur_matches}), and the green curve  the total number of matches as a function of radius (the sum of Eqs. \ref{eq:gen_matches} and  \ref{eq:spur_matches} for a given value of $r$).}
    \label{fig:N_tot_vs_r_plot}
\end{figure}

Figure \ref{fig:N_tot_vs_r_plot} shows the histogram of the total number matches as a function of separation radius between the radio sources and all potential optical counterparts.  The non-zero offsets between the radio and optical positions are assumed to be due to measurement error alone, so the {a priori} infinitesimal probability $dP_{\rm{gen}} $ of a genuine optical match being found at a distance between $r$ and $dr$ for all the radio sources is given by the Rayleigh distribution \citep{deruiter1977, wolstencroft1986}
\begin{equation}
\label{eq:gen_matches}
dP_{\rm{gen}} = \frac{N_{\rm{r,gen}}}{\sigma_{\rm{pos}}^2} r e^{-r^2 / 2 \sigma_{\rm{pos}}^2} dr,
\end{equation}
where $N_{\rm{r,gen}}$ is the number of radio sources with a highly probable genuine match (as determined by the likelihood ratio method described in Section \ref{sec:LR}) and $\sigma_{\rm{pos}}^2 = \sigma_{\rm{r}}^2 + \sigma_{\rm{o}}^2$ ($\sigma_{\rm{r}}$ is the radio positional uncertainty and $\sigma_{\rm{o}}$ is the optical positional uncertainty). In  this work, $N_{\rm{r,gen}} = 4770$, $\sigma_{\rm{r}}=0.37''$ (using the average radio positional uncertainty and assuming a simple model in which the RA and Dec radio positional uncertainties are symmetric), and $\sigma_{\rm{o}} = 0.30''$.  Entering these values into Equation \ref{eq:gen_matches} generated the genuine match curve shown in Figure \ref{fig:N_tot_vs_r_plot}.  On the other hand, the probability $dP_{\rm{spur}}$ that a spurious optical match will be found between $r$ and $dr$ is given by
\begin{equation}
\label{eq:spur_matches}
dP_{\rm{spur}} = N_{\rm{r}} \rho_{\rm{o}} (2 \pi r) dr,
\end{equation}
where $N_{\rm{r}}$ = 6287 is the total number of radio sources in XXL-S and $\rho_{\rm{o}} \approx$ 127793 deg$^{-2}$ is the globally averaged density on the sky of all $z$-band\footnote{The $z$-band was used because the data are deeper and galaxies are normally brighter in the infrared than they are at bluer wavelengths, so it gives access to a greater number of sources.} sources in XXL-S.  Using these values for Equation \ref{eq:spur_matches} produced the spurious match line in Figure \ref{fig:N_tot_vs_r_plot}, which shows that the percentage of spurious matches rapidly increases beyond $\sim$1$''$.  For example, at $\sim$1.3$''$, there are $\sim$50\% spurious matches.  While there are some genuine matches at 
large radii, using a simple radial match increases the contamination from spurious matches.  In order to be able to capture as many genuine matches as possible while avoiding spurious matches, a more sophisticated technique is required.

\subsubsection{Likelihood ratio}
\label{sec:LR}

The likelihood ratio (LR) technique (\citealp{richter1975,deruiter1977,sutherland1992,ciliegi2003}) is useful for distinguishing between the probable genuine optical match to a radio source and unrelated background objects.  The LR is defined as the ratio of the probability of a given optical source being the genuine counterpart of the radio source to the probability of the optical source being an unrelated background object.  For a given potential optical counterpart with magnitude $m$ and positional offset $r$ from the radio source position, the LR is expressed as
\begin{equation}
LR = \frac{q(m) f(r)}{n(m)},
\end{equation}
where $q(m)$ is the expected probability distribution of the true optical counterparts as a function of magnitude, $f(r)$ is the probability distribution function of the combined radio and optical positional uncertainties (i.e. an elliptical Gaussian distribution representing the quadrature sum of the radio and optical positional uncertainties in RA and Dec) and $n(m)$ is the probability distribution of background objects as a function of magnitude.  A detailed discussion on the procedure for calculating $q(m)$, $f(r)$, and $n(m)$ is found in \cite{ciliegi2003}.  The particular method employed is described in Ciliegi et al. (in prep.), and is briefly outlined as follows:

\begin{enumerate}
\item The catalogue of all optical matches within 3$''$ of each radio source was used as the input;
\item LR analysis was performed in seven different optical and NIR bands (BCS $griz$ or DECam $griz$, and VISTA $JHK$) for each potential counterpart;
\item Potential counterparts in a given optical band were considered reliable if $LR > LR_{\rm{th}}$ (the LR threshold value).  The $LR_{\rm{th}}$ value that maximised the sum of the reliability and completeness of the counterpart selection is $LR_{\rm{th}} = 1-Q$ \citep{ciliegi2003,ciliegi2005}, where $Q = \int^{m_{\rm{lim}}} q(m) dm$ is the probability that the genuine optical match for a given ATCA XXL-S radio source has been detected, given the magnitude limit of the optical band (i.e. the fraction of radio sources for which at least one optical counterpart could be found given the sensitivity limit of the optical band).  The value of $Q$ was estimated by taking the ratio of the total number of expected counterpart identifications (the integral of the $q(m)$ distribution) to the total number of radio sources (see \citealp{ciliegi2003}).  A value of $Q=0.8$ (corresponding to $LR_{\rm{th}} = 0.2$) was found to be applicable to all bands (i.e. within 4\% of the individual $Q$ values for each individual band).  If there were no potential counterparts with $LR > LR_{\rm{th}}$ for a given radio source in a given optical band, then that radio source was considered to be unmatched in that band;
\item For a given radio source and a given optical band, if there were two potential counterparts with $LR > LR_{\rm{th}}$, then only the optical source with the highest reliability was chosen as the counterpart in that band. The reliability is the relative likelihood that a potential optical counterpart is the genuine counterpart, given the other potential counterparts and the probability that the genuine counterpart was not detected ($1-Q$).  It is defined as
\begin{equation}
R_j=\frac{LR_j}{\Sigma_i(LR_i)+(1-Q)}, 
\end{equation}
where $LR_j$ is the LR value for the optical source being considered and $\Sigma_i LR_i$ is the sum of the LR values for all potential counterparts \citep{ciliegi2003}; 
\item The seven LR catalogues from each band ($grizJHK$) were merged into one master catalogue;
\item For a given radio source, if two or more potential counterparts with $LR > LR_{\rm{th}}$ existed in different bands in the master catalogue,
    \begin{enumerate}
    \item If all potential counterparts existed in one common band, the counterpart with the highest reliability in that band was chosen;
    \item If none of the potential counterparts existed in at least one common band, the counterpart with the highest reliability in any band was selected.
    \end{enumerate}
\end{enumerate}

Taking these steps resulted in 4770 radio sources that were cross-matched to a reliable optical counterpart.  This is 76\% of all radio sources in the ATCA XXL-S sample, which is consistent with previous radio-optical association studies at the same optical magnitude depth as XXL-S (see Ciliegi et al., in prep.).  The number of potential misidentifications can be estimated by summing the $1-R_j$ values for all sources in each band.  Of the seven bands used, the $J$-band was responsible for the lowest number of misidentifications ($\sim$36) and the DECAM $z$-band was responsible for the highest ($\sim$91).  The average number of misidentifications across all bands is $\sim$55 sources, which is $\sim$1.2\% of the 4770 optically matched radio sources.

\subsection{Final sample}
\label{sec:final_sample}

The  number of radio sources that were cross-matched to optical sources using the LR technique is 4770 (Ciliegi et al., in prep.).  There are 12 sources with AAT spectra that were classified as stars, so these sources were removed from the catalogue.  The final sample  used for the purposes of source classification in this paper is the set of 4758 radio sources that were cross-matched with the LR technique to the non-stellar optical sources in the multiwavelength catalogue from XXL Paper VI and Fotopoulou et al. (in prep.), as described in Section~\ref{sec:LR}.  The rest of this paper refers to this sample, unless otherwise stated.

\subsubsection{Radial separation distribution}
\label{sec:sep_dist_gen}

\begin{figure}
        \includegraphics[width=\columnwidth]{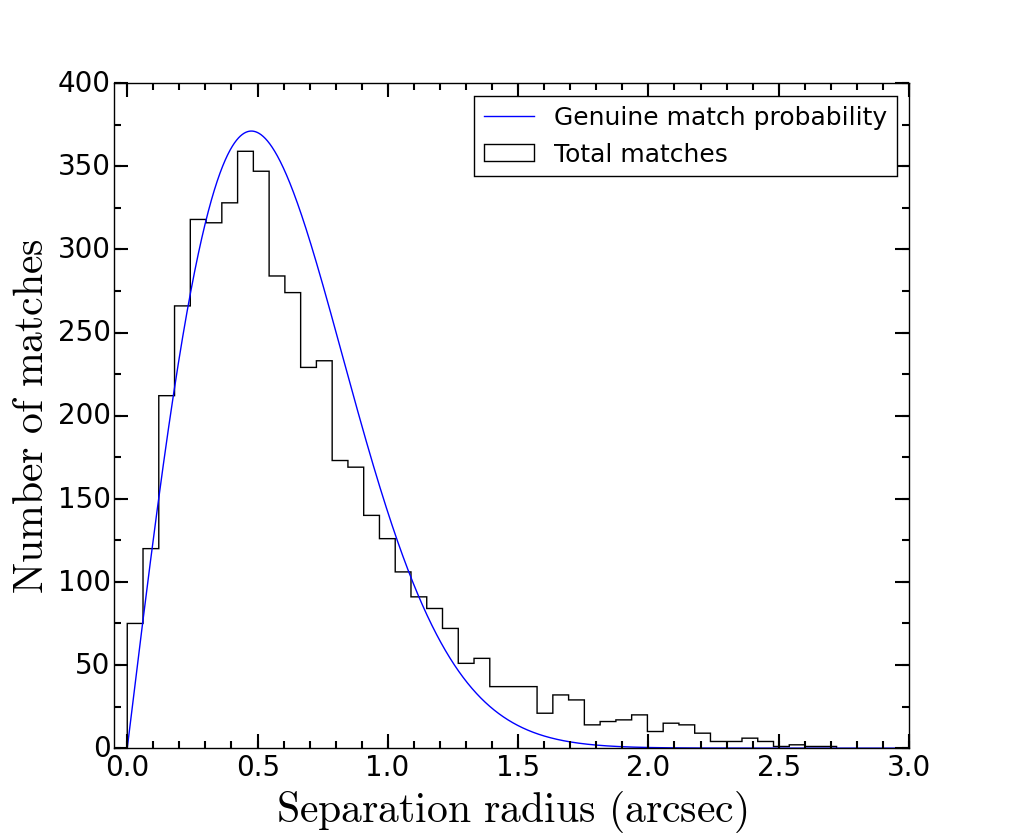}
    \caption{Distribution of the radial separations between the radio sources in XXL-S and their optical counterparts, as determined by the LR method.  The blue curve is the same curve in Figure \ref{fig:N_tot_vs_r_plot} and is given by Equation \ref{eq:gen_matches}.  The radius at the peak of the distribution ($r \approx 0.45''$) is very close to the quadrature sum of the assumed radio and optical positional uncertainties ($\sigma_{\rm{pos}} \approx 0.48''$).}
    \label{fig:N_gen_vs_r_plot}
\end{figure}

Figure \ref{fig:N_gen_vs_r_plot} shows the distribution of the radial separation between the XXL-S radio sources and their optical counterparts and the theoretical curve for the genuine number of matches (Equation \ref{eq:gen_matches}).  The theoretical curve overpredicts the number of matches in the range 0.5$''$ $\lesssim$ $r$ $\lesssim$ 1.1$''$ and underpredicts the number of matches in the range 1.1$''$ $\lesssim$ $r$ $\lesssim$ 2.75$''$, probably because   it does not take into account the varying positional uncertainty of the radio sources as a function of flux density.  The median ATCA XXL-S flux density is $\sim$0.53 mJy.  The average positional uncertainty for sources fainter than this value is $\sim$0.53$''$, and  for sources brighter than this value it is $\sim$0.22$''$.  The average radio positional uncertainty of $\sigma_{\rm{r}}=0.37''$ was assumed for simplicity.  Therefore, the distribution of radial separations contains values of larger radii than would otherwise be expected due to the large population of faint sources, leaving fewer sources at smaller radii. In addition, the LR method is more sophisticated than a simple radial cross-match, so a slight discrepancy between a theoretical curve based on one simple variable and the actual separation distribution found by the LR method is to be expected.  Nevertheless, Figure \ref{fig:N_gen_vs_r_plot} shows that the genuine matches roughly follow the Rayleigh distribution, demonstrating that the LR method is consistent with a priori expectations of cross-matching radio sources to optical counterparts.

\subsubsection{Photometric redshift accuracy}
\label{sec:photoz_accuracy}

The  number of optically matched XXL-S radio sources for which a reliable spectroscopic redshift is available is 1110.  For the remaining 
3648 sources, the extensive photometric coverage of XXL-S was utilised to compute their photometric redshifts as described in XXL Paper VI and Fotopoulou et al. (in prep.).  To summarise, a random forest machine-learning classifier was trained to distinguish between passive, star-forming, starburst, AGN and QSO galaxies according to their broadband colours ($GRIZYJHK$, plus IRAC channel 1) and their $I$-band half-light radius. Given this class assignment, the photometric redshift of each source was computed with a restricted set of templates corresponding to each galaxy class.  

\begin{figure}
        \includegraphics[width=\columnwidth]{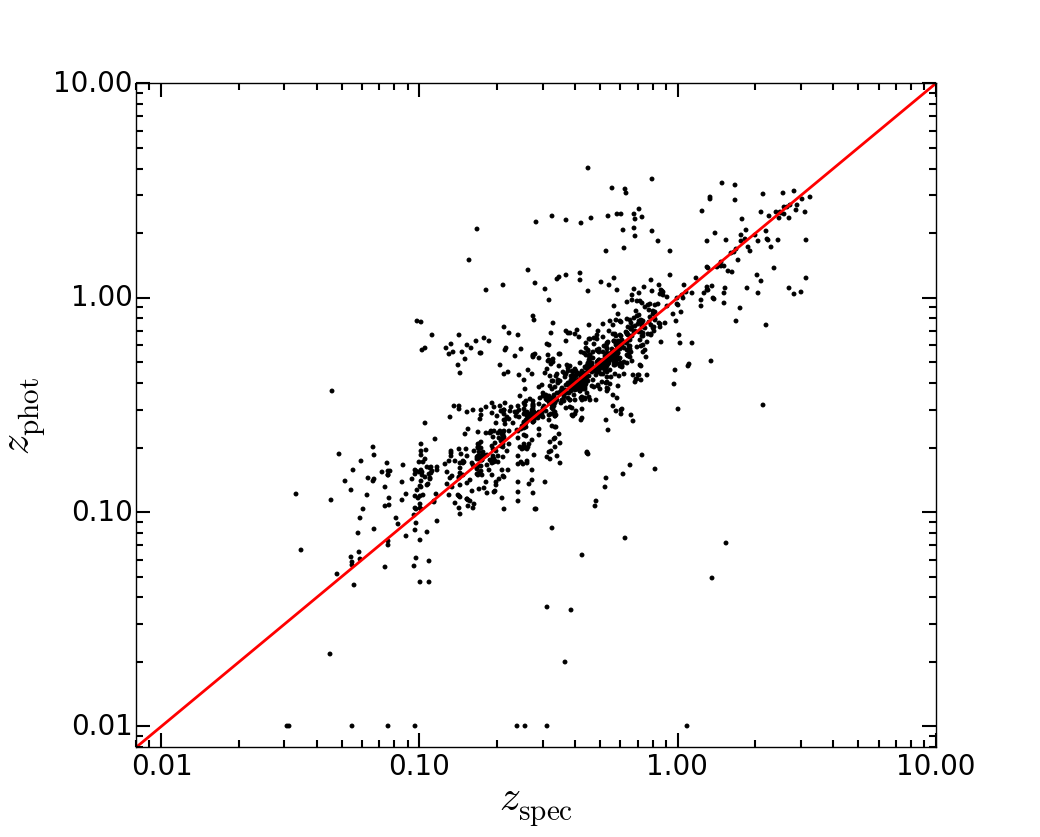}
    \caption{Comparison of $z_{\rm{phot}}$ and $z_{\rm{spec}}$ for the 1110 radio sources with a spectroscopic redshift.  The red line represents  $z_{\rm{phot}} = z_{\rm{spec}}$.  The data points for which $z_{\rm{phot}}$ = 0.01 are those whose  photometric redshift was degenerate enough to cause a poor fit.}
    \label{fig:logzphot_vs_logzspec_plot}
\end{figure}

\begin{figure}
        \includegraphics[width=\columnwidth]{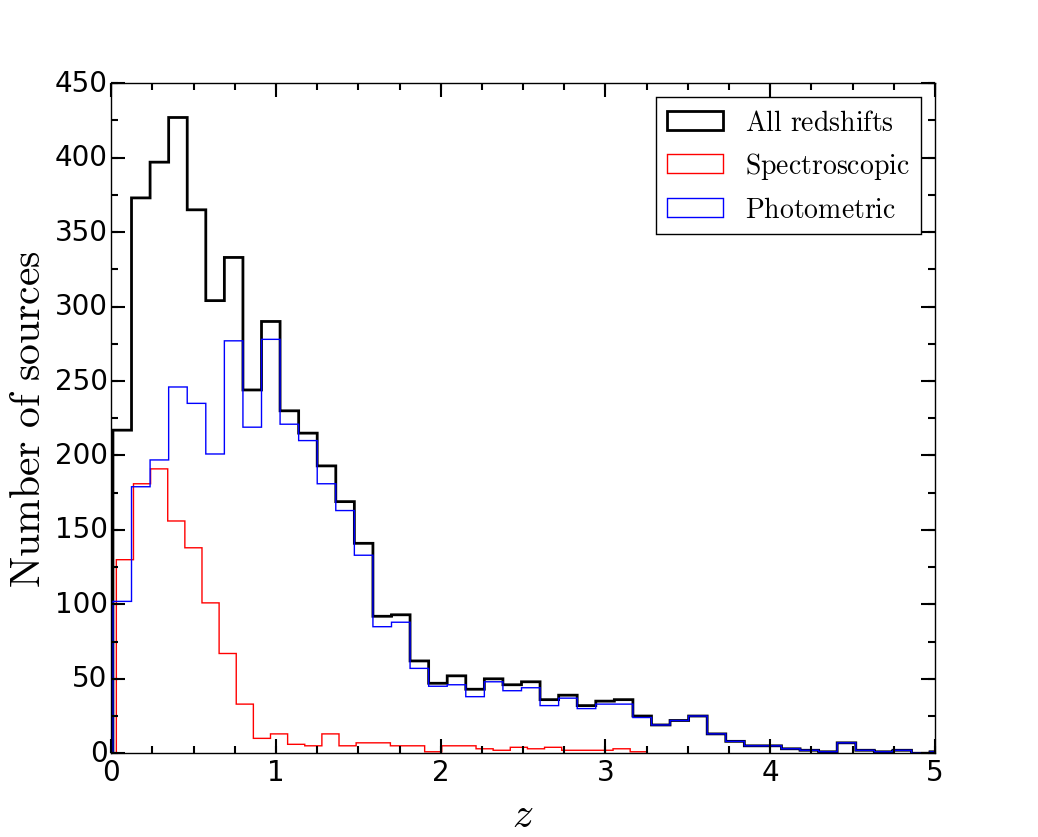}
    \caption{Redshift distribution of all the optically matched radio sources in XXL-S.  The black histogram represents all redshifts (spectroscopic redshifts were used where available).  The red line represents sources with spectroscopic redshifts and the blue line represents the remaining sources with photometric redshifts.}
    \label{fig:redshift_dist_plot}
\end{figure}

The accuracy of photometric redshifts is usually estimated with two measurements: the normalised median absolute deviation (NMAD), defined as
\begin{equation}
\sigma_{\rm{NMAD}} = 1.48 \left[\frac{| z_{\rm{phot}} - z_{\rm{spec}} |}{1 + z_{\rm{spec}}}\right]_{\rm{median}},
\end{equation}
and the fraction of catastrophic outliers $\eta$, defined as the fraction of sources with
\begin{equation}
| z_{\rm{phot}} - z_{\rm{spec}} | > 0.15 (1+z_{\rm{spec}}).
\end{equation}
For all radio sources in XXL-S, $\sigma_{\rm{NMAD}} = 0.062$ and $\eta=17\%$, and sources with $z_{\rm{spec}} > 1$ exhibit $\sigma_{\rm{NMAD}} = 0.156$. For comparison, the $\sim$2900 normal galaxies (i.e.  without X-ray detections) with reliable spectroscopic redshifts in XXL-S show $\sigma_{\rm{NMAD}} = 0.056$ and $\eta=15\%$.  
The combined emission from AGN and their host galaxies creates difficulties for photometric redshift estimations (e.g. \citealp{brodwin2006,salvato2009,assef2010}).  However, the XXL-S photometric redshift performance is close to the expectation given the use of broadband filters and the lack of deep $u$-band and near-infrared observations (Table 4 in \citealp{salvato2009}).  Figure \ref{fig:logzphot_vs_logzspec_plot} shows the comparison between the available spectroscopic redshifts and their corresponding photometric redshifts.  Figure \ref{fig:redshift_dist_plot} shows the redshift distribution of all 4758 optically matched radio sources, using  spectroscopic redshifts where available. The distribution peaks at $z \sim 0.45,$ and features a tail extending to $2 < z < 4$.

\subsubsection{Radio spectral indices and luminosities}
\label{sec:radio_alpha_L}

\begin{figure}
        \includegraphics[width=\columnwidth]{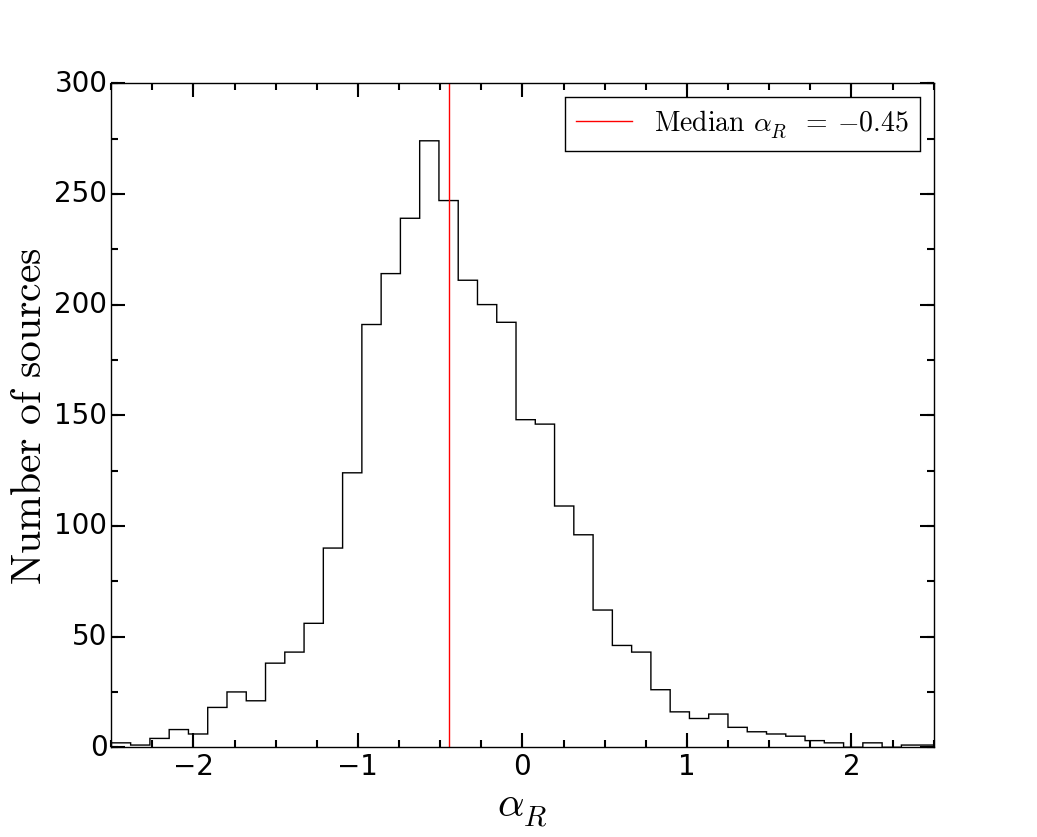}
    \caption{Distribution of ATCA 1$-$3 GHz in-band radio spectral indices ($\alpha_R$) for optically matched XXL-S radio sources for which a spectral index calculation is available.  The median spectral index  is $\alpha_R = -0.45$ (red line).}
    \label{fig:alpha_dist_det_cm_plot}
\end{figure}

\begin{figure}
        \includegraphics[width=\columnwidth]{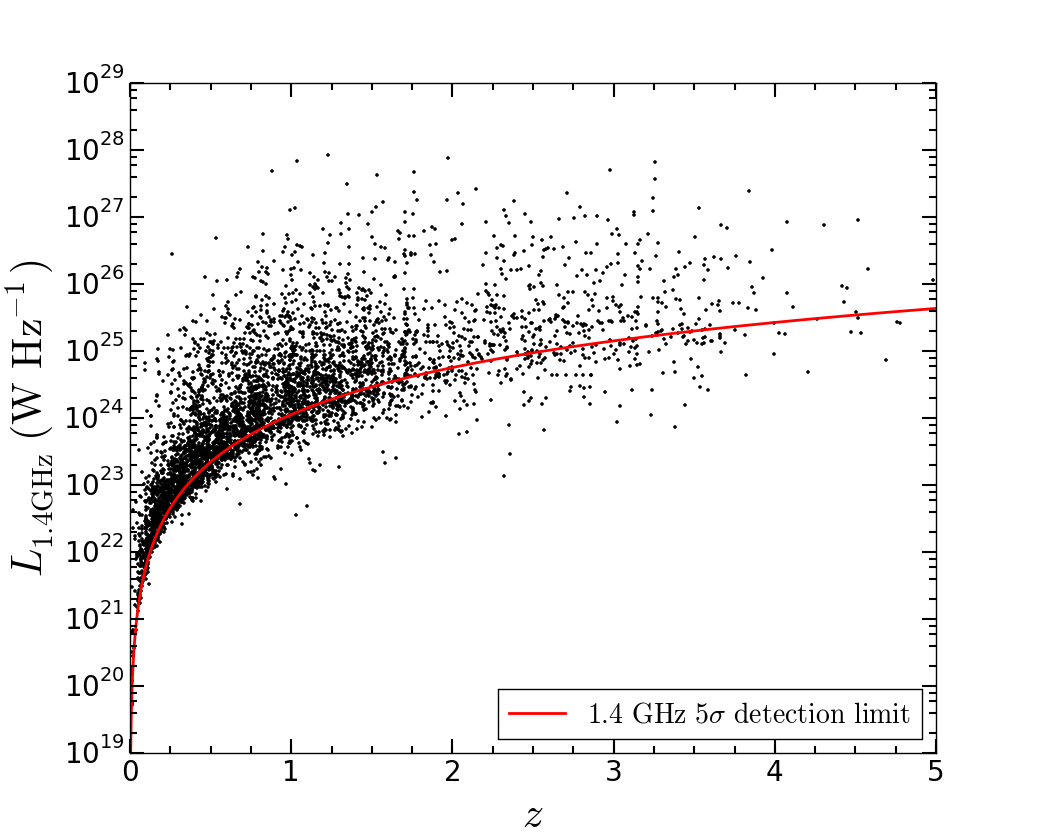}
    \caption{Rest-frame 1.4 GHz radio luminosity ($L_{\rm{1.4GHz}}$) vs. redshift for each XXL-S radio source with an optical counterpart (spectroscopic redshifts were used when available).  The ATCA in-band spectral indices were used to convert the source flux densities from 2.1 GHz to 1.4 GHz.  The red curve is the 5$\sigma$ detection limit at 1.4 GHz, derived from the 2.1 GHz 5$\sigma$ detection limit assuming a spectral index of $-$0.75.}
    \label{fig:L_R_vs_z_plot}
\end{figure}

Figure \ref{fig:alpha_dist_det_cm_plot} shows the distribution of the ATCA 1$-$3 GHz in-band radio spectral indices ($\alpha_R$) for all optically matched XXL-S radio sources for which a spectral index calculation is available (2974 sources). The median value is $\alpha_R \approx -0.45$, which is flatter than the value used in XXL Paper XVIII  for XXL-S radio sources matched to Sydney University Molonglo Sky Survey (SUMSS) 843 MHz sources ($-0.75$).  
Appendix \ref{sec:alpha_calculations} describes the reason for this flatter median $\alpha_R$ value, and  how the spectral indices were calculated and assigned.  For comparison with other studies, the rest-frame 1.4 GHz monochromatic luminosity densities (hereafter luminosities) of the sources were calculated.  First, each 2.1 GHz flux density was converted into a 1.4 GHz flux density ($S_{\rm{1.4GHz}}$) using the $\alpha_R$ value for each source. Then the 1.4 GHz luminosity of each source was computed with the  equation
\begin{equation}
\label{eq:L_R}
L_{\rm{1.4GHz}} = 4 \pi d_L^2 S_{\rm{1.4GHz}} (1+z)^{-(1+\alpha_R)},
\end{equation}
where $d_L$ is the luminosity distance in metres, $z$ is the best redshift (spectroscopic if available, otherwise photometric), and $\alpha_R$ is the ATCA 1$-$3 GHz in-band radio spectral index.
Figure \ref{fig:L_R_vs_z_plot} shows $L_{\rm{1.4GHz}}$ versus redshift for each radio source with an optical counterpart.

\section{Radio source classification}
\label{sec:agn_sfg_class}

This section describes the multiwavelength diagnostics used to identify AGN and SFGs in the final optically matched radio sample.  The diagnostics are based on X-ray luminosities and hardness ratios (Section \ref{sec:x-ray_class}); spectral energy distribution (SED) fitting (Section \ref{sec:sed_class}); MIR colours (Section \ref{sec:mir_class}); optical spectra (Section \ref{sec:spectra_class}); optical colours (Section \ref{sec:opt_colours_class}); and radio luminosities, spectral indices, and morphologies (Section \ref{sec:radio_class}).  The decision tree used to classify the sources is detailed in Section \ref{sec:decision_tree}.

\subsection{X-ray classification}
\label{sec:x-ray_class}

X-ray emission from AGN arises from inverse Compton scattering of lower energy photons by hot electrons and potentially from thermal emission at soft X-ray energies ($E$ < 2 keV) from the inner accretion disk.  In particular, multiple scatterings within the hot electron corona result in a non-thermal power-law X-ray spectra for energies in the range 1 keV $\lesssim E \lesssim$ 300 keV.  The power law has the form $F(E) \propto E^{-\Gamma}$;  $\Gamma$ is the photon index, which has a typical value of $\Gamma$ = 1.4--1.6 for radio-loud AGN and $\Gamma$ = 1.8--2.0 for radio-quiet AGN (e.g. \citealp{reeves2000,piconcelli2005,page2005}). On the other hand, \cite{bauer2004} observed that the X-ray emission from local SFGs can be characterised by luminosities below $L_{0.5-8\rm{keV}} = 3 \times 10^{42}$ erg s$^{-1}$, and \cite{colbert2004} found that SFGs are typically characterised by relatively soft emission with hardness ratios\footnote{In this paper, $HR = (S_{\rm{hard}} - S_{\rm{soft}}) / (S_{\rm{hard}} + S_{\rm{soft}})$} of $HR < -0.1$ (corresponding to effective photon indices of $\Gamma < 1$). Therefore, following \cite{juneau2013}, each X-ray source was classified as an X-ray AGN if it had either $L_{2-10\rm{keV}}$ > 10$^{42}$ erg s$^{-1}$ or $HR > -0.1$.

\begin{figure}
        \includegraphics[width=\columnwidth]{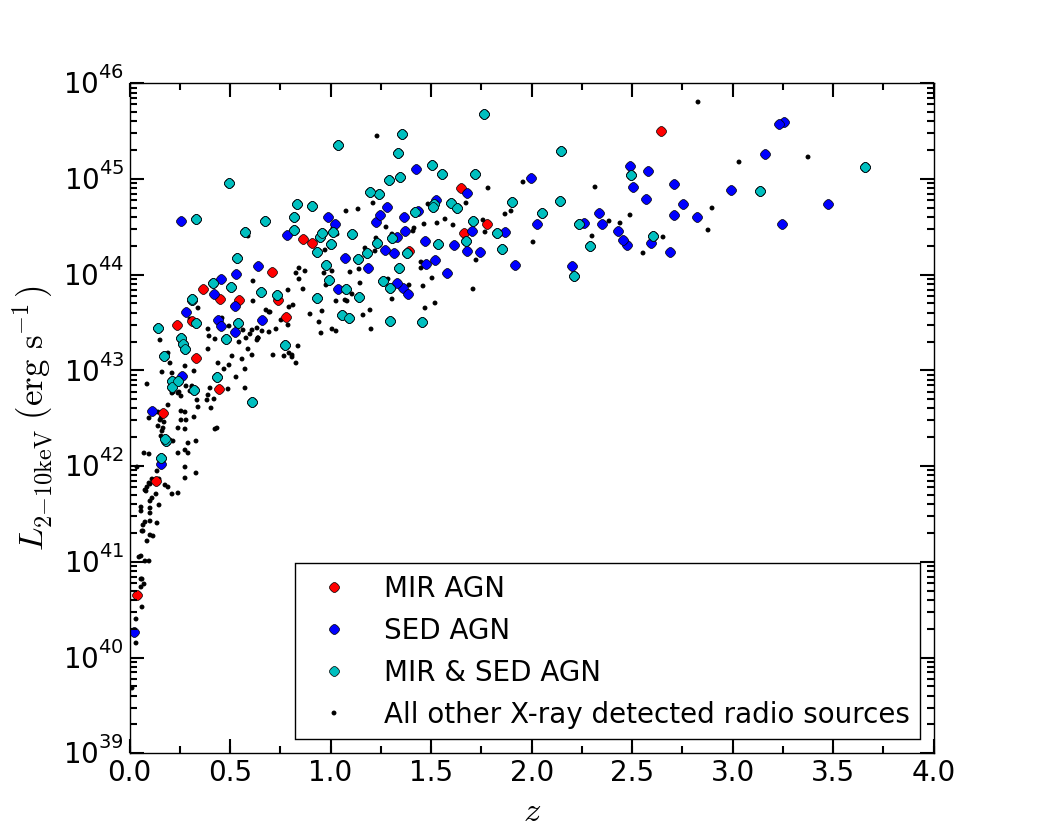}
    \caption{X-ray luminosity in the 2$-$10 keV band ($L_{\rm{2-10keV}}$) vs. redshift for XXL-S radio sources detected in at least one X-ray band.  The red circles are galaxies classified as AGN according to their MIR colours (see Section \ref{sec:mir_class}), the blue circles are sources classified as AGN according to their SED fit (see Section \ref{sec:sed_class}), the cyan circles are both MIR and SED AGN, and the black points are all other X-ray detected radio sources.}
    \label{fig:L_X_vs_z_plot}
\end{figure}

There are 414 optically matched XXL-S radio sources with an X-ray counterpart that had flux measurements in at least one \emph{XMM} band. If a source had flux measurements in both the 0.5$-$2 keV and 2$-$10 keV bands, its X-ray spectral index, $\alpha_X$, was calculated according to 
\begin{equation}
\alpha_X = \frac{\log (S_{2-10\rm{keV}} / S_{0.5-2\rm{keV}})}{\log (E_{2-10\rm{keV}} / E_{0.5-2\rm{keV}})}
\end{equation}
\citep{schmidt1986}, where $E_{0.5-2\rm{keV}}$ and $E_{2-10\rm{keV}}$ are the average energies in the 0.5$-$2 keV and 2$-$10 keV bands, or 1.25 keV and 6.0 keV, respectively.  This was converted into the X-ray photon index, given by
\begin{equation}
\Gamma = \alpha_X + 1
\end{equation}
The median value of $\Gamma$ for sources with a detection in both bands is $\Gamma_{\rm{med}}$ = 1.56.  The X-ray luminosity $L_{\rm{2-10keV}}$ was then calculated for each source according to the  equation from \cite{xue2011}, 
\begin{equation}
L_{\rm{2-10keV}} = 4 \pi d_L^2 S_{2-10\rm{keV}} (1 + z)^{\Gamma - 2},
\end{equation}
where $d_L$ is the luminosity distance in cm, $S_{2-10\rm{keV}}$ is the observed flux in erg s$^{-1}$ cm$^{-2}$ in the 2$-$10 keV band, and $z$ is the best redshift (spectroscopic if available, otherwise photometric).  The observed flux is used because the fluxes in this sample are not high enough to constrain the amount of X-ray absorption. However, applying an absorption correction would not change the number of sources classified as X-ray AGN by a significant amount, and therefore using the observed X-ray fluxes has a negligible effect on the classification results. For sources with a detection in only one band, $\Gamma_{\rm{med}}$ was assumed for the calculation of their luminosities and hardness ratios.  According to \cite{xue2011}, luminosities calculated in this way agree with the values derived from direct spectral fitting to within $\sim$30\% (higher values are  possible for heavily obscured AGN). However, the number of X-ray sources classified as X-ray AGN according to the criteria 
mentioned above is so high (412 out of 414, or 99.5\%) that 
an increase in the accuracy of the X-ray luminosities 
would not have a significant effect on the final classification results. Figure \ref{fig:L_X_vs_z_plot} shows a plot of $L_{\rm{2-10keV}}$ versus redshift for the 414 X-ray detected radio sources.  

\subsection{SED classification}
\label{sec:sed_class}

The photometry of each radio source from optical to millimetre wavelengths was fit to identify possible evidence of AGN activity on the basis of a panchromatic SED-fitting analysis. The SED-fitting code {\sc magphys}\footnote{\url{http://www.iap.fr/magphys/magphys/MAGPHYS.html}} \citep{dacunha2008} and the three-component SED-fitting code {\sc sed3fit}\footnote{\url{http://cosmos.astro.caltech.edu/page/other-tools}} by \cite{berta2013}, which accounts for an additional AGN contribution, were used to do this.

The {\sc magphys} code is designed to reproduce a variety of galaxy SEDs, and relies on the energy balance between the dust-absorbed stellar continuum and the reprocessed dust emission at infrared wavelengths.  However, it does not account for a possible AGN emission component. The {\sc sed3fit} code combines the emission from stars, dust heated by star formation, and an AGN (continuum and torus) component from the library of \cite{feltre2012} and \cite{fritz2006}, resulting in a simultaneous three-component fit. In order to quantify the relative likelihood of a possible AGN component, the approach from \cite{delvecchio2014} was followed: each source's SED was fit with both the {\sc magphys} and {\sc sed3fit} codes. The reduced $\chi^2$ values of the SED fits with and without the AGN were calculated.  If the reduced $\chi^2$ of the fit with the AGN was significantly smaller than that of the fit without the AGN (at the $\geq$ 99\% confidence level on the basis of a Fisher test, as found in \citealp{bevington2003}), then that source was considered to have a significant AGN component in its SED.

\begin{figure}
        \includegraphics[width=\columnwidth]{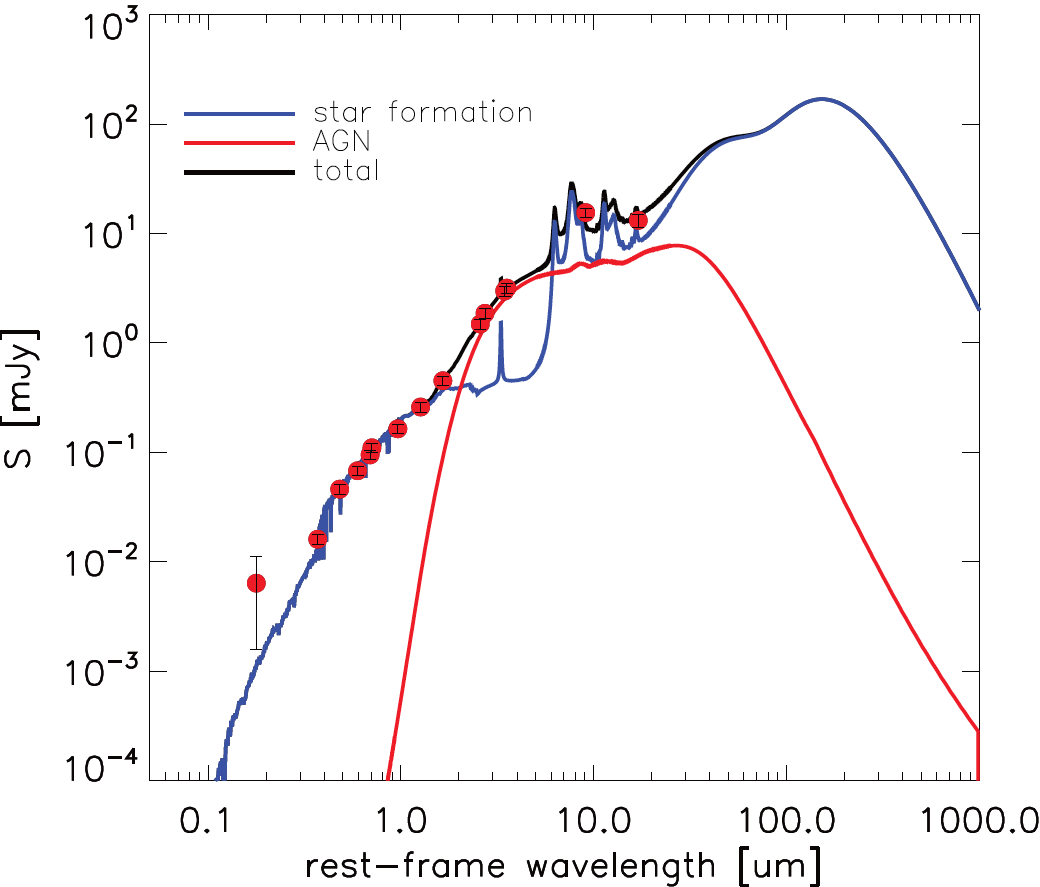}
    \caption{Example of a source that required a significant AGN component in its SED fit.  The blue line is the stellar component, the red line is the AGN component, and the black line is the total SED.}
    \label{fig:sed_example}
\end{figure}

This approach resulted in the identification of 988 radio sources that are considered AGN on the basis of their SED fits.  The remaining 3770 sources were classified as non-AGN according to their SED fit.  Figure \ref{fig:sed_example} shows an example of a source at $z=0.301$ that required a significant AGN component to fit its SED. 

\subsection{Mid-infrared classification}
\label{sec:mir_class}

At MIR wavelengths of $\lambda$ = 5--10 $\mu$m, the MIR SEDs of SFGs are dominated by broadband emission from polycyclic aromatic hydrocarbons (PAHs), which exhibit emission features mainly at 3.3, 6.2, 7.7, 8.6, and 11.3 $\mu$m.  At longer wavelengths of $\lambda$~>~20$\mu$m, the SEDs are dominated by thermal continuum emission from larger warm dust particles \citep{draine2003}.  If an AGN is present, they can also contain higher temperature dust at 50~$-$~300~K, whose radiation peaks at $\lambda$ $\sim$ 20 $\mu$m.  AGN can also destroy PAH molecules via their strong UV radiation, thus damping PAH emission.  This results in a decrease in the relative contribution of PAH to MIR continuum emission for warmer galaxies, or lower PAH equivalent widths \citep{elbaz2011}.  Therefore, examining the MIR emission of galaxies at wavelengths corresponding to PAH emission features and the blackbody radiation of warm dust can assist in identifying AGN.  The MIR AGN indicators used in this paper are explained in the next three subsections.  A source was considered a MIR AGN if at least one of the MIR criteria was met.

\subsubsection{WISE colour-colour diagrams}

In general, AGN exhibit redder MIR colours than normal galaxies. For example, the WISE AGN wedges in \cite{mateos2012} are defined to include all objects with MIR colours expected for power-law SEDs with spectral index $\alpha_{\rm{IR}} \leq -0.3$.  Assuming that X-ray luminosity in the 2$-$10 keV band is a good tracer of AGN activity, \cite{mateos2012} demonstrated that their WISE selection has one of the highest reliabilities  of all the MIR AGN indicators in the literature, with minimal contamination from high-redshift SFGs.  Therefore, the \cite{mateos2012} wedges for $W1-W2$ versus $W2-W3,$ and $W1-W2$ versus $W3-W4$ were chosen to select MIR AGN.

The $W1-W2$ versus $W2-W3$ wedge is defined by the intersection of the following regions \citep{mateos2012}: 
\begin{enumerate}
\item Top edge: $W1 - W2 < 0.315 (W2 - W3) + 0.796$
\item Left edge: $W1 - W2 > -3.172 (W2 - W3) + 7.624$
\item Bottom edge: $W1 - W2 > 0.315 (W2 - W3) - 0.222$
\end{enumerate}

\begin{figure}
        \includegraphics[width=\columnwidth]{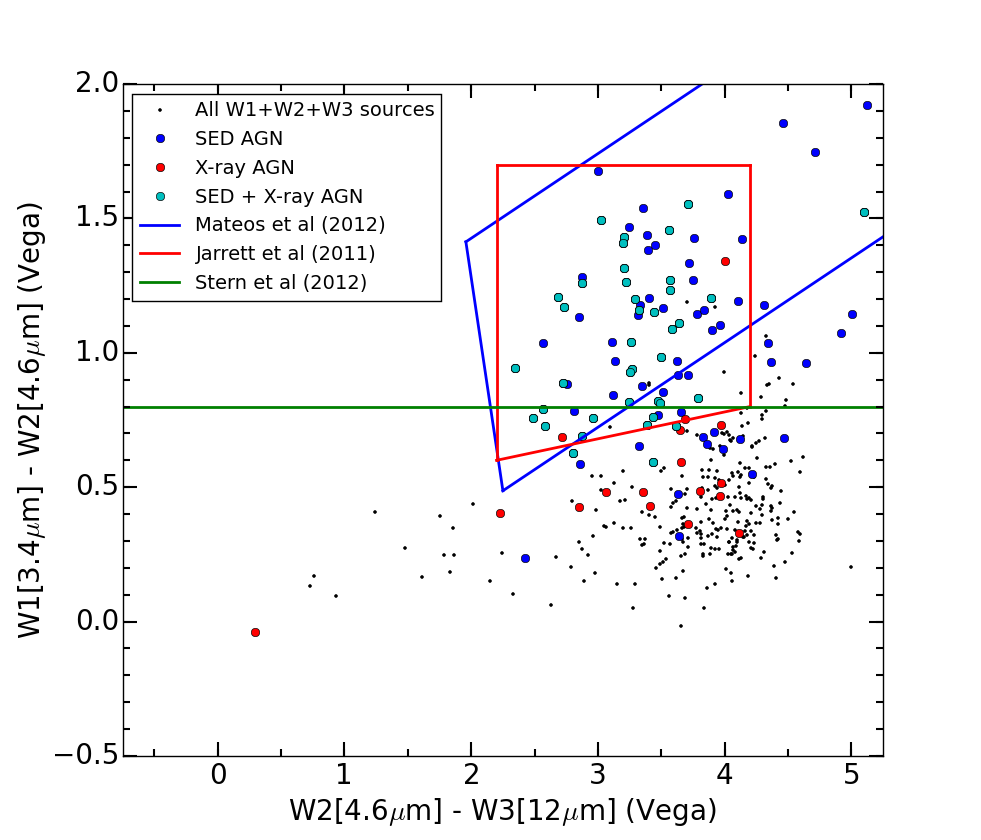}
    \caption{WISE $W1-W2$ vs. $W2-W3$ diagram for XXL-S radio sources.  The blue circles are galaxies that have been classified as AGN according to their SED fit, the red circles  galaxies that have been classified as X-ray AGN, and the cyan circles  both SED and X-ray AGN.  The blue \cite{mateos2012} wedge is the indicator used in this work.  For comparison, the \cite{jarrett2011} box (red) and the \cite{stern2012} line (green) are shown.}
    \label{fig:W1-W2_vs_W2-W3_plot}
\end{figure}

Figure \ref{fig:W1-W2_vs_W2-W3_plot} shows the $W1-W2$ versus $W2-W3$ plot for the XXL-S radio sources with $W1$, $W2$, and $W3$ data available. Of the 393 sources with WISE bands 1--3 available, 71 were classified as MIR AGN.

The $W1-W2$ versus $W3-W4$ wedge is described by the intersection of the following regions \citep{mateos2012}:
\begin{enumerate}
\item Top edge: $W1 - W2 = 0.5 (W3 - W4) + 0.979$
\item Left edge: $W1 - W2 = -2 (W3 - W4) + 4.33$
\item Bottom edge: $W1 - W2 = 0.5 (W3 - W4) - 0.405$
\end{enumerate}

\begin{figure}
        \includegraphics[width=\columnwidth]{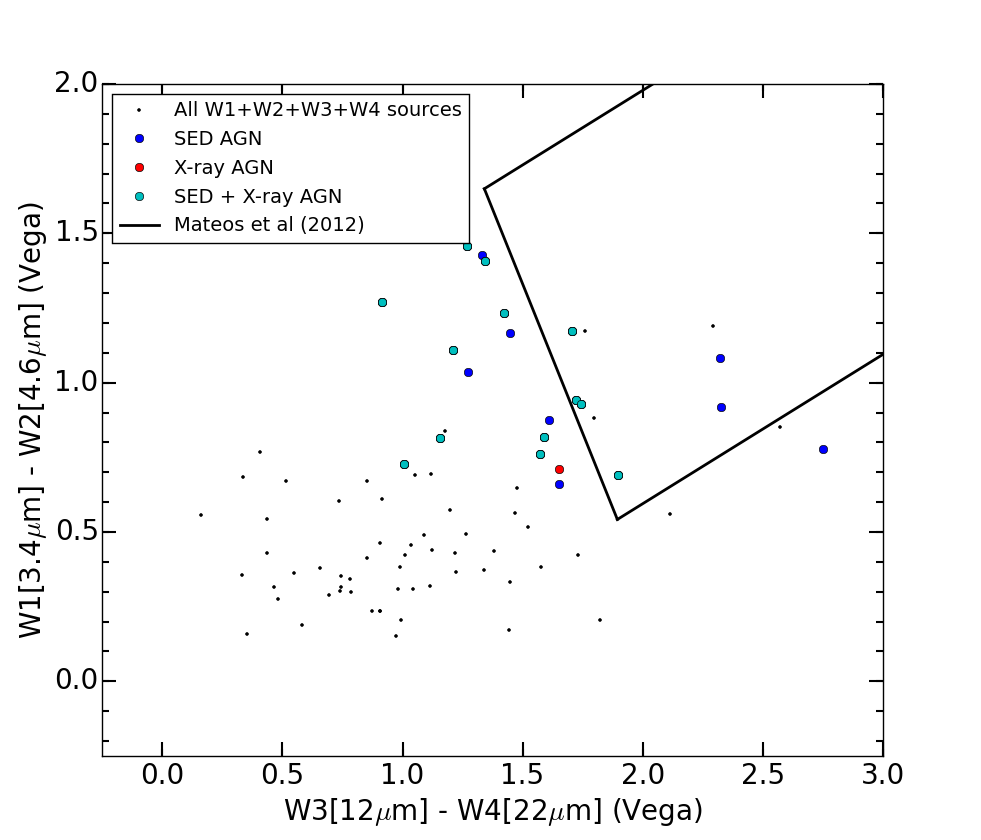}
    \caption{WISE $W1-W2$ vs. $W3-W4$ diagram for XXL-S radio sources.  The blue circles are galaxies that have been classified as AGN according to their SED fit, the red circles  galaxies that have been classified as X-ray AGN, and the cyan circles  both SED and X-ray AGN.}
    \label{fig:W1-W2_vs_W3-W4_plot}
\end{figure}

Figure \ref{fig:W1-W2_vs_W3-W4_plot} shows the WISE $W1-W2$ versus $W3-W4$ diagram for the XXL-S radio sources with $W1$, $W2$, $W3,$ and $W4$ data available.  Of the 82 radio sources with WISE bands 1--4 available, 9 were considered MIR AGN.

\subsubsection{WISE colour-magnitude diagram}

Only a small percentage (8\%) of the XXL-S optically matched radio sources had at least WISE bands 1--3 available, yet 1375 had at least WISE bands 1 and 2. Therefore, a method was sought to include these sources in the analysis.  Previous studies of MIR AGN selection  relied on a simple colour cut in $W1-W2$ (e.g. \citealp{stern2012}).  Those samples were brighter than the MIR sources in the XXL-S optically matched sample, so in order to maintain reliability for fainter sources, a magnitude-dependent $W1-W2$ colour selection is needed.  \cite{assef2013} derived such a selection for WISE AGN in the NDWFS Bo\"{o}tes field, and thus the AGN selection criteria for $W1$ and $W2$ from that work are used in this paper.  

\begin{figure}
        \includegraphics[width=\columnwidth]{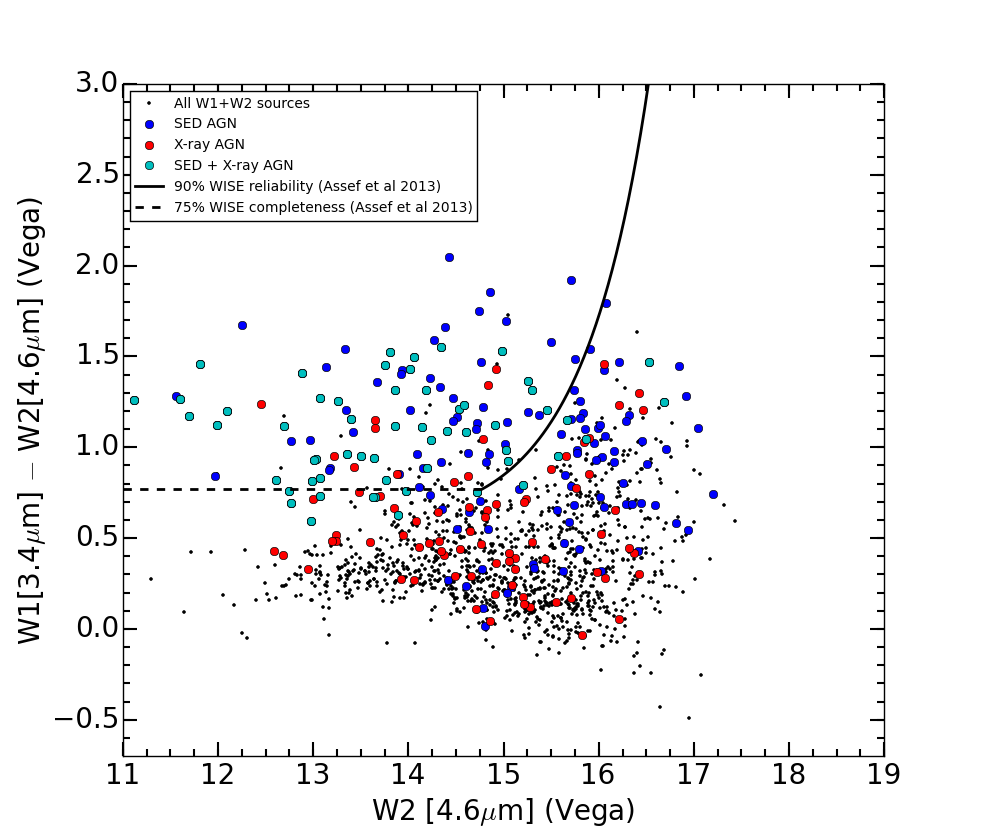}
    \caption{WISE $W1 - W2$ vs. $W2$ 
    for XXL-S radio sources.  The blue circles are galaxies that have been classified as AGN according to their SED fit, the red circles  galaxies that have been classified as X-ray AGN, and the cyan circles  both SED and X-ray AGN.  The solid curve is the 90\% WISE reliability curve (Equation \ref{eq:assef}) and the dashed line is the 75\% WISE completeness line ($W1 - W2$ = 0.77), both from \cite{assef2013}.}
    \label{fig:W1-W2_vs_W2_plot}
\end{figure}

Figure \ref{fig:W1-W2_vs_W2_plot} shows the WISE colour-magnitude diagram ($W1 - W2$ versus $W2$)
for XXL-S radio sources with $W1$ and $W2$ data available.  The 75\% completeness line ($W1-W2$ = 0.77, shown as the dashed line in Figure \ref{fig:W1-W2_vs_W2_plot}), and the 90\% reliability curve (solid curve in Figure \ref{fig:W1-W2_vs_W2_plot}) were used to select MIR AGN.  The latter is described by
\begin{equation}
\label{eq:assef}
W1 - W2 > \alpha \ \rm{exp}[\beta (W2 - \gamma)^2]
,\end{equation}
where $\alpha$ = 0.662, $\beta$ = 0.232, and $\gamma$ = 13.97 \citep{assef2013}.
The two lines intersect at $W2$ = 14.777.  
This magnitude limit and the colour cut of $W1-W2$ = 0.77 is very similar to the \cite{stern2012} colour cut ($W1-W2 \geq$ 0.8 for $W2$ < 15.05).  Therefore, if the brighter MIR sources ($W$2 < 14.777) had $W1 - W2$ > 0.77, they were classified as MIR AGN.  If the fainter MIR sources ($W2$ > 14.777) were above the curve described in Equation \ref{eq:assef}, they were also classified as MIR AGN.  Of the 1375 sources with $W1$ and $W2$ data available, 129 were identified as MIR AGN.

\subsubsection{IRAC colour-magnitude diagram}

\begin{figure}
        \includegraphics[width=\columnwidth]{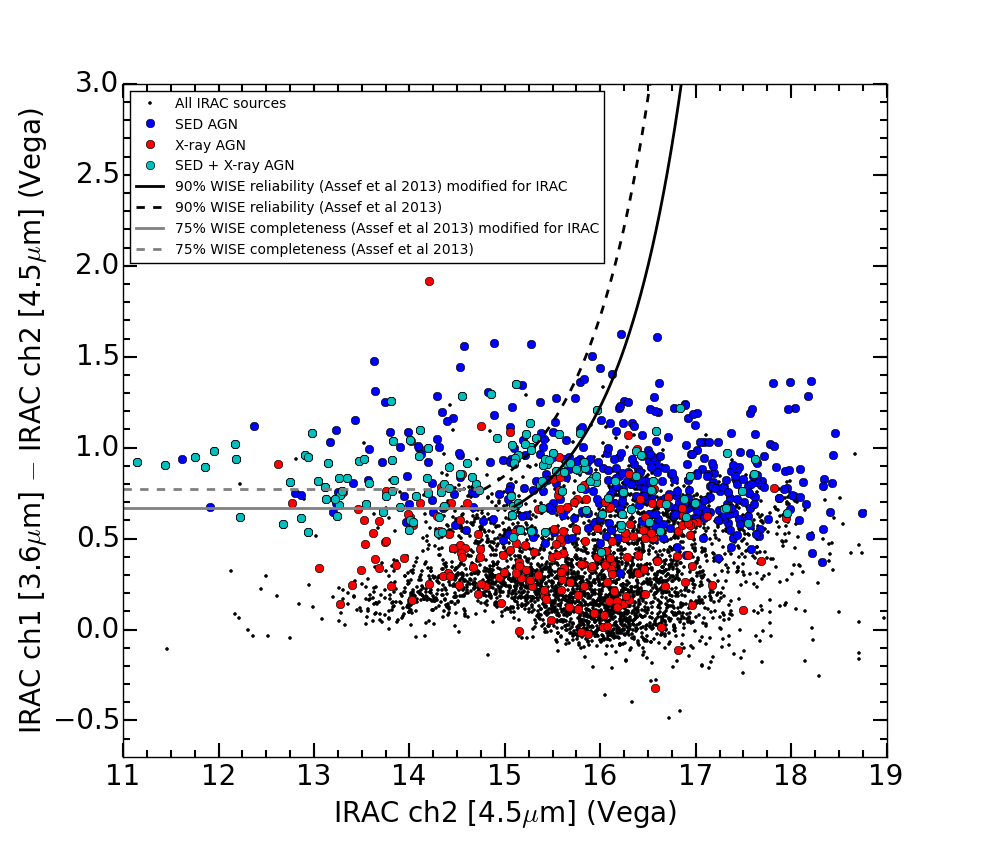}
    \caption{$IRAC1 - IRAC2$ vs. $IRAC2$ for XXL-S radio sources.  The red circles are galaxies that have been classified as AGN according to their SED fit, the blue circles  galaxies that have been classified as X-ray AGN, and the cyan circles  both SED and X-ray AGN.  The solid lines are the \cite{assef2013} lines modified for use with the IRAC data ($IRAC1 - IRAC2$ = 0.67 and Equation \ref{eq:assef_irac}).  For comparison, the original \cite{assef2013} curves for WISE are shown as dashed lines.}
    \label{fig:IRAC1-IRAC2_vs_IRAC2_plot}
\end{figure}

IRAC colours using all four channels are commonly used to select AGN (e.g. \citealp{donley2012}).  However, only IRAC channels 1 and 2 (hereafter $IRAC1$ and $IRAC2$) are available for XXL-S.  Nevertheless, the central wavelengths and spectral response functions for $IRAC1$ and $IRAC2$ are very similar to $W1$ and $W2$, respectively.  Additionally, IRAC data is deeper than WISE, and therefore there are more sources with IRAC data available (4023  of the 4758 optically matched radio sources, as opposed to 1375 with $W1$ and $W2$).  Therefore, in order to make the most of the IRAC MIR data available for XXL-S, a colour-magnitude diagram similar to the one presented in the previous section was 
adopted for the $IRAC1$ and $IRAC2$ data.  Figure \ref{fig:IRAC1-IRAC2_vs_IRAC2_plot} shows the IRAC colour-magnitude diagram ($IRAC1 - IRAC2$ versus $IRAC2$) for sources with IRAC data available in XXL-S.

In order to find the appropriate adaptation of the \cite{assef2013} criteria to the IRAC data, the IRAC and WISE colours were compared.  The median value for $W1-IRAC1$ is $-0.2$, and  for $W2-IRAC2$ it is $-0.3$ (these offsets are roughly constant with magnitude).  Therefore, the median difference between $W1~-~W2$ and $IRAC1-IRAC2$ is 0.1.  Inserting these values into Equation \ref{eq:assef} resulted in
\begin{equation}
\label{eq:assef_irac}
IRAC1 - IRAC2 = \alpha \ \rm{exp}[\beta (\emph{IRAC}2 - 0.3 - \gamma)^2] - 0.1
,\end{equation}
where $\alpha$, $\beta$, and $\gamma$ have the same values as for the WISE data.

Since $(IRAC1-IRAC2) - (W1-W2) = -0.1$, the 75\% completeness line was modified to be $IRAC1 - IRAC2$ = 0.67.  This line and the curve described in Equation \ref{eq:assef_irac} intersect at $IRAC2$ = 15.077.  Any source brighter than this was classified as a MIR AGN if it had $IRAC1 - IRAC2$ > 0.67, and any source fainter than this was considered a MIR AGN if its $IRAC1-IRAC2$ value was above the curve in Equation \ref{eq:assef_irac}.  Of the 4023 optically matched XXL-S radio sources with IRAC data available, 251 were classified as MIR AGN.

\subsubsection{Overlap between MIR AGN indicators}

\begin{figure}
        \includegraphics[width=\columnwidth]{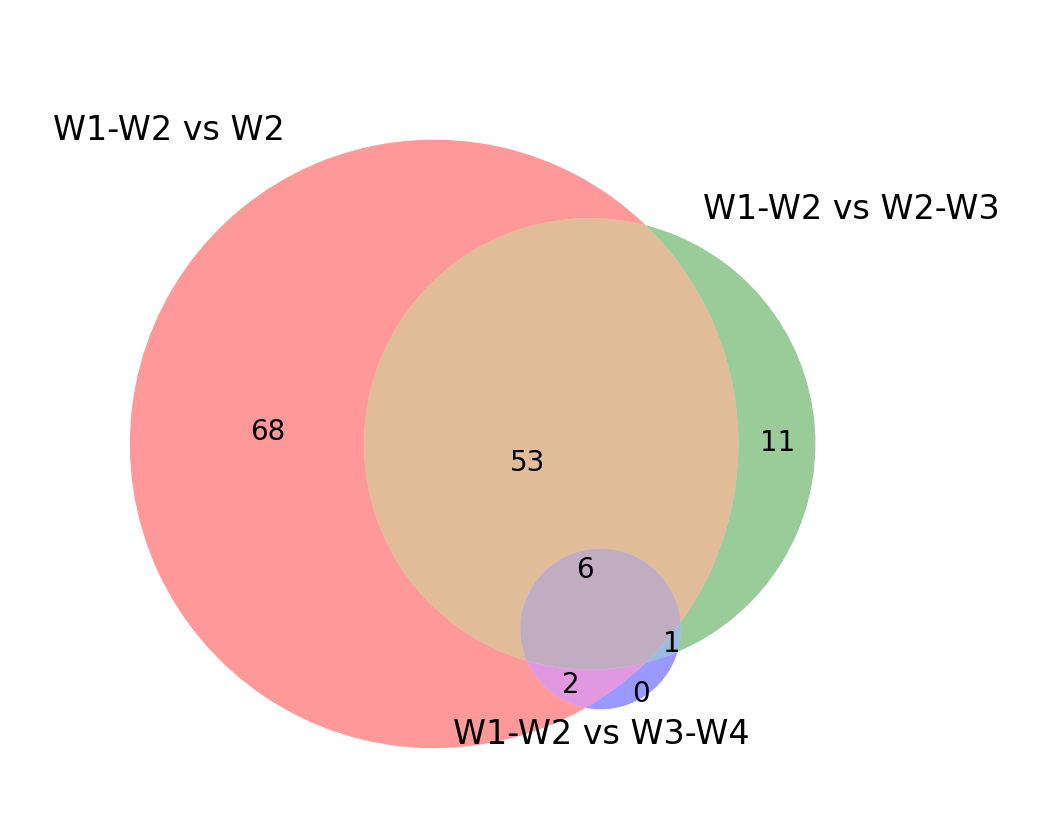}
    \caption{Venn diagram showing the overlap between the AGN classified by the three WISE AGN indicators.}
    \label{fig:venn_wise_agn_plot}
\end{figure}

\begin{figure}
        \includegraphics[width=\columnwidth]{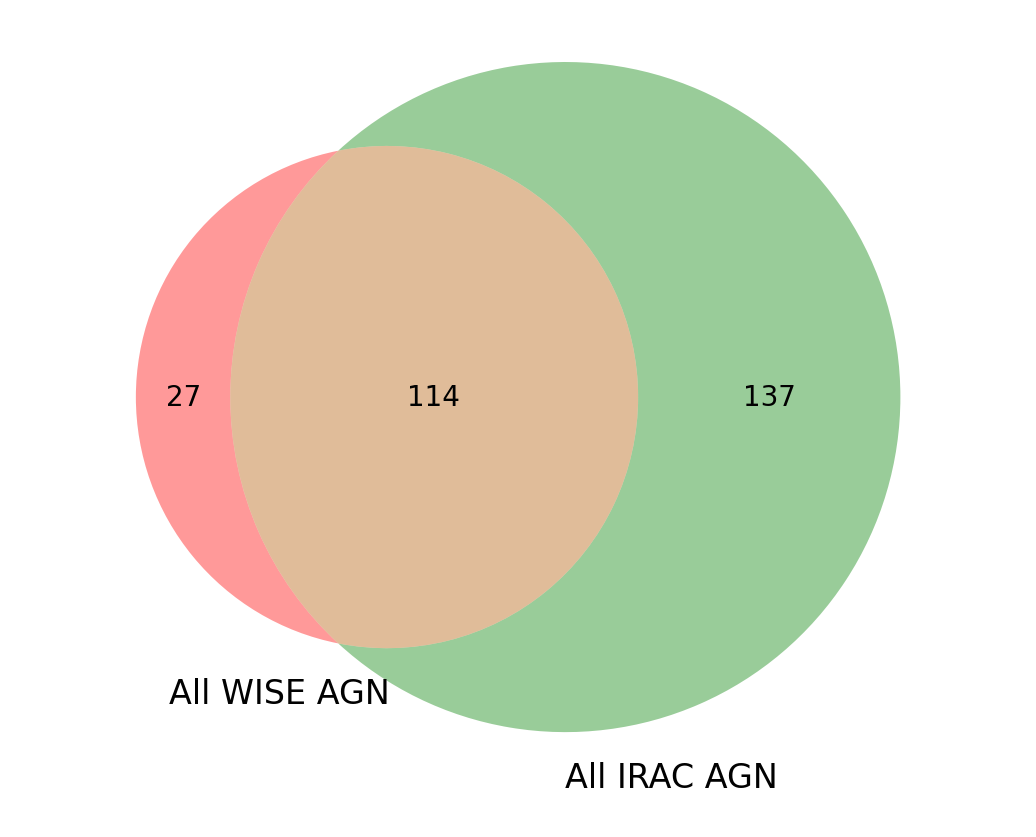}
    \caption{Venn diagram showing the overlap between the AGN identified by any of the three WISE AGN indicators and the IRAC AGN indicator.}
    \label{fig:venn_irac_wise_agn_plot}
\end{figure}

In summary, 278 optically matched radio sources were identified as AGN on the basis of their MIR properties (141 according to the WISE indicators and 251 according to the IRAC indicator). Figure \ref{fig:venn_wise_agn_plot} shows the overlap between the three WISE AGN indicators. It demonstrates that the $W1-W2$ versus $W2$ MIR AGN indicator misses only 12 galaxies that are classified as MIR AGN by the other WISE indicators. In other words, the majority of sources only have  $W1$ and $W2$ data available, but this is sufficient to capture most of the WISE AGN in the field.  Figure \ref{fig:venn_irac_wise_agn_plot} shows that there are 114 MIR sources that were identified as AGN according to any of the WISE and IRAC indicators.  This means that 114 out of 141 (80.9\%) of the WISE MIR AGN have $IRAC1-IRAC2$ colours such that they were also classified as AGN according to the $IRAC$-modified \cite{assef2013} indicator. These Venn diagrams show that the IRAC and WISE AGN indicators are consistent with each other.

\subsection{Optical spectra classification}
\label{sec:spectra_class}

The two AAT observing runs described in Section \ref{sec:spectra} added 6473 reliable spectroscopic redshifts to the data available for XXL-S, 1098 (23.3\%) of which corresponded to non-stellar optical counterparts of XXL-S radio sources.  Twelve additional sources had spectroscopic redshifts available from the other non-AAT surveys referred to in Section \ref{sec:spectra}.  In total, 1110 XXL-S radio sources have spectroscopic redshifts. 

\subsubsection{BPT diagram}

The Baldwin-Phillips-Terlevich (BPT) diagram \citep{baldwin1981} was designed to distinguish between AGN and SFGs by utilising the ratios between emission lines generated by different ionisation mechanisms. 
They found that defining AGN and SFG regions in a log([\ion{O}{III}]/H$\beta$) versus log([\ion{N}{II}]/H$\alpha$) plot was an excellent way to identify AGN. However, for XXL-S sources with $z \gtrsim 0.36$, the [\ion{N}{II}] line ($\lambda$=658.527 nm) is outside the effective wavelength range of the AAOmega data ($\sim$375-893 nm), so only sources with $z \lesssim 0.36$ were able to have the BPT diagnostic examined.

There are 522 sources with $z < 0.36$ for which spectra are available.  The software \textsc{robospect} \citep{waters2013} was used to measure the equivalent widths (EW) of the [\ion{O}{III}], H$\beta$, [\ion{N}{II}], and H$\alpha$ lines in the spectra for each of these sources.  A signal-to-noise ratio of $S/N > 3$ was required for each line to qualify a source for the BPT diagram.  Of the 522 spectra with $z < 0.36$, 156 had [\ion{O}{III}], H$\beta$, [\ion{N}{II}], and H$\alpha$ emission line EWs with $S/N > 3$.  The errors in log([\ion{O}{III}]/H$\beta$) and log([\ion{N}{II}]/H$\alpha$) were computed according to
\begin{equation}
\label{eq:OIII_Hb_line_errors}
\sigma_{\rm{log([\ion{O}{III}]/H}\beta\rm{)}} = \sqrt{\left(\frac{\sigma_{\rm{EW([\ion{O}{III}])}}}{\rm{EW([\ion{O}{III}])}}\right)^2 + \left(\frac{\sigma_{\rm{EW(H}\beta\rm{)}}}{\rm{EW(H}\beta\rm{)}}\right)^2} \Big/ \rm{ln}(10)
\end{equation}
\begin{equation}
\label{eq:NII_Ha_line_errors}
\sigma_{\rm{log([\ion{N}{II}]/H}\alpha\rm{)}} = \sqrt{\left(\frac{\sigma_{\rm{EW([\ion{N}{II}])}}}{\rm{EW([\ion{N}{II}])}}\right)^2 + \left(\frac{\sigma_{\rm{EW(H}\alpha\rm{)}}}{\rm{EW(H}\alpha\rm{)}}\right)^2} \Big/ \rm{ln}(10),
\end{equation}
where $\sigma_{\rm{EW([\ion{O}{III}])}}$, etc., are the uncertainties in the equivalent widths of the corresponding lines.

\begin{figure}
        \includegraphics[width=\columnwidth]{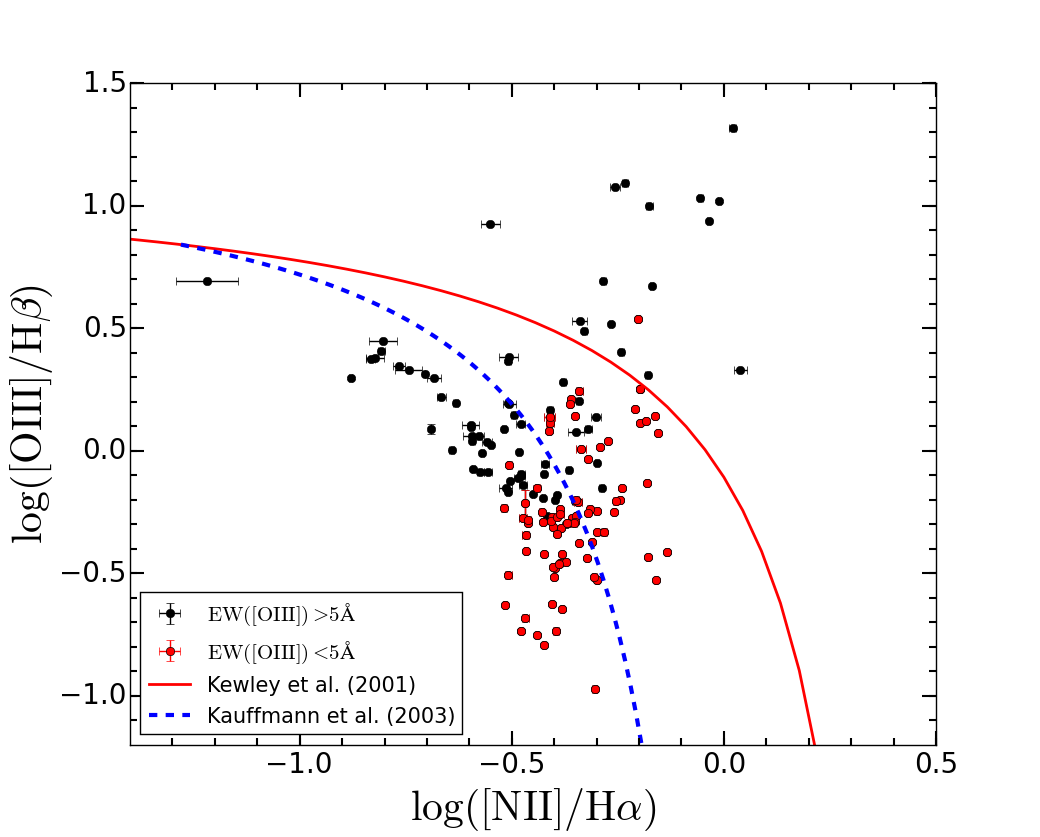}
    \caption{BPT diagram for radio sources in XXL-S.  The red points are those for which EW([\ion{O}{III}]) < 5 $\AA$, and the black points are those for which EW([\ion{O}{III}]) > 5 $\AA$.  The error bars were calculated according to Equations \ref{eq:OIII_Hb_line_errors} and \ref{eq:NII_Ha_line_errors}.  The solid red line is the extreme starburst line from \cite{kewley2001b} and the dashed blue line is the star-forming galaxy line from \cite{kauffmann2003}.}
    \label{fig:bpt_plot}
\end{figure}

Figure \ref{fig:bpt_plot} shows the BPT diagram for the XXL-S radio sources at $z < 0.36$. The solid red line in the figure is the SFG--AGN dividing line from \cite{kewley2001b} and the dashed blue line is the updated version of that demarcation from \cite{kauffmann2003}. Taking into account the errors in log([\ion{O}{III}]/H$\beta$) and log([\ion{N}{II}]/H$\alpha$), any sources that are above the solid red line are considered AGN (HERG or LERG), and any that are below the dashed blue line are consistent with pure SFGs. Sources in between the two lines are considered composite objects with contributions from both AGN and star formation \citep{kauffmann2003}.  Therefore, these sources are unclassifiable by the BPT diagram.  There are 17 sources that are considered AGN and 92 consistent with SFGs according to this diagram, with 47 unclassifiable sources.  Since it is still possible for LERGs and SFGs to exhibit  high-excitation lines, no separation between HERGs, LERGs, and SFGs could be made using the BPT diagram alone.  The equivalent width of the [\ion{O}{III}] line is necessary for this task.

\subsubsection{[OIII] line luminosity}

The [\ion{O}{III}] line ($\lambda$=500.824 nm) is a useful tool for identifying AGN  as its EW traces the electron density of the narrow-line region surrounding the black hole and the ionising photon flux from the accretion disk \citep{baskin2005}.  It has been used as a powerful indicator to distinguish between HERGs and LERGs in previous studies of these AGN populations.  For example, \cite{hardcastle2013} found that for their sample of $\sim$2500 radio-loud AGN ($S_{\rm{1.4GHz}} > 0.5$ mJy) with spectroscopic redshifts, classifying HERGs as having EW([\ion{O}{III}]) $> 5$ \AA $ $ and LERGs as having EW([\ion{O}{III}]) $< 5$ \AA $ $ resulted in very similar HERG and LERG samples to those obtained via manual visual classification of the same galaxies using the strength of lines such as [\ion{O}{III}], [\ion{N}{II}], [MgII], [CIII], [CIV], or Ly$\alpha$ as HERG and LERG indicators.  In addition, \cite{ching2017} found that the EW([\ion{O}{III}]) $= 5$ \AA $ $ dividing line between HERGs and LERGs is very consistent with their visual spectral classification scheme for the $\sim$11000 radio sources in the LARGESS survey with reliable spectroscopic redshifts. \cite{best2012} also employed this cut to distinguish between HERGs and LERGs.

Due to the effective wavelength range of AAOmega, only the sources with $z \lesssim 0.78$ were able to have their [\ion{O}{III}] lines examined.  There are 971 sources at $z < 0.78$ with spectra. Requiring the [\ion{O}{III}] line to have a $S/N > 3$ resulted in 428 sources.  Of these, 268 have EW([\ion{O}{III}]) $> 5$ \AA, and are therefore classified as HERGs.  The remaining 160 sources have EW([\ion{O}{III}]) $< 5$ \AA, so they could either be LERGs or SFGs depending on their optical spectral templates (Section \ref{sec:opt_spec_temp_class}), optical colours (Section \ref{sec:opt_colours_class}), or radio properties (Section \ref{sec:radio_class}).

\subsubsection{Optical spectral templates}
\label{sec:opt_spec_temp_class}

\begin{figure*}
\begin{centering}
        \includegraphics[width=15.5cm]{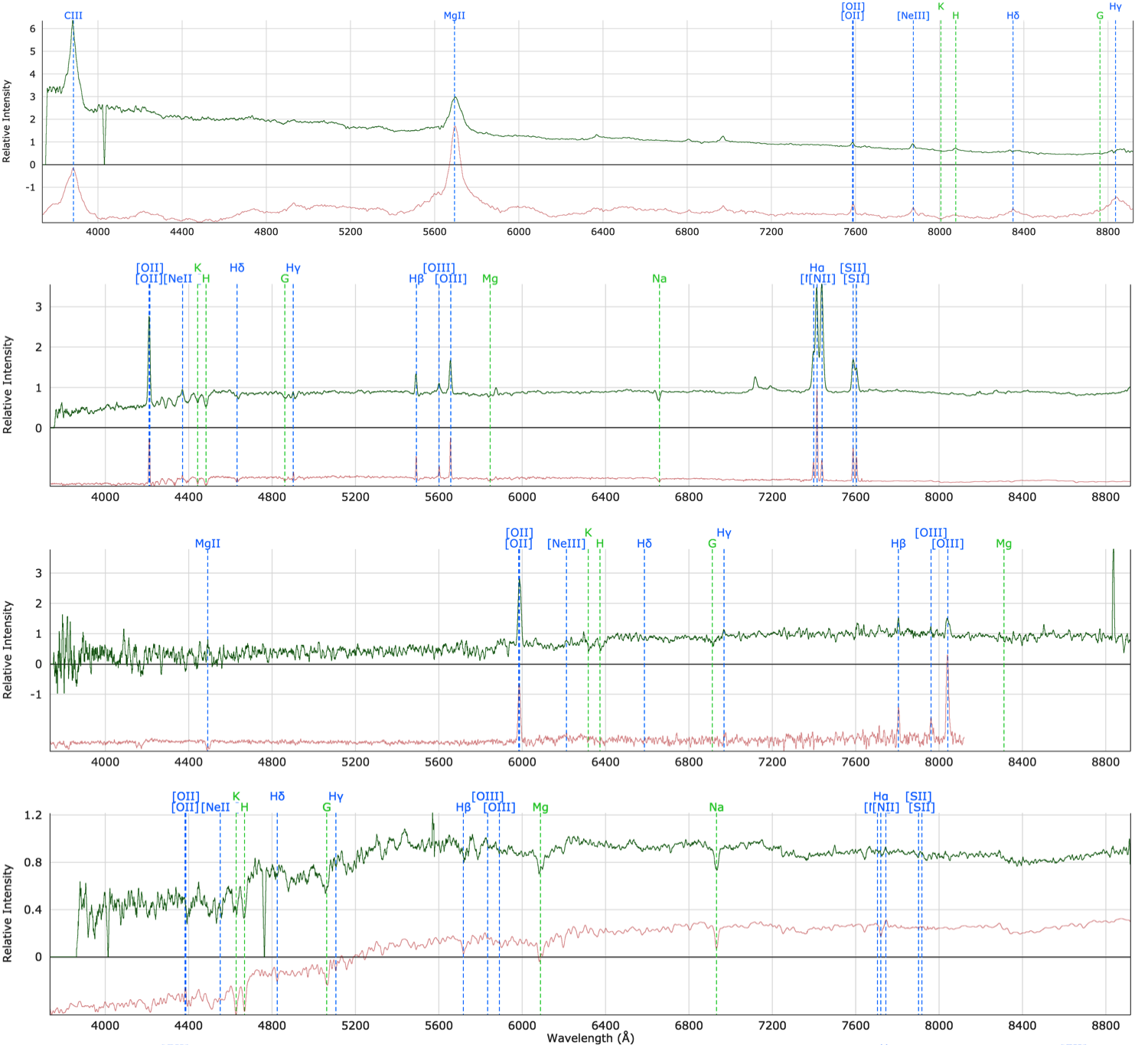}
    \caption{Example spectra (green) with corresponding \textsc{marz} best fit templates (red), absorption lines (green dashed lines), and emission lines (blue dashed lines).  The spectra have been smoothed for clarity. The best fit spectral templates are (from top to bottom) quasar, late-type emission galaxy, high redshift star-forming galaxy, and early-type absorption galaxy (ETAG).}
    \label{fig:spectra}
\end{centering}
\end{figure*}

For the 954 spectra that did not qualify for the BPT diagram, the full wavelength range of each of the spectra were examined in order to determine the source types responsible for the observed spectral characteristics. The spectra were required to have a continuum $S/N > 3$, resulting in 871 sources eligible for this analysis.  The software \textsc{marz} \citep{hinton2016} was used to find the best fit spectral template for each of these sources. \textsc{marz} uses seven galaxy templates, but just two are used in this work to explicitly classify the radio sources: quasar and early-type absorption galaxy (ETAG). Figure \ref{fig:spectra} shows examples of spectra that correspond to these templates and two other \textsc{marz} templates. Quasars were classified as HERGs, and ETAGs were classified as LERGs if none of the other indicators showed HERG activity.

\subsection{Optical rest-frame colours}
\label{sec:opt_colours_class}

For the radio sources with no spectra available or for which the spectra were not usable, optical colours were used for further classification if the X-ray, MIR, SED, or optical spectral analyses did not indicate HERG activity. \cite{ilbert2010} used the rest-frame $M_{\rm{NUV}} - M_r$ colours (not corrected for dust extinction) to separate the quiescent from the star-forming populations in a sample of 196000 galaxies classified by morphology and SED fits out to $z=2$ in the $\sim$2 deg$^2$ COSMOS field. The $M_{\rm{NUV}} - M_r$ colour is a good indicator of the current versus past star formation activity (e.g. \citealp{martin2007,arnouts2007}), but the $M_{\rm{NUV}} - M_u$ colours display a more easily identifiable separation between quiescent galaxies and SFGs.  This is manifested in quiescent galaxies having, for a given $g-i$ colour, larger relative NUV~$-$~$u$ colour offsets 
compared to SFGs than they do for NUV~$-$~$r$ colours.  In other words, for a given $g-i$ colour, there is a larger fractional offset between quiescent galaxies and SFGs for NUV~$-$~$u$ than there is for NUV~$-$~$r$.

The physical reason for this is that the $u$-band covers wavelengths slightly shorter than the 4000 \AA $ $ break ($3000~\lesssim~\lambda~\lesssim~4000$~\AA) as well as the [OII] line at 3727 \AA $ $, while the NUV covers shorter wavelengths ($1800~\lesssim~\lambda~\lesssim~2800$~\AA).  This results in quiescent galaxies having larger values of $M_{\rm{NUV}}~-~M_u$ because their ultraviolet luminosities drop off rapidly after the 4000 \AA $ $ break.  On the other hand, SFGs have smaller $M_{\rm{NUV}} - M_u$ values because they do not display strong 4000 \AA $ $ breaks and they possess a much larger population of young OB stars that emit large amounts of ultraviolet radiation in the NUV.

\begin{figure}
        \includegraphics[width=\columnwidth]{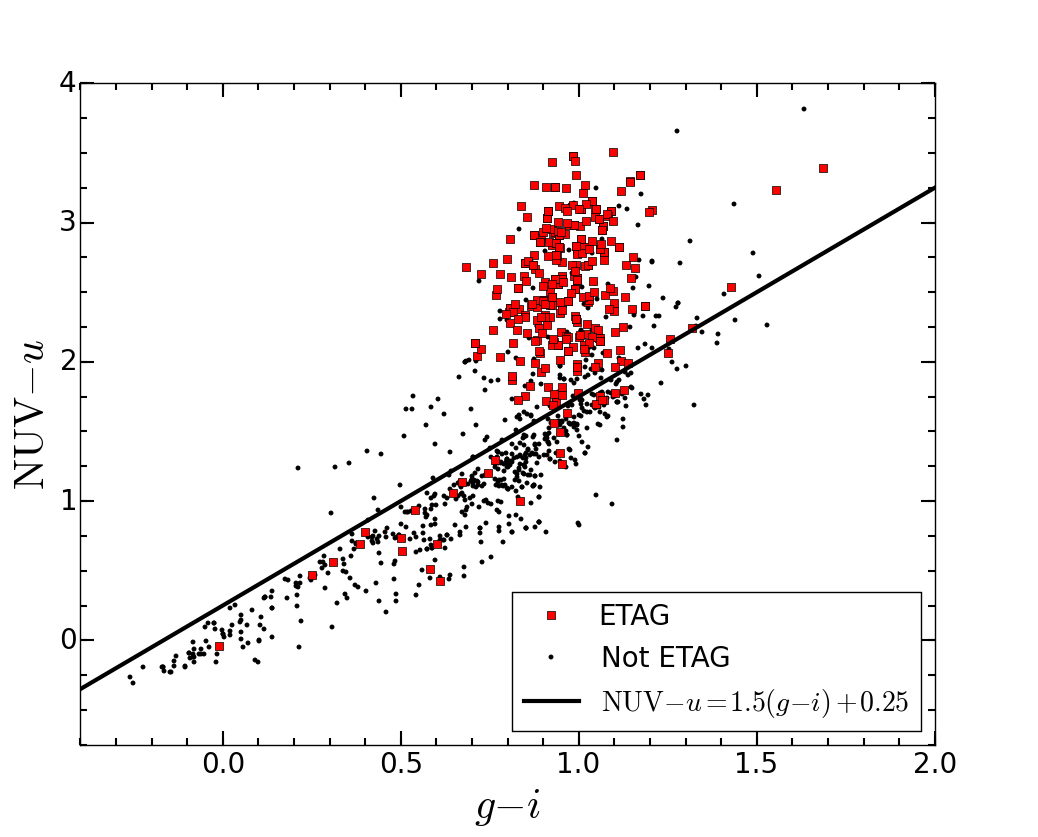}
    \caption{NUV $-$ $u$ vs. $g-i$ for all radio sources in XXL-S that possess a spectrum with continuum $S/N > 3$.  The colours have not been corrected for dust extinction.  The red squares represent the galaxies classified as early-type absorption galaxies (ETAGs) according to their spectra and the black points represent all the galaxies that were not classified as ETAGs according to their spectra.  The ETAGs tend to be found in the colour space defined by $\mathrm{NUV} - u > 2.0$ and $g-i < 1.2$.}
    \label{fig:NUV_minus_u_vs_g_minus_i_spectra_plot}
\end{figure}

This is demonstrated in Figure \ref{fig:NUV_minus_u_vs_g_minus_i_spectra_plot}, which shows the $M_{\rm{NUV}}~-~M_u$ versus $M_g~-~M_i$ (hereafter NUV~$-$~$u$ versus $g-i$) plot for all radio sources in XXL-S with a spectrum available with continuum $S/N > 3$. The rest-frame colours were calculated from the \textsc{magphys} SED fits, and they have not been corrected for dust extinction.  The sources classified as ETAGs according to \textsc{marz} tend to be found in the colour space defined by $\mathrm{NUV} - u > 2.0$ and $g-i < 1.2$, and all other galaxies tend to be found outside that space.  Of all   ETAGs, 82.5\% are found in that colour space.  Of all  galaxies that reside in that colour space, 76.9\% are ETAGs.  In addition, Figure \ref{fig:NUV_minus_u_vs_g_minus_i_spectra_plot} shows that there is a clear trend for the non-ETAG galaxies to exist on or below a linear relation between $\mathrm{NUV} - u$ and $g-i$.  Of all non-ETAG galaxies, 73.0\% exist below the line defined by $\mathrm{NUV} - u = 1.5(g-i) + 0.25$, shown as the black line in Figure \ref{fig:NUV_minus_u_vs_g_minus_i_spectra_plot}.  This indicates that most of the galaxies that have at least a moderate level of star formation in the sample have colours that fulfil $\mathrm{NUV} - u < 1.5(g-i) + 0.25$.  Galaxies that lie above this line have redder UV colours for a given $g-i$, suggesting that they exhibit less UV emission and therefore lower levels of star formation.  Of the ETAGs, 91.2\% of them exist above the line, and 59.0\% of all galaxies with spectra above the line are ETAGs.

Therefore, sources are considered red (with levels of star formation consistent with quiescent galaxies) if they lie above the line defined by~$\mathrm{NUV}-u=1.5(g-i)+0.25$.  All other galaxies are considered blue and star-forming.  These colour cuts are similar to the $UVJ$ colours used by previous authors to separate quiescent and star-forming galaxies (\citealp{muzzin2013} and references therein).  How the NUV $-$ $u$ versus $g - i$ colours affected the final classification of each source depended on its radio properties.

\subsection{Radio classification}
\label{sec:radio_class}

A source was considered a radio AGN if at least one of the following radio criteria  was met.

\subsubsection{Luminosity}

In normal galaxies, the radio continuum can be used to trace star formation via the relativistic electrons accelerated around magnetic fields in supernovae and the electrons interacting with ions in HII regions \citep{condon1992}.  The 1.4 GHz radio luminosity of normal galaxies does not normally exceed $L_{\rm{1.4GHz}}$~=~10$^{24}$~W Hz$^{-1}$ (e.g. \citealp{sadler2002,best2005}).  This luminosity corresponds to a star formation rate (SFR) of $SFR_{\rm{1.4GHz}} \sim$ 600 M$_{\odot}$ year$^{-1}$ assuming the 1.4 GHz SFR calibration of \cite{murphy2011}.  Some extreme SFGs have higher SFRs than this.  For example, \cite{barger2014} found that high-$z$ submillimetre galaxies in the GOODS-N field have a characteristic maximum SFR of $\sim$2000 M$_{\odot}$ year$^{-1}$.  This SFR corresponds to a luminosity of $L_{\rm{1.4GHz}}$ = 10$^{24.5}$ W Hz$^{-1}$. Therefore, a radio source is considered a radio AGN if it has $L_{\rm{1.4GHz}}$~>~10$^{24.5}$~W Hz$^{-1}$.  The number of sources that meet this criterion is 1789.

\subsubsection{Spectral index}

Radio continuum emission at frequencies below 30 GHz is dominated by non-thermal synchrotron emission from relativistic electrons, with thermal emission comprising only $\sim$10\% of the total emission at 1$-$3 GHz \citep{condon1992,peel2011}.  Both star formation and AGN activity produce synchrotron radiation, which exhibits a typical spectral index of $\alpha_R \sim -0.75$.  Therefore, it is difficult to disentangle AGN from SFGs at frequencies around 1 GHz on the basis of radio spectral index alone.  However, \cite{thompson2006} demonstrated that the range of radio spectral indices expected in starburst galaxies for the 1$-$3 GHz range is $-1.0 < \alpha_R < -0.5$.  Steep spectra ($\alpha_R < -1.0$) are generated by old electrons, whereas flat spectra ($\alpha_R > -0.5$) correspond to multiple components in the line of sight or a recent outburst of electrons (e.g. \citealp{kellermann1966,kellermann1969}), both of which only AGN can produce.  Therefore, applying a similar criterion to that used by \cite{rawlings2015}, if a radio source has a 1$\sigma_{\alpha_R}$ upper bound of $\alpha_R < -1.0$ or a 1$\sigma_{\alpha_R}$ lower bound of $\alpha_R > -0.5$, it is considered a radio AGN.  The number of sources identified as radio AGN according to their radio spectral indices is 766. 

\subsubsection{Morphology}

\begin{figure*}
        \includegraphics[width=\textwidth]{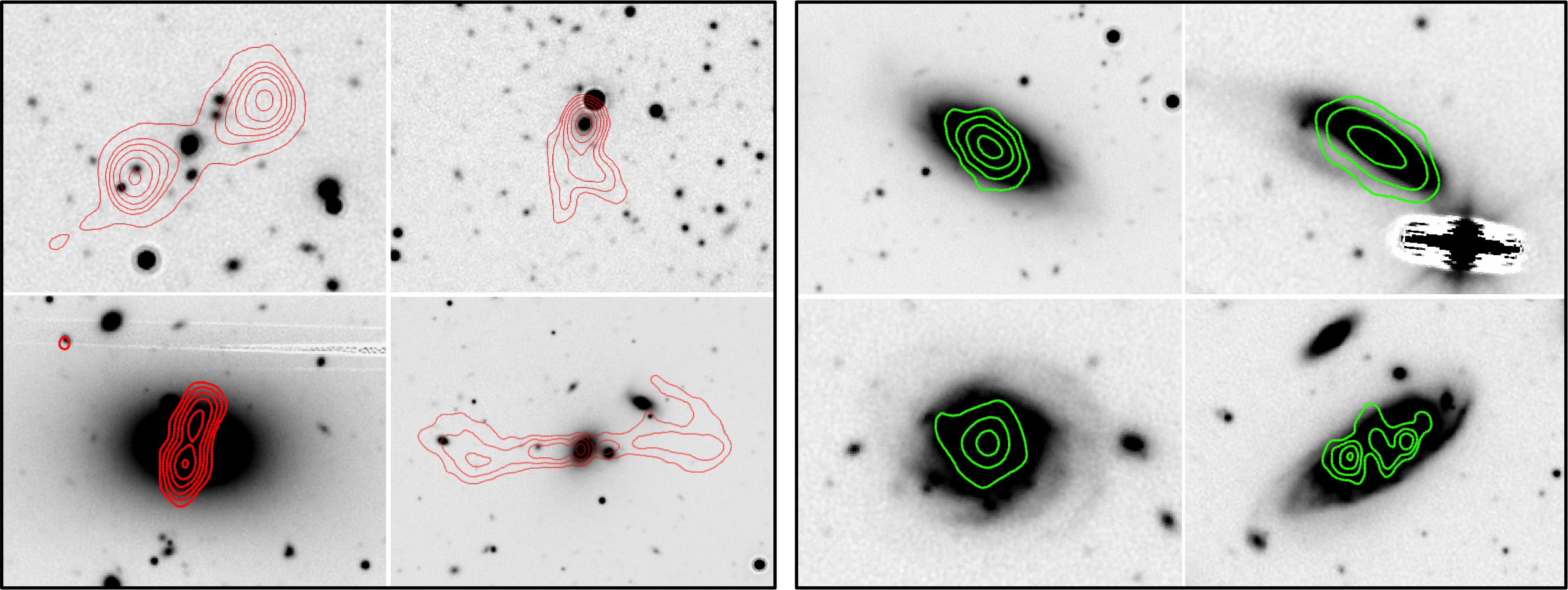}
    \caption{\textbf{Left panel:} Examples of radio sources that were classified as radio AGN according to their radio morphology.  The background image is the $z$-band cutout at the position of the radio source, and the red contours are the 2.1 GHz radio contours.  The contours are at levels of 5$\sigma$, 10$\sigma$, 20$\sigma$, 40$\sigma$, 80$\sigma$, and 160$\sigma$, where $\sigma$ is the local rms noise. \textbf{Right panel:} Examples of radio sources that were not considered AGN according to their radio morphology.  The background image is the $z$-band cutout at the position of the radio source, and the green contours are the 2.1 GHz radio contours.  For the top row, the contours are at levels of 5$\sigma$, 10$\sigma$, 20$\sigma$, and 40$\sigma$.  For the galaxy at bottom left, the contours are at levels of 5$\sigma$, 10$\sigma$, and 15$\sigma$.  For the bottom right galaxy, the contours are at levels of 4$\sigma$, 5$\sigma$, and 6$\sigma$.}
    \label{fig:agn_sfg_morph_examples_plot}
\end{figure*}

The radio morphology of each of the complex radio sources, including those with multiple components (see Sections 3.6--3.8 in XXL Paper XVIII) was inspected.  Nearly all of these sources are resolved. If the structure of a source's radio contours displayed double lobes, jets, extended emission away from the host galaxy, or elongation perpendicular to the major axis of the optical host galaxy, then the source was classified as a radio AGN (see the left panel of Figure \ref{fig:agn_sfg_morph_examples_plot} for examples).  The number of sources that were considered AGN according to their radio morphology is 401. There are six nearby disk galaxies for which the origin of the radio morphology cannot be determined, as revealed by the fact that the overall orientation of the radio contours is consistent with that of the host galaxy and the contours do not extend well beyond the disk (see the right panel of Figure \ref{fig:agn_sfg_morph_examples_plot} for examples).  Only if none of the other AGN indicators were positive were these galaxies classified as SFGs.  Of the six galaxies without definite radio AGN morphology, three have final classifications as HERGs (one radio-loud and two radio-quiet), two are classified as SFGs, and one is an unclassified  source.

\subsubsection{Luminosity excess}

\begin{figure}
        \includegraphics[width=\columnwidth]{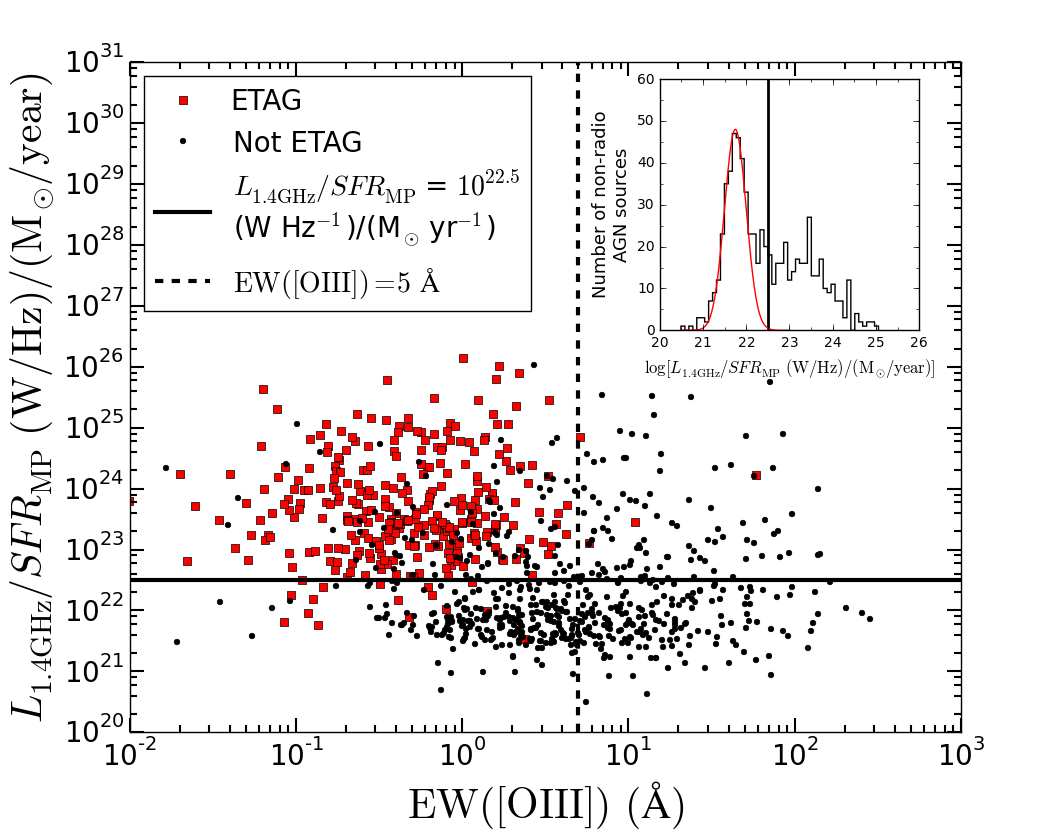}
    \caption{$L_{\rm{1.4GHz}}/SFR_{\rm{MP}}$ vs. EW([\ion{O}{III}]) for all XXL-S radio sources that possess a spectrum with continuum $S/N > 3$.  The vertical dashed line represents EW([\ion{O}{III}]) $=$ 5$ $ \AA $ $ (the dividing line between HERGs and LERGs/SFGs) and the horizontal solid line represents $L_{\rm{1.4GHz}}/SFR_{\rm{MP}}~=~10^{22.5}$ (W Hz$^{-1}$)/(M$_{\odot}$ year$^{-1}$).  The red squares represent the galaxies classified as early-type absorption galaxies (ETAGs) according to their spectra and the black points represent all other sources with the [\ion{O}{III}] line available.  The ETAGs tend to be found in the space defined by $L_{\rm{1.4GHz}}/SFR_{\rm{MP}}>10^{22.5}$ (W Hz$^{-1}$)/(M$_{\odot}$ year$^{-1}$) and EW([\ion{O}{III}]) $<$ 5$ $ \AA $ $ (the upper left quadrant), and the vast majority of non-ETAG sources are outside this space.  This suggests that the galaxies in the upper left quadrant of the plot are likely to be LERGs.  The inset shows the distribution of log[$L_{\rm{1.4GHz}}/SFR_{\rm{MP}}$] for sources that have not been identified as radio AGN by the other three radio indicators.  The red curve is the Gaussian fit to the portion of the distribution that is assumed to be representative of the SFG population, and the vertical black line represents log[$L_{\rm{1.4GHz}}/SFR_{\rm{MP}}$] = 22.5.  The latter value is $3\sigma$ above the SFG population's Gaussian peak.}
    \label{fig:L_1400MHz_div_SFR_MP_vs_OIII_plot}
\end{figure}

The $SFR_{\rm{1.4GHz}}$ calibration of \cite{murphy2011} implies that the radio luminosity of a galaxy is directly proportional to its SFR. Therefore, if a galaxy has a radio luminosity greater than  is expected from star formation, it is reasonable to assume that the excess luminosity is due to the presence of a radio AGN.  In order to determine what level of radio excess would be appropriate to select radio AGN, the ratio of $L_{\rm{1.4GHz}}$ to the SFR computed by the \textsc{magphys} SED fitting ($SFR_{\rm{MP}}$) was plotted versus EW([\ion{O}{III}]), which is shown in Figure \ref{fig:L_1400MHz_div_SFR_MP_vs_OIII_plot}.  The vertical dashed line represents EW([\ion{O}{III}]) $= 5$ $\AA$, so none of the sources to the left of this line are HERGs according to the classification scheme.  The red squares in the figure represent the galaxies whose spectra indicated that they are ETAGs.  Nearly all (95.1\%) of the ETAGs with EW([\ion{O}{III}])~$<~5$~$\AA$ have $L_{\rm{1.4GHz}}/SFR_{\rm{MP}}$ $> 10^{22.5}$ (W Hz$^{-1}$)/(M$_{\odot}$ year$^{-1}$), while 60.3\% of non-ETAGs with EW([\ion{O}{III}])~$<~5$~$\AA$ are below this line. Furthermore, the fact that 62.1\% of the galaxies with $L_{\rm{1.4GHz}}/SFR_{\rm{MP}}$~$>~10^{22.5}$ (W Hz$^{-1}$)/(M$_{\odot}$ year$^{-1}$) and EW([\ion{O}{III}])~$<~5$~$\AA$ are ETAGs demonstrates that non-HERG galaxies with radio luminosity excess tend to reside in early-type galaxies.  This can only be explained by the presence of a radio AGN. The one caveat is that this analysis applies to sources with $z<0.78$ with a spectrum available.  However, the far-IR--radio correlation shows only a slight decrease up to $z \sim 2$ \citep{garrett2002,sargent2010,magnelli2015,delhaize2017}, below which 88\% of the sample in this paper is found, which suggests that  there is no observational reason why the relationship between radio excess and EW([\ion{O}{III}]) would change significantly as a function of redshift.

In addition, the radio excess distribution of the sources that were not  identified as radio AGN by the previous three radio indicators suggests that at least two source populations contribute to the distribution (see inset in Figure \ref{fig:L_1400MHz_div_SFR_MP_vs_OIII_plot}).  It is assumed that the portion of the distribution responsible for the peak at log($L_{\rm{1.4GHz}}/SFR_{\rm{MP}}$) = 21.75 represents the SFG population due to the relatively low radio excess values.  A Gaussian function was fit to this part of the distribution, which is shown by the red curve in the inset in Figure \ref{fig:L_1400MHz_div_SFR_MP_vs_OIII_plot}.  The standard deviation of this Gaussian function is 0.25 in log($L_{\rm{1.4GHz}}/SFR_{\rm{MP}}$) space.  Therefore, the radio excess level that is $3\sigma$ above the Gaussian peak is log($L_{\rm{1.4GHz}}/SFR_{\rm{MP}}$) = 22.5, which is the solid vertical black line in the inset. Therefore, in order to select previously unidentified radio AGN in the most reliable way possible, all sources with a radio excess of $L_{\rm{1.4GHz}}/SFR_{\rm{MP}}$ $> 10^{22.5}$ (W Hz$^{-1}$)/(M$_{\odot}$ year$^{-1}$) were considered radio AGN.  This indicator identified 3418 radio AGN, 1390 of which were not identified by the other three radio indicators.

\subsection{Decision tree}
\label{sec:decision_tree}

Figure \ref{fig:decision_tree} shows the decision tree by which the optically matched radio sources were classified.  If a source displayed evidence of AGN activity via the X-ray, MIR, SED, or optical spectra indicators, then it was considered a HERG.  If one of the radio indicators determined that the source contained a radio AGN, it was classified as a radio-loud (RL) HERG.  If a radio AGN was not detected, then this suggested that the origin of the radio emission in the galaxy is star formation (not AGN activity), and so these sources were classified as radio-quiet (RQ) HERGs (e.g. \citealp{padovani2011,bonzini2013}).

If HERG activity was not detected, the optical spectrum of the source, if available, was examined. If the source was in the AGN region of the BPT diagram, then it was classified as a LERG. If it was in the SFG region of the BPT diagram, it was classified as a LERG if a radio AGN was detected, and it was classified as a SFG otherwise.  If the source was  between the \cite{kauffmann2003} and \cite{kewley2001b} lines in the BPT diagram, its NUV~$-$~$u$ versus $g-i$ colours were checked.  If it was at $z>0.36$ (meaning the BPT diagnostic was unavailable), its best fit optical spectral template was examined.  If its spectrum was determined to be that of an ETAG, then it was classified as a LERG.  If not, then its NUV~$-$~$u$ versus $g-i$ colours were also checked.

If a source had colours that fulfilled NUV~$-$~$u$~>~1.5($g-i$)~+~0.25, the galaxy was considered red.  Of these galaxies, if a source met one of the radio AGN criteria, it was classified as a LERG. If it did not exhibit signs of a radio AGN, it was left as an unclassified source. A source was considered blue if its colours were ~$\mathrm{NUV} - u < 1.5(g-i) + 0.25$.  If a blue galaxy showed no signs of a radio AGN, then it was classified as an SFG.  Otherwise, it was left as an unclassified AGN.

\begin{figure}
        \includegraphics[width=\columnwidth]{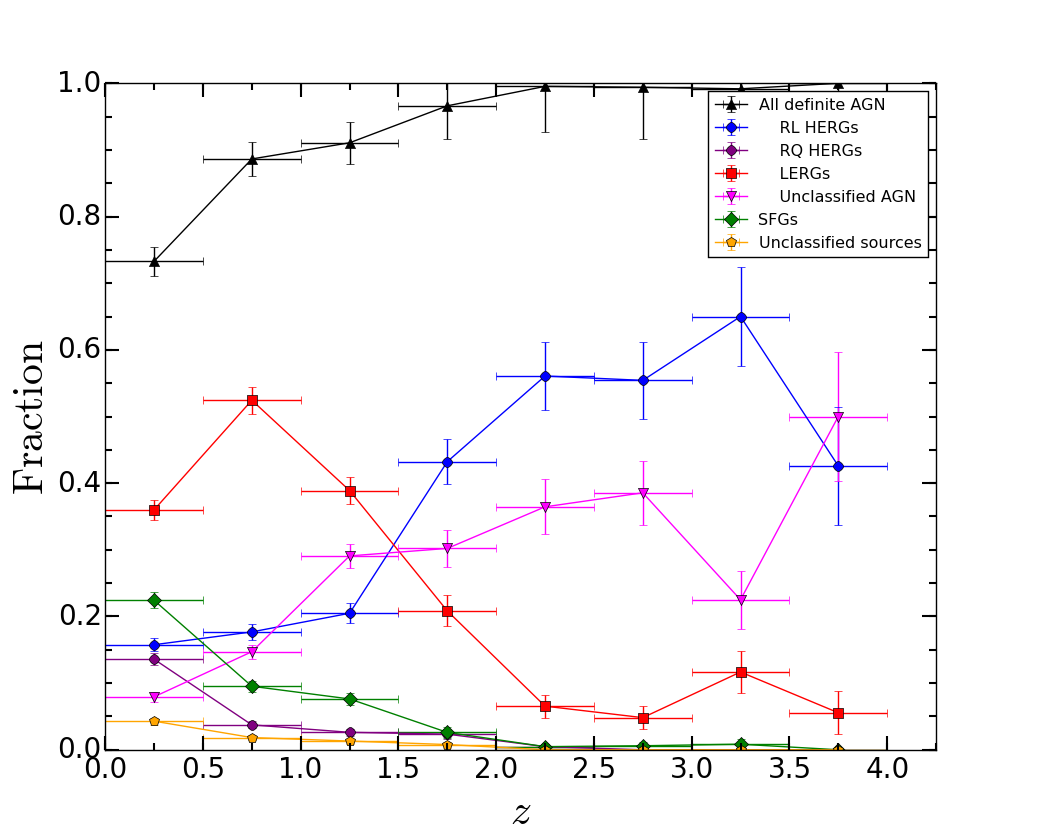}
    \caption{Relative fraction of XXL-S source types as a function of redshift.  The black triangles represent all definite radio AGN (whether classified or not), the blue circles represent RL HERGs, the purple circles represent RQ HERGs, the red squares represent LERGs, the magenta triangles represent unclassified AGN (which potentially include LERGs and RL HERGs), the green diamonds represent SFGs, and the orange pentagons represent unclassified sources (which potentially include LERGs, RQ HERGs, and SFGs).  The y-axis error bars for each population are Poissonian and are given by Equation \ref{eq:agn_sfg_frac_err}, calculated for each redshift bin.  The errors along the x-axis are the redshift bin widths.}
    \label{fig:agn_sfg_frac_vs_z_plot}
\end{figure}

\begin{figure*}
        \includegraphics[width=\textwidth]{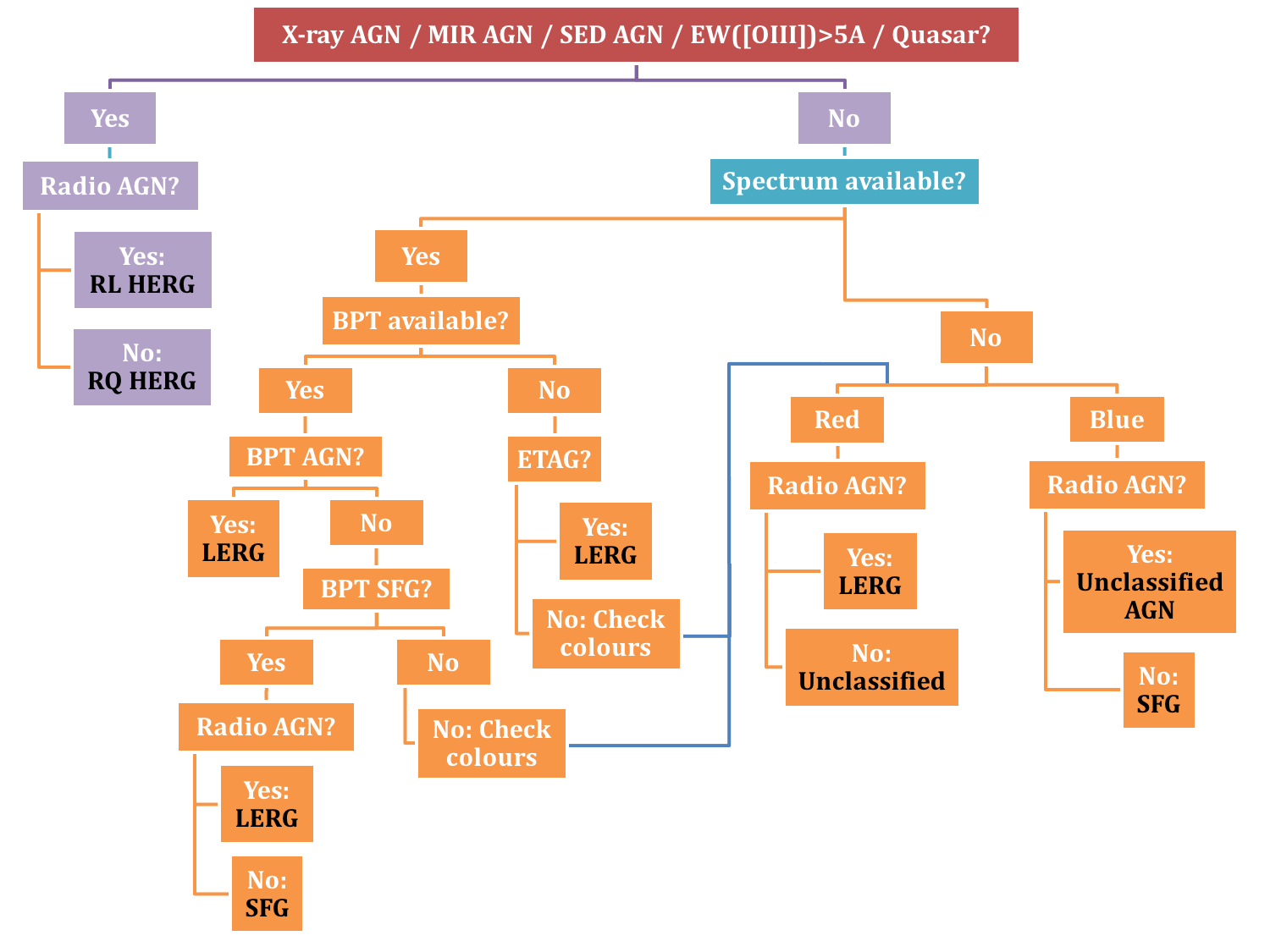}
    \caption{Decision tree used to  classify the optically matched radio sources in XXL-S.  The black text indicates the final classification for each path.}
    \label{fig:decision_tree}
\end{figure*}

The classification process resulted in 1729 LERGs (36.3\%), 1159 RL HERGs (24.4\%), 296 RQ HERGs (6.2\%), and 558 SFGs (11.7\%), with 910 unclassified AGN (19.1\%) and 106 unclassified sources (2.2\%). Unclassified AGN potentially include LERGs and RL HERGs, while unclassified sources potentially include LERGs, RQ HERGs, and SFGs. These results are summarised in Table \ref{tab:class_results}, with the redshift dependence shown in Figure \ref{fig:agn_sfg_frac_vs_z_plot}.

\begin{table}
\centering
\caption{Results of classification of XXL-S radio sources.  Unclassified AGN potentially include LERGs and RL HERGs, while unclassified sources potentially include LERGs, RQ HERGs, and SFGs.}
\begin{tabular}{c c c}
Source type & Number & Fraction of final sample \\
\hline
\hline
LERGs & 1729 & 36.3\% \\
RL HERGs & 1159 & 24.4\% \\
RQ HERGs & 296 & 6.2\% \\
SFGs & 558 & 11.7\% \\
Unclassified AGN & 910 & 19.1\% \\
Unclassified sources & 106 & 2.2\% \\
\end{tabular}
\label{tab:class_results}
\end{table}

\section{Discussion}
\label{sec:discussion}

\subsection{AGN selection effects}

A number of different factors lead to AGN being detected in one wavelength regime but not another.  If a galaxy is particularly dusty, the MIR signal from the galactic dust may overpower the MIR signal from the dusty torus near the AGN, causing the relative MIR contribution from the AGN to be low and resulting in MIR colours that are outside of the domain of the MIR AGN indicators.  It is also possible for a normal galaxy to have an AGN with a relatively small amount of dust, leading to a weak MIR signal from the AGN.  For example, \cite{hickox2009}, who describes host galaxy properties of AGN in the $\sim$9 deg$^2$ NOAO Deep Wide-Field Survey (NDWFS) field \citep{jannuzi1999}, found that most of their X-ray AGN are not selected by the IRAC MIR colour-colour criteria in \cite{stern2005}.  They concluded that the MIR emission in the IRAC bands from those host galaxies probably mask a relatively low-luminosity IR-faint AGN \citep{gorjian2008}.  In addition, the X-ray emission from an AGN may be obscured by the dusty torus, resulting in an AGN that is classified as a MIR AGN but not an X-ray AGN. Compton-thick ($n_H \gtrsim 10^{24}$ cm$^{-2}$) AGN are difficult to detect even in hard X-rays.  For example, \cite{mendez2013} found that $10-25\%$ of MIR selected AGN in the Prism Multi-object Survey \citep{coil2011} are not classified as X-ray AGN, depending on the depth of the X-ray data.  Radio AGN are not affected by dust, so they are still detectable in dusty galaxies. Since not all AGN can be detected in the same way, it is important to examine the overlap (or lack thereof) between AGN selected by different techniques.

\subsubsection{Comparison of AGN classification techniques}
\label{sec:rel_comp}

The reliability and completeness of each of the AGN selection techniques presented in this work are an important quantification of the uniqueness of the AGN populations detected by each indicator.  That is, the reliability and completeness of each indicator probe the degree to which other indicators can detect the overall AGN signal in the same objects.  In this analysis, the reliability and completeness for a given indicator are defined as follows.  Reliability is the percentage of sources detected by the indicator that are detected by another given indicator.  Completeness is the percentage of all XXL-S AGN that are detected by the given indicator.  All XXL-S AGN include LERGs, HERGs, and unclassified radio AGN.

Table \ref{tab:rel_comp_AGN} summarises the reliability and completeness of each AGN indicator.  The reliability and completeness for the [\ion{O}{III}] indicator is of limited use as only 428 (9.0\%) of the 4758 XXL-S optically matched radio sources have a spectrum with the [\ion{O}{III}] emission line at $S/N>3$ available.  Therefore, the [\ion{O}{III}] completeness applies only to those 428 sources (but the [\ion{O}{III}] completeness relative to all XXL-S AGN is still displayed in Table \ref{tab:rel_comp_AGN} for consistency).  In the table, the reliability of each indicator is read from left to right.  For example, the percentage of radio AGN that are SED AGN is 24.2\%.  In other words, the reliability of radio AGN in terms of SED AGN is 24.2\%. The completeness of each indicator is read from top to bottom.  For example, MIR AGN probe 6.8\% of all AGN in XXL-S.  In other words, the completeness of MIR AGN in terms of all AGN in XXL-S is 6.8\%.

Radio AGN are limited in their ability to probe other AGN populations at the $\sim$30\% level.  In other words, $\sim$70\% of radio AGN can be found only in the radio.  This is mostly due to the LERGs in the sample, which do not show signs of AGN other than at radio wavelengths.  In addition, there is a large fraction of [\ion{O}{III}]-only AGN, which are bright and at relatively low redshift ($z<0.78$).  This demonstrates the importance of spectra for finding HERGs.

\begin{table*}
\centering
\caption{Reliability and completeness of AGN selection techniques.  For reliability, the table should be read from left to right.  For example, the percentage of X-ray AGN that are MIR AGN is 25.7\% and the percentage of radio AGN that are any other kind of AGN is 30.5\%. For completeness, the table should be read top to bottom.  For example, X-ray AGN probe 16.1\% of SED AGN and MIR AGN probe 23.5\% of [\ion{O}{III}] AGN.}
\begin{adjustbox}{width=18cm}
\begin{tabular}{c c c c c c c}
 & X-ray & MIR & SED & OIII & Radio & Any\\
\hline
\hline
X-ray & -- & 25.7\% (106/412) & 38.6\% (159/412) & 21.1\% (87/412) & 74.5\% (307/412) & 88.6\% (365/412)\\
MIR & 38.1\% (106/278) & -- & 67.6\% (188/278) & 22.7\% (63/278) & 64.0\% (178/278) & 88.5\% (246/278)\\
SED & 16.1\% (159/988) & 19.0\% (188/988) & -- & 5.4\% (53/988) & 93.0\% (919/988) & 97.3\% (961/988)\\
{[}\ion{O}{III}] & 32.5\% (87/268) & 23.5\% (63/268) & 19.8\% (53/268) & -- & 40.7\% (109/268) & 66.0\% (177/268)\\
Radio & 8.1\% (307/3791) & 4.7\% (178/3791) & 24.2\% (919/3791) & 2.9\% (109/3791) & -- & 30.5\% (1155/3791)\\
\hline
All & 10.1\% (412/4087) & 6.8\% (278/4087) & 24.2\% (988/4087) & 6.6\% (268/4087) & 92.8\% (3791/4087) & --\\
\end{tabular}
\end{adjustbox}
\label{tab:rel_comp_AGN}
\end{table*}

\subsubsection{X-ray versus MIR versus radio selection of AGN}

\begin{figure}
        \includegraphics[width=\columnwidth]{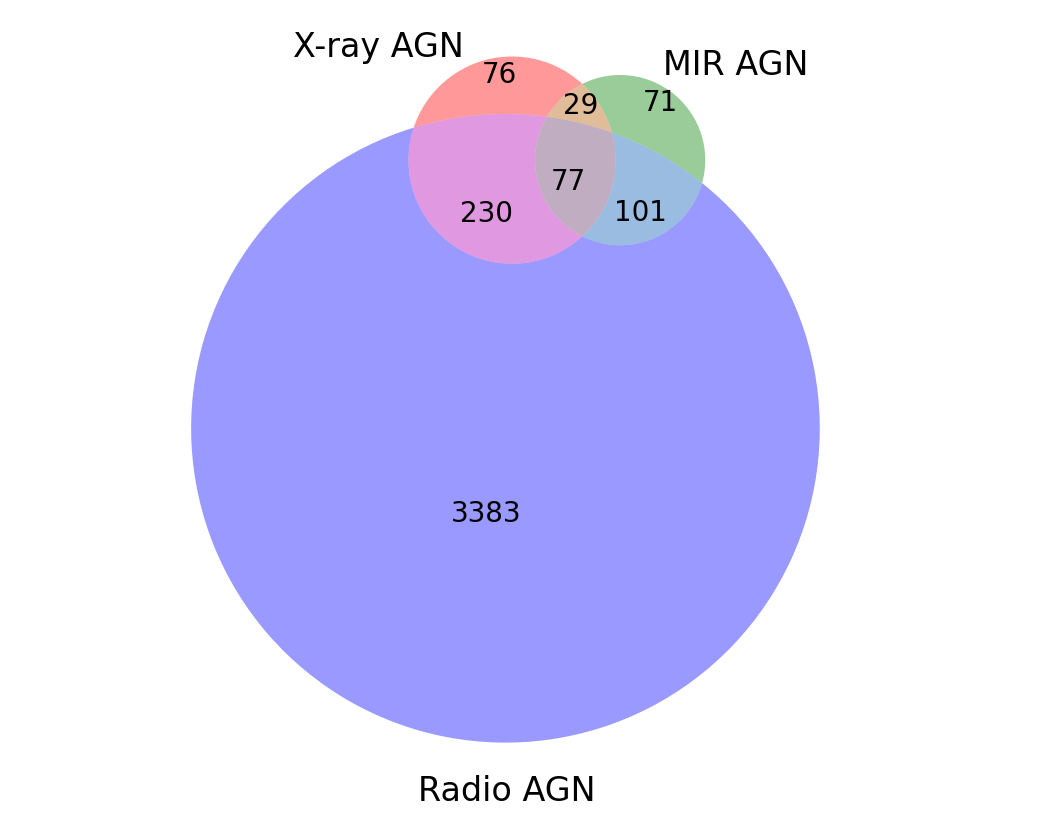}
    \caption{Venn diagram showing overlap between X-ray AGN, MIR AGN, and radio AGN for the optically matched radio sources in XXL-S.}
    \label{fig:venn_xray_mir_radio_agn_plot}
\end{figure}

Figure \ref{fig:venn_xray_mir_radio_agn_plot} displays the overlap between X-ray AGN, MIR AGN, and radio AGN in XXL-S.  The AGN for each wavelength regime are defined as having at least one positive in any of the corresponding indicators.  The sample in this paper is radio selected, so there are many more radio AGN than other classes of AGN.  A large portion (69.4\%) of the radio AGN represented in the diagram are classified as AGN by the radio indicators only (i.e. LERGs).  This illustrates the prominence of the LERG population. The remaining radio AGN are HERGs that were classified according to their SED fits, EW([\ion{O}{III}]), or quasar spectral template, which are not taken into account in the diagram.

Of all radio AGN, 10.8\% $\pm$ 0.5\% are also X-ray AGN or MIR AGN.  This can be compared to the overlap found by \cite{hickox2009}.  Even though their sample was optically selected (so there are X-ray and MIR AGN that are not radio detected) and even though they used a radio AGN threshold of $L_{\rm{1.4GHz}}$ > 10$^{23.8}$ W Hz$^{-1}$, they found a similar overlap: 11.7\% $\pm$ 3.4\% of the radio AGN in their sample are also X-ray AGN or MIR AGN.

Considering the other AGN in XXL-S, 25.7\% of X-ray AGN are also MIR AGN, and 38.1\% of MIR AGN are also X-ray AGN.  Even though XXL-S is radio selected, the latter percentage is similar to that found for MIR-selected samples.  For example, \cite{donley2012} constructed an IRAC-selected MIR AGN sample for the COSMOS field and found that 38\% of the MIR AGN had X-ray counterparts, which rose to 52\% with deeper X-ray data.  This is nearly identical to what \cite{mateos2012} found for the Bright Ultrahard \emph{XMM-Newton} survey (BUXS): 38.5\% of the MIR AGN selected with the WISE bands 1--3 were detected in hard X-rays, which rose to 49.8\% if deeper X-ray observations were used.  Given the X-ray and MIR depths available for XXL-S, the overlap between X-ray and MIR AGN is consistent with that found for other surveys.

\subsubsection{X-ray versus MIR versus SED selection of AGN}

\begin{figure}
        \includegraphics[width=\columnwidth]{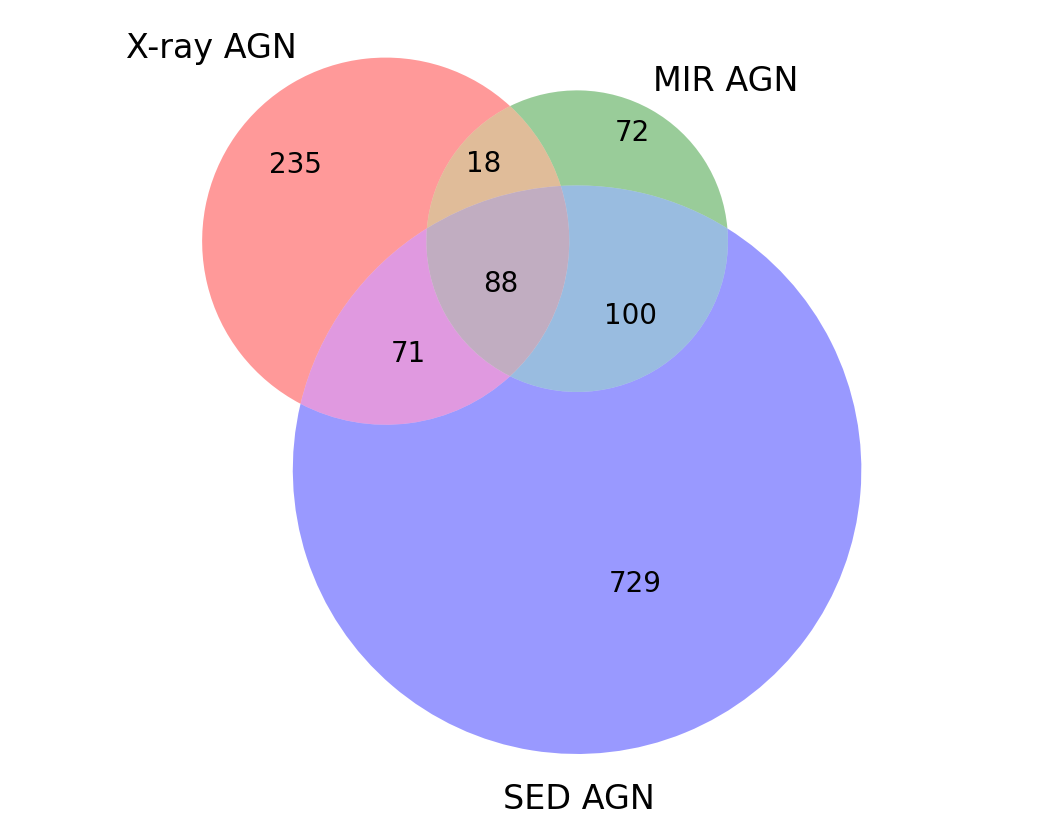}
    \caption{Venn diagram showing overlap between X-ray AGN, MIR AGN, and SED AGN for the optically matched radio sources in XXL-S.}
    \label{fig:venn_xray_mir_sed_agn_plot}
\end{figure}

The SED fits of the sources give an extra dimension of information regarding the selection effects of different AGN identification techniques.  Figure \ref{fig:venn_xray_mir_sed_agn_plot} displays the overlap between X-ray AGN, MIR AGN, and SED AGN.  The MIR and SED AGN have the most overall overlap (67.6\% of MIR AGN are also SED AGN),  probably because the vast majority of the total energy of an AGN signal in the optical and IR bands occurs at MIR wavelengths.  The SED AGN that are not MIR AGN show that the SED fitting process is more sensitive to the MIR bump (hot dust component).  In other words, a significant AGN component in an SED fit may cause the MIR colours of a source to change but not fall within the AGN regions of one of the MIR indicators presented in this paper.  On the other hand, the MIR AGN that are not SED AGN demonstrate that there is some contamination of normal galaxies in the AGN regions of the MIR indicators.

Furthermore, 38.6\% of the X-ray AGN are also SED AGN.  This is similar to what \cite{marchesi2016} found in the \emph{Chandra COSMOS-Legacy} Survey for an X-ray selected sample: $\sim$32\% of the COSMOS X-ray sources with photometric redshifts required an AGN component to their SED fits.  This large fraction of X-ray only AGN is likely due to the fact that X-ray data are less affected by the host-galaxy contamination compared to optical and MIR data, or that there are some AGN with low amounts of dust.  This is reflected in Figures \ref{fig:W1-W2_vs_W2_plot} and \ref{fig:IRAC1-IRAC2_vs_IRAC2_plot}, where at least some of the X-ray AGN are found in the normal galaxy regions.

\subsubsection{X-ray/MIR/SED versus [OIII] versus radio selection of AGN}

\begin{figure}
        \includegraphics[width=\columnwidth]{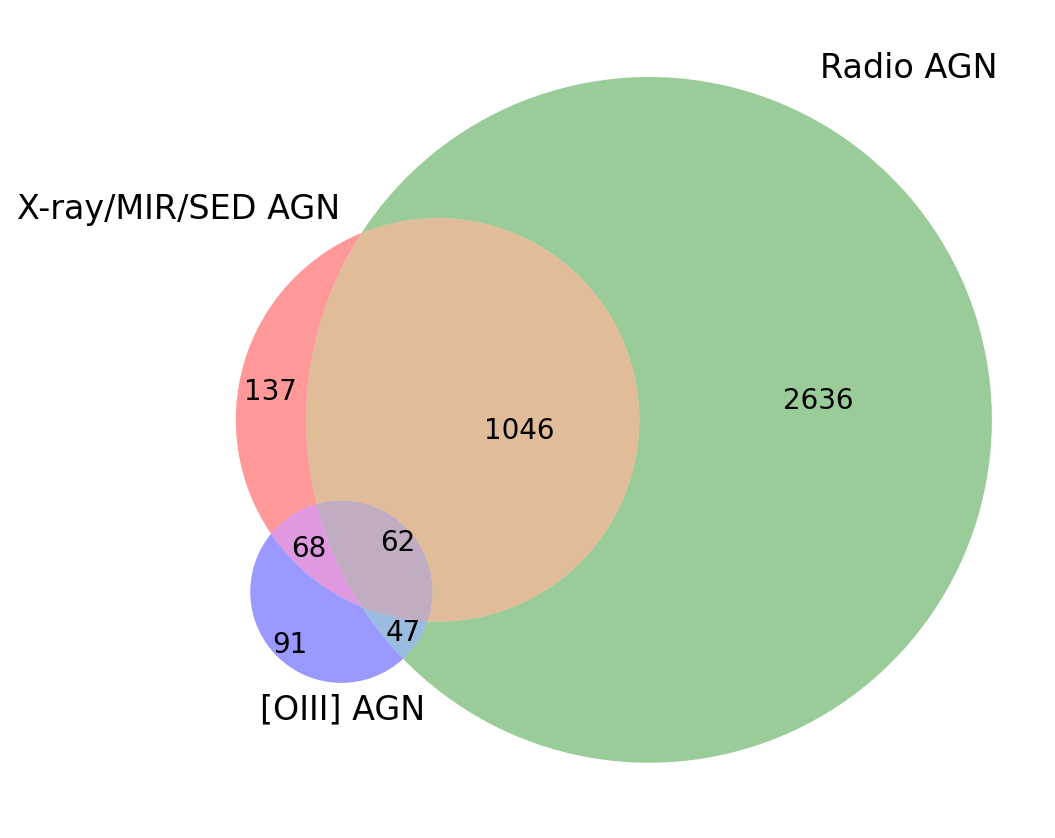}
    \caption{Venn diagram showing overlap between X-ray/MIR/SED AGN, radio AGN, and [\ion{O}{III}] AGN for the optically matched radio sources in XXL-S.  A large portion (69.9\%) of the radio AGN are radio-only AGN (i.e. LERGs).}
    \label{fig:venn_xray_mir_sed_radio_OIII_agn_plot}
\end{figure}

Figure \ref{fig:venn_xray_mir_sed_radio_OIII_agn_plot} shows the overlap between sources classified as either X-ray/MIR/SED AGN, sources classified as radio AGN, and sources classified as [\ion{O}{III}] AGN (EW([\ion{O}{III}])  $>5$ \AA).  About 66\% of the [\ion{O}{III}] AGN are also classified as X-ray/MIR/SED/radio AGN.  Although spectral coverage of the 4758 optically matched XXL-S radio sources is only $\sim$23\% complete, this clearly implies that at least some HERGs can only be identified by the signatures in their optical spectra.  In fact, all 91 of the sources classified as AGN solely by their [\ion{O}{III}] emission lines are RQ HERGs.  This is a further reflection of the observation that different selection techniques identify different populations of AGN.

\subsubsection{Contamination of LERG and SFG populations}

Notwithstanding differences in survey depths and selection techniques, the broad similarities between the overlap of different classes of AGN in XXL-S and that of other multiwavelength samples of AGN suggests that the HERGs and LERGs in XXL-S have been robustly classified and the overall picture of the different classes of AGN representing distinct populations of sources is generally valid.

However, the spectral coverage of XXL-S is $\sim$77\% incomplete.  If it were more complete, more HERGs would be detected out of the population of sources classified by colours.  Accordingly, it is useful to estimate the number of potential HERG sources that exist in the populations classified as LERGs and SFGs.  It is assumed that HERGs are missed due to an unavailable optical spectrum.  Of the 445 HERGs (either RL or RQ) that are red (the colour that corresponds to 99.0\% of the LERG population), 415 (93.3\%) are identified as either X-ray, MIR, or SED AGN.  Therefore, 30 (6.7\%) of the red HERGs could only be identified by their spectra (EW([\ion{O}{III}]) > 5 \AA $ $ or a quasar spectral template).  This means that $\sim$6.7\% ($\sim$93 sources) of the 1377 LERGs without spectra (i.e. identified by colours only) or $\sim$5.4\% of the whole LERG population could potentially be HERGs.  On the other hand, of the 261 HERGs (either RL or RQ) that are blue and not classifiable as radio AGN (corresponding to the properties of 99.6\% of the SFG population), 176 (67.4\%) are identified as either X-ray, MIR, or SED AGN.  So 85 (32.6\%) blue HERGs not identifiable as radio AGN could only be identified by their spectra.  This implies that out of the SFG population without spectra (385 sources), $\sim$32.6\% ($\sim$125 sources) or $\sim$22.5\% of the whole SFG population could potentially be HERGs.  This is a high fraction of potential contaminants, but in terms of the HERG population only $\sim$125 sources ($\sim$8.6\%) have been potentially missed.  This is the best that can be done with the available data.  Of the LERG and SFG populations combined, $\sim$9.5\% ($\sim$218 sources) are potential HERGs.

\subsection{The faint radio population}

\begin{figure}
        \includegraphics[width=\columnwidth]{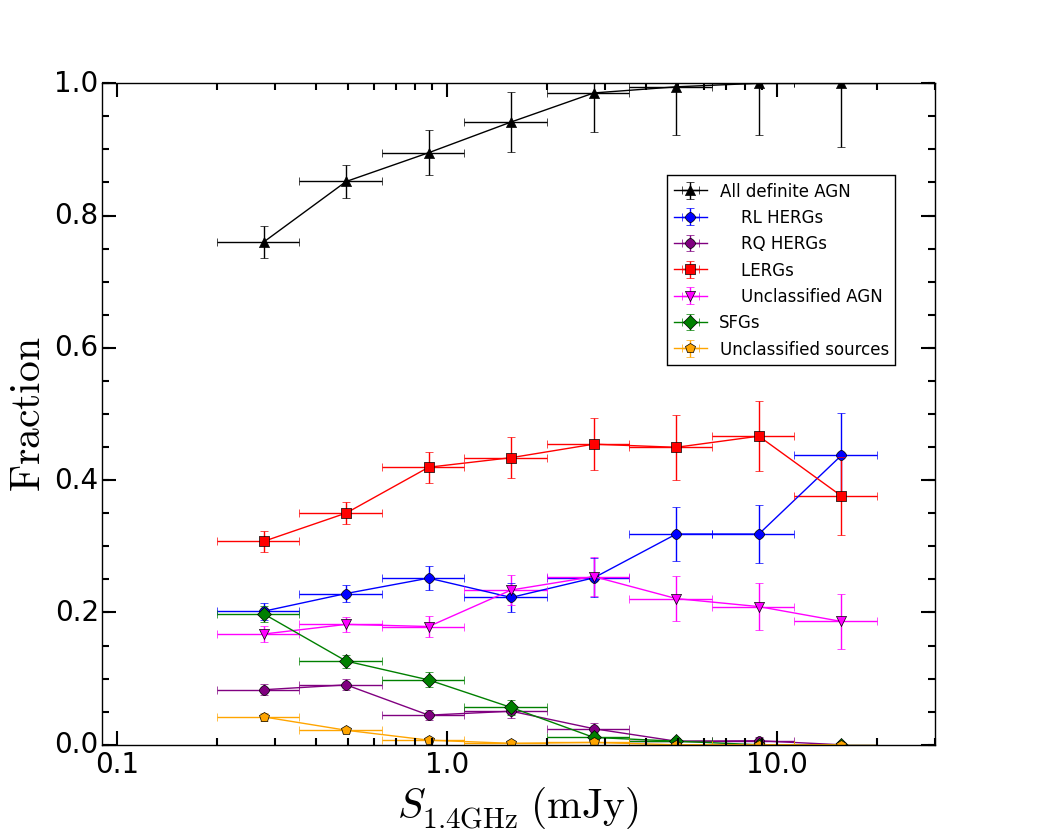}
    \caption{Relative fraction of radio source types as a function of $S_{\rm{1.4GHz}}$.  The black triangles represent all definite radio AGN (whether classified or not), the blue circles  RL HERGs, the purple circles  RQ HERGs, the red squares  LERGs, the magenta triangles  unclassified AGN (which potentially include LERGs and RL HERGs), the green diamonds  SFGs, and the orange pentagons  unclassified sources (which potentially include LERGs, RQ HERGs, and SFGs).  The y-axis error bars for each population are Poissonian and are given by Equation \ref{eq:agn_sfg_frac_err}, calculated for each $S_{\rm{1.4GHz}}$ bin. The errors along the x-axis are the $S_{\rm{1.4GHz}}$ bin widths.}
    \label{fig:agn_sfg_frac_vs_S_plot}
\end{figure}

Figure \ref{fig:agn_sfg_frac_vs_S_plot} shows a plot of the relative fraction of radio source types as a function of $S_{\rm{1.4GHz}}$ (similar to Figure 6 in \citealp{bonzini2013}).  The plot only considers radio sources above the median 5$\sigma$ ATCA XXL-S limit of $\sim$0.2 mJy.  The Poissonian errors of each fraction $\sigma_{\rm{frac}}$ (along the y-axis) are given by
\begin{equation}
\label{eq:agn_sfg_frac_err}
\sigma_{\rm{frac}} = \frac{\sqrt{N_{\rm{pop}}}}{N_{\rm{tot}}}
,\end{equation}
where $N_{\rm{pop}}$ is the number of sources in a given population in a given $S_{\rm{1.4GHz}}$ bin and $N_{\rm{tot}}$ is the total number of sources in that $S_{\rm{1.4GHz}}$ bin. The area of the sky covered by the XXL-S radio survey at a given flux density is flux density dependent (see Figure 4 in XXL Paper XVIII), so each source is weighted by the inverse of the fraction of area covered in the radio mosaic at its corresponding flux density.

The figure shows that LERGs are more numerous than all HERGs between 1 mJy < $S_{\rm{1.4GHz}}$ < 10 mJy, but for $S_{\rm{1.4GHz}} \lesssim 1$ mJy, the combined fractions of RL and RQ HERGs become roughly equal to the LERG fraction, each accounting for $\sim$30\% of the total radio source population. The RL HERG fraction decreases from $\sim$45\% at $S_{\rm{1.4GHz}} \approx 15$ mJy to $\sim$20\% at $S_{\rm{1.4GHz}} \approx 2.75$ mJy, and stays at this level down to $S_{\rm{1.4GHz}} = 0.2$ mJy.  The RQ HERG fraction is virtually non-existent for $S_{\rm{1.4GHz}} > 3.5$ mJy, but steadily increases to $\sim$10\% down to $S_{\rm{1.4GHz}} = 0.2$ mJy. The unclassified AGN (which are a mix of LERGs and RL HERGs) rise steadily between 2 mJy < $S_{\rm{1.4GHz}}$ < 20 mJy to $\sim$25\%, then flatten to a constant level of $\sim$15\% for $S_{\rm{1.4GHz}}$ $\lesssim$ 1 mJy.  The unclassified sources (which are a mix of LERGs, RQ HERGs, and SFGs) make up a very small fraction of the total source population, accounting for $\sim$5\% down to $S_{\rm{1.4GHz}} = 0.2$ mJy.

All AGN (which include RL and RQ HERGs, LERGs, and unclassified AGN) make up $\sim$100\% of the radio population at $S_{\rm{1.4GHz}} > 3$ mJy (virtually no SFGs exist in that flux density regime). At $S_{\rm{1.4GHz}} \approx 0.3$ mJy (the faintest flux density bin), AGN make up $\sim$75\% of the radio source population. SFGs start becoming an appreciable fraction ($\sim$10\%) of the radio source population at $S_{\rm{1.4GHz}}<1$ mJy, and they reach up to $\sim$20\% of the total down to 0.2 mJy.  At the same flux density, \cite{bonzini2013} found that SFGs make up  $\sim$40\% of the radio source population and AGN  $\sim$60\%.  This is a significant disagreement (>3$\sigma$) for the SFGs.  However, \cite{bonzini2013} classified 100\% of the radio sources in their sample.  They did not use spectra and SED fitting (which pick out AGN in galaxies that otherwise appear normal), and if any source did not exhibit at least one of their AGN indicators, it was classified as an SFG. Therefore, their classification method overestimates the relative fraction of SFGs.

The XXL-S results can also be compared to that of the 3 GHz data (converted to 1.4 GHz) for the low to moderate radiative luminosity AGN (MLAGN; LERG equivalent) and the moderate to high radiative luminosity AGN (HLAGN; HERG equivalent) sources in the COSMOS field (see Figure 12 in \citealp{smolcic2017}).
The HLAGN fraction in COSMOS, like the XXL-S HERG fraction, stays roughly constant at a mean value of $\sim$20--30\% for 0.2 mJy < $S_{\rm{1.4GHz}}$ < 1 mJy.  On the other hand, the XXL-S LERG fraction is $\sim$40\% at 1 mJy and decreases to $\sim$30\% at $\sim$0.3 mJy, whereas the MLAGN fraction in COSMOS is $\sim$70\% between 0.65 mJy < $S_{\rm{1.4GHz}}$ < 1 mJy and stays between 40--50\% between 0.2 mJy < $S_{\rm{1.4GHz}}$ < 0.55 mJy.  If, however, the unclassified AGN in XXL-S were mostly LERGs, the XXL-S LERG fraction would be 15--20\% higher, making both LERG fractions more similar.  The XXL-S SFG fraction is higher ($\sim$10\%) than the COSMOS SFG fraction at 1 mJy ($\sim$0\%), but at 0.3 mJy, both fractions are similar ($\sim$20\%).

Despite slight discrepancies, the faint ($\lesssim$1 mJy) radio source population in XXL-S broadly agrees with that of \cite{bonzini2013} and \cite{smolcic2017} over the same flux density range, showing a decline in the relative fraction of AGN and a steady rise of the SFG fraction with decreasing 1.4 GHz flux density.  

\subsection{Host galaxy properties of LERGs, HERGs, and SFGs}

\begin{figure*}
        \includegraphics[width=\textwidth]{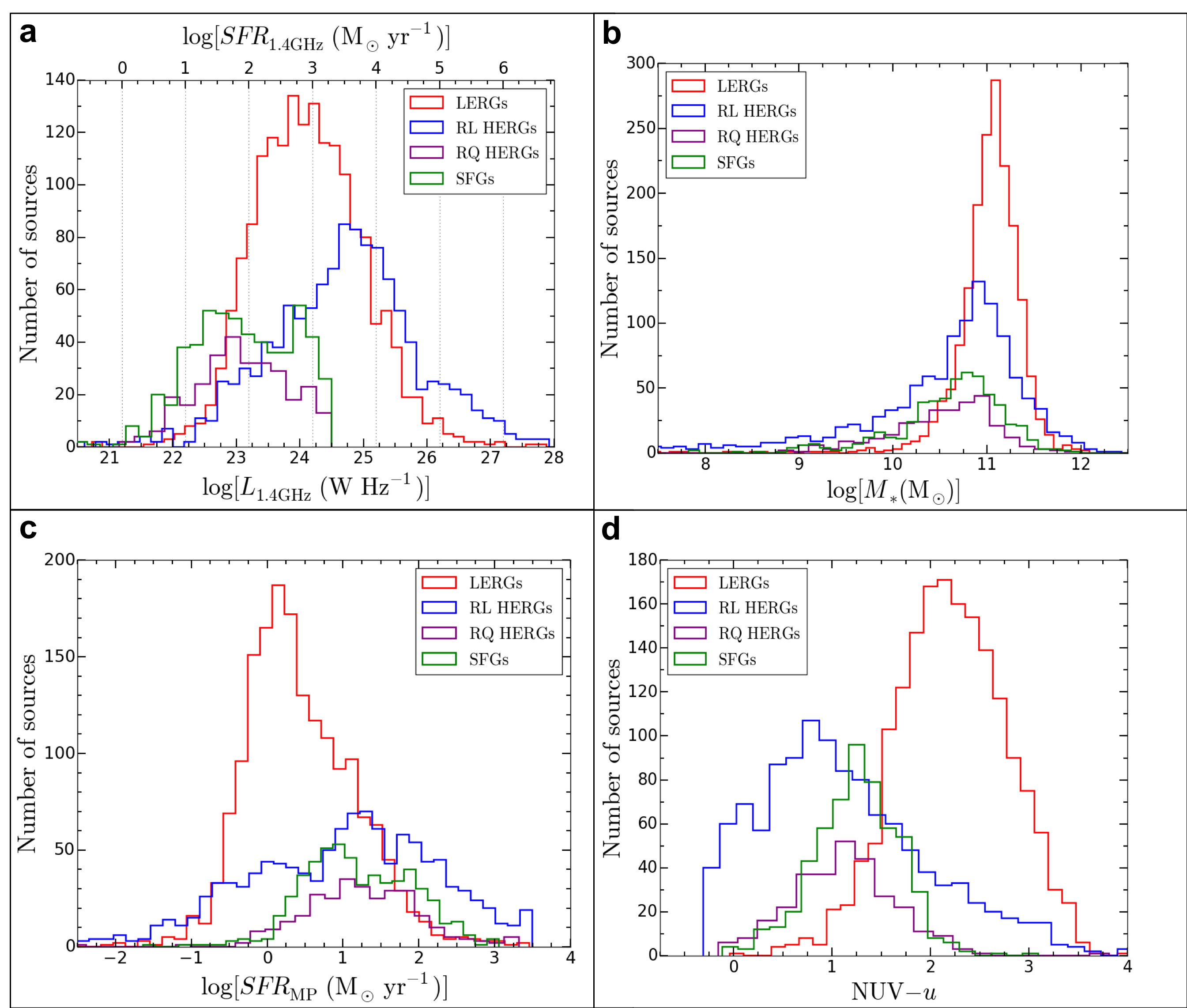}
    \caption{\textbf{(a)} Rest-frame 1.4 GHz radio luminosity ($L_{\rm{1.4GHz}}$, bottom axis) and 1.4 GHz SFR ($SFR_{\rm{1.4GHz}}$, top axis) distributions for the LERGs, HERGs, and SFGs in XXL-S.  LERGs and RL HERGs are found at all radio luminosities, but RL HERGs tend to have higher luminosities.  The SFGs and RQ HERGs are cut off at at $L_{\rm{1.4GHz}}<10^{24.5}$ W Hz$^{-1}$ because of the radio luminosity requirement for radio AGN.  The high $SFR_{\rm{1.4GHz}}$ values for LERGs and RL HERGs ($\gtrsim$10$^{3.3}$ M$_{\odot}$ yr$^{-1}$) are due to AGN contamination in their radio emission.  \textbf{(b)} \textsc{magphys} stellar mass ($M_*$) distribution for the LERGs, HERGs, and SFGs in XXL-S.  The host galaxies of LERGs have median $M_* \approx 10^{11.06}$ M$_{\odot}$, whereas HERGs and SFGs tend to be hosted by galaxies of lower median stellar mass (median $M_* \approx$ $10^{10.6}$ - $10^{10.8}$ M$_{\odot}$).  \textbf{(c)} \textsc{magphys} star formation rate ($SFR_{\rm{MP}}$) distribution for the LERGs, HERGs, and SFGs in XXL-S.  LERGs exhibit low median SFRs ($\sim$2 M$_{\odot}$ yr$^{-1}$) compared to that of HERGs and SFGs ($\sim$12--17 M$_{\odot}$ yr$^{-1}$).  \textbf{(d)} NUV $-$ $u$ colour distribution for the LERGs, HERGs, and SFGs in XXL-S.  The colours were derived from the \textsc{magphys} SED fitting.  LERGs have redder median colours (NUV~$-$~$u$~$\approx$~2.21) than RL HERGs (NUV~$-$~$u$~$\approx$~0.99), RQ HERGs (NUV~$-$~$u$~$\approx$~1.12), and SFGs (NUV~$-$~$u$~$\approx$~1.25).}
    \label{fig:dists_LERGs_HERGs_SFGs_plot}
\end{figure*}

Table \ref{tab:LERG_HERG_SFG_properties} summarises the median properties of the LERGs, HERGs, and SFGs in the optically matched XXL-S radio source sample.  The decision tree used to classify the sources (Figure \ref{fig:decision_tree}) allows for an independent derivation of the HERG host galaxy properties since the method used to identify HERGs does not make use of any previously discovered properties of HERG host galaxies.  However, the LERGs and SFGs without classifiable spectra were selected largely on the basis of the properties of previously constructed LERG and SFGs samples (e.g. \citealp{best2012}).  In other words, LERGs are assumed to be red radio AGN and SFGs are assumed to be blue and not to conform to the radio AGN criteria as defined in Section \ref{sec:radio_class}.  Therefore, this caveat should be considered when evaluating the differences between the properties of XXL-S HERGs, LERGs, and SFGs.

Figure \ref{fig:dists_LERGs_HERGs_SFGs_plot}a shows the 1.4 GHz luminosity ($L_{\rm{1.4GHz}}$) and 1.4 GHz SFR ($SFR_{\rm{1.4GHz}}$) distributions for each type of radio source.  The $L_{\rm{1.4GHz}}$ values were calculated according to Equation \ref{eq:L_R} in this paper and the $SFR_{\rm{1.4GHz}}$ values were calculated according to Equation 17 from \cite{murphy2011}, converted to units of W Hz$^{-1}$:
\begin{equation}
SFR_{\rm{1.4GHz}} = \left(\frac{L_{\rm{1.4GHz}}}{1.575 \times 10^{21} \ \rm{W} \ \rm{Hz}^{-1}}\right) M_{\odot} \ \rm{yr}^{-1}.
\end{equation}
A straight conversion between $L_{\rm{1.4GHz}}$ and $SFR_{\rm{1.4GHz}}$ is not valid for RL AGN, but allows a comparison to the optically derived \textsc{magphys} SFR (see Figure \ref{fig:dists_LERGs_HERGs_SFGs_plot}c) because the excess between the \textsc{magphys} SFR and the 1.4 GHz SFR indicates AGN contamination in the radio. The high $SFR_{\rm{1.4GHz}}$ values for LERGs and RL HERGs ($\gtrsim$10$^3$ M$_{\odot}$ yr$^{-1}$) are due to the radio emission from the AGN, not star formation.  The figure shows that LERGs and RL HERGs are found at all radio luminosities, but there are more RL HERGs at higher luminosities ($L_{\rm{1.4GHz}} > 10^{26}$ W Hz$^{-1}$), which is also what \cite{best2012} found.  The SFGs and RQ HERGs show a sharp cutoff at $L_{\rm{1.4GHz}}<10^{24.5}$ W Hz$^{-1}$ because of the radio luminosity criterion for radio AGN.  The median $L_{\rm{1.4GHz}}$ value for LERGs ($\sim$10$^{24.1}$ W Hz$^{-1}$) is about one order of magnitude larger than that of SFGs and RQ HERGs ($\sim$10$^{23.1}$ W Hz$^{-1}$), but approximately a factor of four below that of RL HERGs ($\sim$10$^{24.7}$ W Hz$^{-1}$).

\begin{table}
\centering
\caption{Median properties of LERGs, HERGs, and SFGs in XXL-S.  The estimation of $\rm{log[}$$SFR_{\rm{1.4GHz}}\rm{]}$ does not apply to LERGs and RL HERGs because their radio emission is contaminated by AGN activity.}
\begin{adjustbox}{width=8.9cm}
\begin{tabular}{c c c c c}
Property & LERGs & RL HERGs & RQ HERGs & SFGs\\
\hline
\hline
log[$L_{\rm{1.4GHz}}$ (W Hz$^{-1}$)] & 24.11 & 24.72 & 23.06 & 23.07\\
log[$M_*$ ($\rm{M}_{\odot}$)] & 11.06 & 10.78 & 10.62 & 10.70\\
log[$SFR_{\rm{MP}}$ ($\rm{M}_{\odot}$ year$^{-1}$)] & 0.33 & 1.08 & 1.22 & 1.14\\
log[$SFR_{\rm{1.4GHz}}$ ($\rm{M}_{\odot}$ year$^{-1}$)] & & & 1.87 & 1.87\\
NUV $-$ $u$ & 2.21 & 0.99 & 1.12 & 1.25\\
\hline
\end{tabular}
\end{adjustbox}
\label{tab:LERG_HERG_SFG_properties}
\end{table}

The stellar mass ($M_*$) and SFR ($SFR_{\rm{MP}}$) distributions, both of which were derived as a result of the SED fits that \textsc{magphys} performed, are shown for each population in Figures \ref{fig:dists_LERGs_HERGs_SFGs_plot}b and \ref{fig:dists_LERGs_HERGs_SFGs_plot}c, respectively.  The best fit stellar masses were used, but for the SFRs, the median $SFR_{\rm{MP}}$ values were used instead because the lack of far-infrared constraints can result in a simultaneous overestimation of the SFR and an accurate fit to the SED at optical and near-infrared wavelengths.  Nevertheless, some sources with poor photometry have high $SFR_{\rm{MP}}$ values (>1000 M$_{\rm{\odot}}$ yr$^{-1}$), but there are just 119 of these sources,   $\sim$98\% of which were classified as LERGs, HERGs, or unclassified AGN.  LERGs tend to occupy galaxies with high stellar masses (median $M_* \approx 10^{11.06}$ M$_{\odot}$) and low SFRs (median $SFR_{\rm{MP}}$ $\approx$ 2 M$_{\odot}$ yr$^{-1}$).  On the other hand, HERGs and SFGs tend to occupy galaxies of lower stellar masses (median $M_* \approx$ $10^{10.6}$ -- $10^{10.8}$ M$_{\odot}$) and higher SFRs (median $SFR_{\rm{MP}}$ $\approx$ 12--17 M$_{\odot}$ yr$^{-1}$). The similarity in SFR for HERGs and SFGs is a reflection of the fact that this is a radio-selected sample, and therefore does not probe the whole SFG population in XXL-S.

\begin{figure}
        \includegraphics[width=\columnwidth]{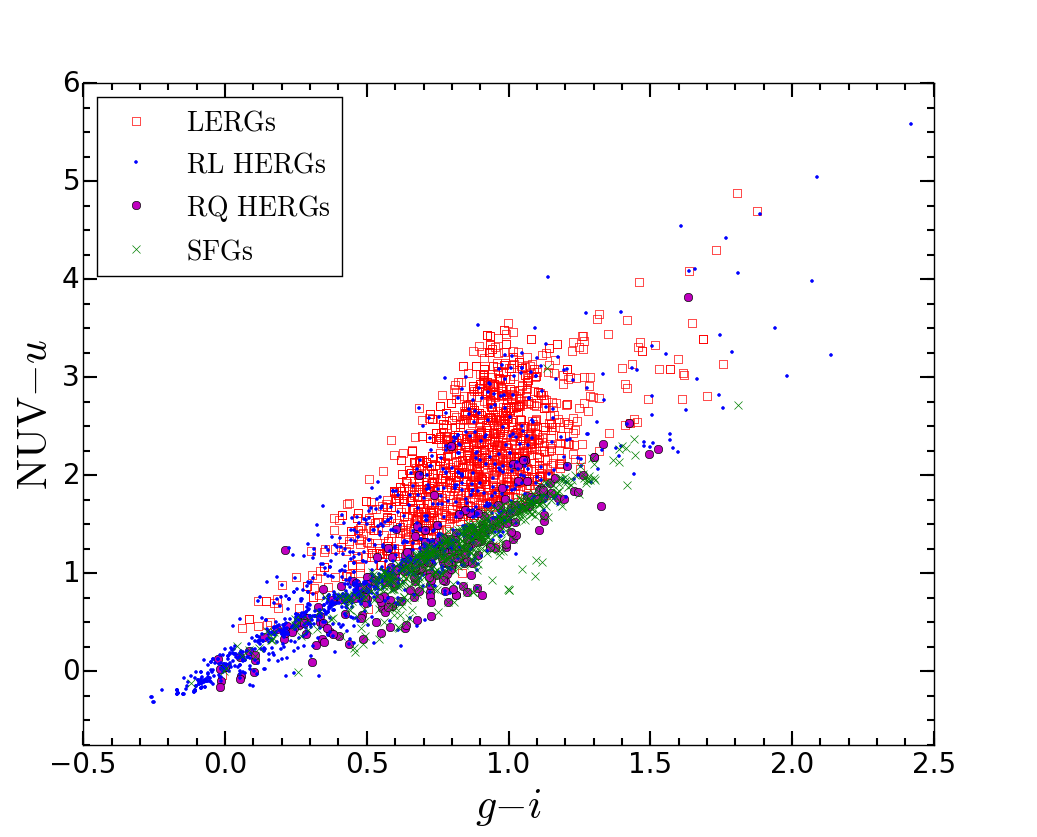}
    \caption{NUV~$-$~$u$ vs. $g$~$-$~$i$ for LERGs (open red squares), RL HERGs (blue dots), RQ HERGs (filled purple circles), and SFGs (green crosses) in XXL-S.  Many ($\sim$70\%) of the HERGs exist in the same colour space as the SFGs.  Only a small percentage (9.7\%) of the HERGs are inside of the quiescent galaxy area of the plot (defined by NUV~$-$~$u$~>~2 and $g$~$-$~$i$~<~1.2), whereas 59.9\% of LERGs exist inside that area.}
    \label{fig:NUV_minus_u_vs_g_minus_i_LERGs_HERGs_SFGs_plot}
\end{figure}

Figure \ref{fig:dists_LERGs_HERGs_SFGs_plot}d shows the NUV~$-$~$u$ colour distribution for LERGs, HERGs, and SFGs.  The colours were also derived from the \textsc{magphys} SED fitting.  LERGs tend to have redder colours (median NUV~$-$~$u$~$\approx$~2.21) than RL HERGs (median NUV~$-$~$u$~$\approx$~0.99), RQ HERGs (median NUV~$-$~$u$~$\approx$~1.12), and SFGs (median NUV~$-$~$u$~$\approx$~1.25). Figure \ref{fig:NUV_minus_u_vs_g_minus_i_LERGs_HERGs_SFGs_plot} shows the NUV~$-$~$u$ versus $g$~$-$~$i$ colours for each population.  Only a small percentage (9.9\%) of all HERGs are inside of the quiescent galaxy area of the plot (defined by NUV~$-$~$u$~>~2 and $g$~$-$~$i$~<~1.2), while 59.9\% of the LERGs exist there. Most (69.4\%) of the HERGs are blue, but the fraction of RQ HERG hosts that are blue  (88.2\%) is higher than the blue fraction of RL HERG hosts (64.6\%). 

LERGs and HERGs have a wide range of properties. For example, there are some LERGs that  form stars just as rapidly as some HERGs ($\sim$100 M$_{\odot}$ yr$^{-1}$), and there are some HERGs that are very red (NUV~$-$~$u$~>~3). However, the overall properties of the LERG population are consistent with the idea that LERGs tend to exhibit lower radio luminosities and occupy the most massive, reddest, quiescent galaxies.  In addition, the properties of the RL HERG population are consistent with sources that are more radio luminous and hosted by less massive, bluer galaxies that tend to form stars as frequently as normal SFGs.  The host galaxies of RQ HERGs are also characterised by bluer colours, lower masses, and ongoing star formation. These results are similar to what numerous other authors have found (e.g. \citealp{kauffmann2003,best2005,janssen2012,best2012,hardcastle2013,best2014,pracy2016,ching2017}). 

\section{Summary and conclusions}
\label{sec:conclusions}

This work presented the results of the optical counterpart matching, classification, and characterisation of the 2.1 GHz radio sources in the 25 deg$^2$ XXL-S field, which is thus far the largest area radio survey conducted down to rms flux densities of $\sim$41 $\mu$Jy beam$^{-1}$.  The likelihood ratio (LR) technique was applied to the XXL-S multiwavelength catalogue in order to find the most likely optical hosts of the radio sources.  The technique found reliable counterparts for 4770 radio sources, 4758 of which were included in the final sample.  These optically matched radio sources were then 
classified according to X-ray, MIR, SED, and radio diagnostics, as well as optical spectral templates, emission lines and colours.
A decision tree was employed to determine a final classification for each source based on the results of each diagnostic.

The overlap between X-ray, MIR, and radio AGN in XXL-S is very similar to that of other multiwavelength studies of radio galaxies and AGN (e.g. \citealp{hickox2009}), which, despite slight differences in the classification methods, validates the idea that different classes of AGN represent distinct populations of sources.  Given the depth of the radio survey and the multiwavelength catalogue, and the low completeness of the spectral coverage, the contamination of the combined LERG and SFG population by HERGs is estimated to be relatively low ($\sim$10\%).  The fractions of sub-mJy ($S_{\rm{1.4GHz}} < 1$ mJy) sources that are AGN and SFGs are similar to the fractions found by \cite{bonzini2013} and \cite{smolcic2017} over the same flux density range, with $\sim$75\% definite radio AGN and $\sim$20\% SFGs down to 0.2 mJy.

The median properties of the HERGs in XXL-S were found to agree with what other similar studies have found, while the selection of the LERGs and SFGs without spectra was based on the properties of previously constructed samples of those populations (e.g. \citealp{best2005,janssen2012,best2012,hardcastle2013,best2014,pracy2016,ching2017}).  Taking this caveat into account, the LERGs tend to be found in the most massive galaxies ($M_* \approx 10^{11.06}$ M$_{\odot}$) with low SFRs ($\sim$2 M$_{\odot}$ yr$^{-1}$) and redder colours (NUV~$-$~$u$~$\approx$~2.2), whereas the HERGs and SFGs exist in galaxies of lower mass ($M_* \approx 10^{10.7}$ M$_{\odot}$), higher SFRs ($\sim$15 M$_{\odot}$ yr$^{-1}$) and bluer colours (NUV~$-$~$u$~$\lesssim$~1.25).  The host galaxies of RQ HERGs have similar properties to RL HERGs, but RQ HERGs have a higher fraction of blue hosts.  LERGs and RL HERGs are found at all radio luminosities, but RL HERGs tend to be more radio luminous (median $L_{\rm{1.4GHz}} \sim 10^{24.7}$ W Hz$^{-1}$) than LERGs (median $L_{\rm{1.4GHz}} \sim 10^{24.1}$ W Hz$^{-1}$).  This is consistent with the idea that LERGs tend to exist in the most massive quiescent galaxies with low star formation rates, indicating that they are fuelled by the hot (X-ray emitting) phase of the IGM typically found in clusters, which limits their radiative output to radio wavelengths.  It is also consistent with the idea that HERGs are associated with galaxies that have undergone recent star formation, indicating that their accretion material is cold gas from the IGM.


\section*{Acknowledgements}

AB acknowledges UWA for funding support from a University Postgraduate Award PhD scholarship and ICRAR for additional support. MH acknowledges UWA for a Research Collaboration Award grant to collaborate with the COSMOS and XXL team at the University of Zagreb.  AK acknowledges financial support from CAASTRO through project number CE110001020, partly received during the execution of this project.  VS acknowledges funding from the European Union's Seventh Framework programme under grant agreement 333654 (CIG, `AGN feedback'). VS, JD, and ID acknowledge funding from the European Union's Seventh Framework programme under grant agreement 337595 (ERC Starting Grant, `CoSMass'). VS acknowledges support from the ICRAR Visiting Fellowship For Senior Women In Astronomy 2015.  All the authors thank the referee for the helpful comments that improved the clarity of the paper. The Saclay group acknowledges long-term support from the Centre National d'Etudes Spatiales (CNES).
XXL is an international project based around an XMM Very Large Programme surveying two 25 deg$^2$ extragalactic fields at a depth of $\sim$5 $\times$ 10$^{-15}$ erg cm$^{-2}$ s$^{-1}$ in the [0.5--2] keV band for point-like sources. The XXL website is http://irfu.cea.fr/xxl. Multi-band information and spectroscopic follow-up of the X-ray sources are obtained through a number of survey programmes, summarised at http://xxlmultiwave.pbworks.com/.  The Australia Telescope Compact Array is part of the Australia Telescope National Facility which is funded by the Australian Government for operation as a National Facility managed by CSIRO.  Based in part on data acquired through the Australian Astronomical Observatory, via programmes A/2013A/018, A/2013B/001, and A/2016B/107.  
This project used data obtained with the Dark Energy Camera (DECam), which was constructed by the Dark Energy Survey (DES) collaboration. Funding for the DES Projects has been provided by the U.S. Department of Energy, the U.S. National Science Foundation, the Ministry of Science and Education of Spain, the Science and Technology Facilities Council of the United Kingdom, the Higher Education Funding Council for England, the National Center for Supercomputing Applications at the University of Illinois at Urbana-Champaign, the Kavli Institute of Cosmological Physics at the University of Chicago, Center for Cosmology and Astro-Particle Physics at the Ohio State University, the Mitchell Institute for Fundamental Physics and Astronomy at Texas A\&M University, Financiadora de Estudos e Projetos, Funda\c{c}\~{a}o Carlos Chagas Filho de Amparo, Financiadora de Estudos e Projetos, Funda\c{c}\~{a}o Carlos Chagas Filho de Amparo \`{a} Pesquisa do Estado do Rio de Janeiro, Conselho Nacional de Desenvolvimento Cient\'{i}fico e Tecnol\'{o}gico and the Minist\'{e}rio da Ci\^{e}ncia, Tecnologia e Inova\c{c}\~{a}o, the Deutsche Forschungsgemeinschaft, and the Collaborating Institutions in the Dark Energy Survey. The Collaborating Institutions are Argonne National Laboratory, the University of California at Santa Cruz, the University of Cambridge, Centro de Investigaciones En\'{e}rgeticas, Medioambientales y Tecnol\'{o}gicas--Madrid, the University of Chicago, University College London, the DES-Brazil Consortium, the University of Edinburgh, the Eidgen\"{o}ssische Technische Hochschule (ETH) Z\"{u}rich, Fermi National Accelerator Laboratory, the University of Illinois at Urbana-Champaign, the Institut de Ci\`{e}ncies de l'Espai (IEEC/CSIC), the Institut de F\'{i}sica d'Altes Energies, Lawrence Berkeley National Laboratory, the Ludwig-Maximilians Universit\"{a}t M\"{u}nchen and the associated Excellence Cluster Universe, the University of Michigan, the National Optical Astronomy Observatory, the University of Nottingham, the Ohio State University, the University of Pennsylvania, the University of Portsmouth, SLAC National Accelerator Laboratory, Stanford University, the University of Sussex, and Texas A\&M University.
Based in part on observations at Cerro Tololo Inter-American Observatory, National Optical Astronomy Observatory (NOAO Prop. IDs: 0616, 0618 and PI: C. Lidman), which is operated by the Association of Universities for Research in Astronomy (AURA) under a cooperative agreement with the National Science Foundation.




\clearpage

\bibliographystyle{aa}
\bibliography{paper_2_aa}




\clearpage

\appendix

\section{radio spectral index calculations}
\label{sec:alpha_calculations}

XXL-S radio sources that are unresolved in the full 2 GHz ATCA bandwidth (the `full-band') were considered for radio spectral index analysis. In order to calculate the in-band radio spectral indices of these sources, radio mosaics of XXL-S for three ATCA sub-bands at central frequencies of 1417 MHz, 2100 MHz, and 2783 MHz, all with bandwidths of 683 MHz, were constructed.  The steps taken to image every pointing in each of these three sub-bands were identical to those for the full-band (see Sections 2.3 and 2.4 in XXL Paper XVIII), with the following exceptions: robust parameters, image dimensions corresponding to the radii at which the primary beam responses are $\sim$8\% for the given central frequencies, a \texttt{SELFCAL} interval of 0.33 minutes (which was found to minimise the image artefacts around bright sources), and a \texttt{CONVOL} beam size of 6.36$''$ $\times$ 5.16$''$ with a position angle of 2.5$^{\circ}$.  Different robust parameters (corresponding to different visibility weighting distributions) were necessary in order to make the synthesised beam sizes for each sub-band as similar as possible before convolving the images. Table \ref{tab:sb_freq_robust} lists the central frequencies, robust parameters, and the radii the individual pointings were imaged out to for each sub-band.  Once each pointing was imaged in each sub-band, the \textsc{miriad} \citep{sault1995} task \texttt{LINMOS} was used to combine and mosaic all the images together for each sub-band.

\begin{table}
\caption{Central frequencies, robust parameters, and image radii of individual pointings for the three 683 MHz sub-bands.}
\begin{tabular}{c c c}
Central frequency & Robust parameter & Image radius\\
(MHz) & & (arcmin)\\
\hline
\hline
1417 & -2.0 & 31.4\\
2100 & 0.0 & 22.7\\
2783 & 0.3 & 16.7\\
\hline
\end{tabular}
\label{tab:sb_freq_robust}
\end{table}

As for the full-band, \textsc{blobcat} was used to detect all sources in each sub-band.  A detection threshold of $3.05\sigma$, where $\sigma$ is the local rms value, was used.  The plot of the ratio of integrated flux density to peak flux density ($S_{\rm{int}}/S_{\rm{p}}$) versus signal-to-noise ratio ($S/N$) for 2100 MHz and 2783 MHz showed no abnormalities, but the plot for 1417 MHz showed a very slight systematic downward curve (see Figure \ref{fig:1417_S_int_S_p_vs_SNR_plot}), similar to the curve described in Appendix A of XXL Paper XVIII, but less severe.  A simple method used to correct the 1417 MHz $S_{\rm{int}}$ values was executed.  An interpolation function was computed for the median $S_{\rm{int}}/S_{\rm{p}}$ values in log-spaced $S/N$ bins.  Each $S_{\rm{int}}$ was corrected according to this function to change the value of $S_{\rm{int}}$ such that at all $S/N$, the value of $S_{\rm{int}}/S_{\rm{p}}$ was approximately equal to its value at high $S/N$ ($\sim$1.02 in this case). After this correction for 1417 MHz was made, all the $S_{\rm{p}}$ values in each sub-band were corrected for bandwidth smearing in the same way as described in Section 3.5 of XXL Paper XVIII.

\begin{figure}
        \includegraphics[width=\columnwidth]{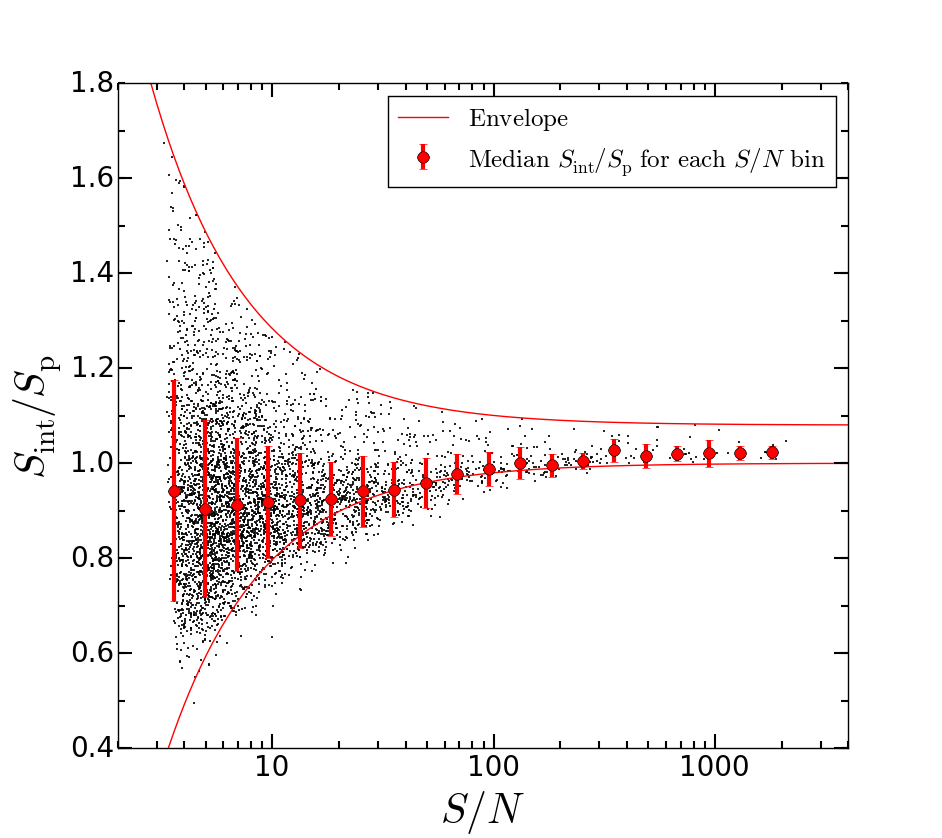}
    \caption{$S_{\rm{int}}/S_{\rm{p}}$ vs. $S/N$ for the 1417 MHz sub-band detections.  The red curves show the envelope by which the full-band sources were determined to be resolved or unresolved (Equations 3 and 4 in XXL Paper XVIII).  The red circles represent the median $S_{\rm{int}}/S_{\rm{p}}$ values and the error bars represent the standard deviation in $S_{\rm{int}}/S_{\rm{p}}$ for each $S/N$ bin.}
    \label{fig:1417_S_int_S_p_vs_SNR_plot}
\end{figure}

Next, the three sub-band source catalogues from \textsc{blobcat} were individually cross-matched to the 6287 full-band radio sources in the XXL-S source catalogue (XXL Paper XVIII). Only radio sources that are unresolved in the full-band were considered for an in-band radio spectral index calculation, resulting in 4661 eligible radio sources.  The full-band sources were not cross-matched to 843 MHz sources from the Sydney University Molonglo Sky Survey (SUMSS) catalogue because only 28 ATCA XXL-S sources were classified as radio AGN on the basis of radio spectral index alone, and none of them has a SUMSS counterpart within 52.8$''$ (the major axis of the SUMSS beam at the declination of XXL-S).

At this point, each eligible full-band radio source had either detections in all three sub-bands, detections in two sub-bands, or fewer than two detections.  The spectral indices were calculated for the 3827 radio sources with at least two sub-band detections. If exactly two detections were available, the spectral indices were calculated in the normal way using
\begin{equation}
\alpha_R = \frac{\log(S_{\rm{high}}/S_{\rm{low}})}{\log(\nu_{\rm{high}}/\nu_{\rm{low}})},
\end{equation}
where $S_{\rm{high}}$ is the peak flux density measured at the higher frequency $\nu_{\rm{high}}$ and $S_{\rm{low}}$ is the peak flux density measured at the lower frequency $\nu_{\rm{low}}$.  If there were three detections, a linear regression was fit to the log($S$) versus log($\nu$) data. Since spectral indices are calculated in log space, the uncertainties on the spectral indices were calculated according to the  formula (adapted from \citealp{ciliegi2003})
\begin{equation}
\label{eq:alpha_errors}
\sigma_{\alpha_R} = \frac{\sqrt{\sum_i (\sigma_{S_{i\rm{p}}} / S_{i\rm{p}})^2}}{\rm{ln}(\nu_{\rm{high}}) - \rm{ln}(\nu_{\rm{low}})},
\end{equation}
where the sum is over all sub-band peak flux densities $S_{i\rm{p}}$ available for the source, $\sigma_{S_{i\rm{p}}}$ is the uncertainty in the corresponding $S_{i\rm{p}}$, and $\nu_{\rm{high}}$ and $\nu_{\rm{low}}$ are respectively the highest and lowest frequencies involved in the calculation.

\begin{figure}
        \includegraphics[width=\columnwidth]{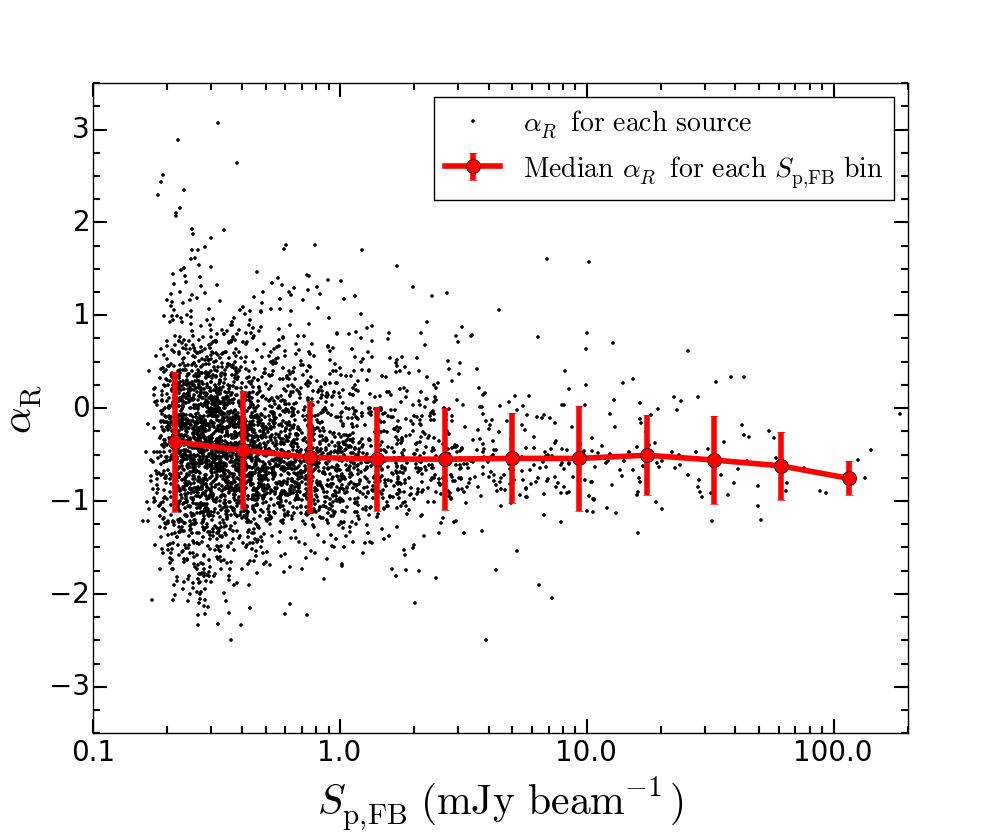}
    \caption{ATCA in-band radio spectral index ($\alpha_R$) vs. full-band peak flux density ($S_{\rm{p,FB}}$) for all XXL-S radio sources for which spectral index calculations are available (black points).  The red circles with connecting red line show the median $\alpha_R$ values for each $S_{\rm{p,FB}}$ bin.  The width of each $S_{\rm{p,FB}}$ bin is $\sim$0.273 in log space.  The median $\alpha_R$ for the faintest bin at $S_{\rm{p,FB}} \approx 0.22$ mJy is $-0.36$, whereas at the brightest bin ($S_{\rm{p,FB}} \approx 115$ mJy) the median $\alpha_R$ is $-0.76$. This clearly shows a trend toward flatter spectral indices for sources with $S_{\rm{p,FB}}~\lesssim~0.5$ mJy.}
    \label{fig:alpha_vs_S_all_plot}
\end{figure}

\begin{figure}
        \includegraphics[width=\columnwidth]{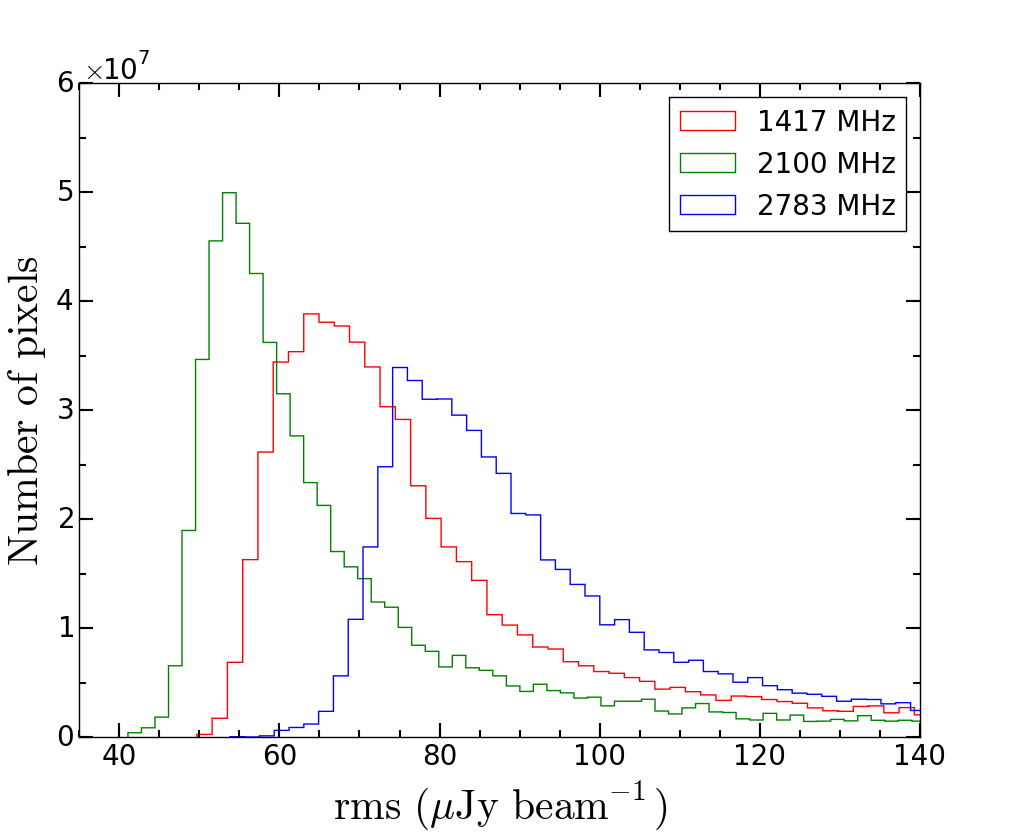}
    \caption{Distributions of the rms noise in the 1417 MHz (red), 2100 MHz (green), and 2783 MHz (blue) ATCA sub-bands.  Table \ref{tab:sb_alpha_vs_SpFB} shows the peak rms values for each sub-band.  The higher peak rms value for 2783 MHz means that low $(S/N)_{\rm{FB}}$ sources with a detection in that band tend to have systematically flatter spectral indices than sources without a 2783 MHz detection.}
    \label{fig:sb3_noise_hist_plot}
\end{figure}

Figure \ref{fig:alpha_vs_S_all_plot} shows the plot of $\alpha_R$ versus full-band peak flux density ($S_{\rm{p,FB}}$) for the 3827 XXL-S radio sources with least two sub-band detections.  The median $\alpha_R$ values for each $S_{\rm{p,FB}}$ bin are displayed in Table \ref{tab:sb_alpha_vs_SpFB}.  The survival analysis performed in XXL Paper XVIII for all XXL-S radio sources with $S_{\rm{p,FB}}$ < 5 mJy indicated a median $\alpha_R = -0.66^{+0.18}_{-0.07}$, which is consistent with all but the lowest $S_{\rm{p,FB}}$ bin given in Table \ref{tab:sb_alpha_vs_SpFB}.  The data in the table clearly show a trend toward flatter spectral indices for sources with $S_{\rm{p,FB}}~\lesssim$~0.5 mJy, which roughly corresponds to a full-band signal-to-nose ratio of $(S/N)_{\rm{FB}} \approx 10$.  The flattening at low $(S/N)_{\rm{FB}}$ is not inherent to the radio source population in XXL-S, but due to a selection effect caused by the different rms noise distributions in each sub-band mosaic, which are shown in Figure \ref{fig:sb3_noise_hist_plot}.  Table \ref{tab:sb_peak_rms} shows the peak rms values for each sub-band, indicating that the 2783 MHz sub-band has the highest peak rms value.  Consequently, at low $(S/N)_{\rm{FB}}$, if a source has a 2783 MHz detection the 3.05$\sigma$ detection threshold selects sources with flatter spectral indices.  In other words, for a given 1417 MHz or 2100 MHz rms value, only sources with spectral indices flat enough to allow their 2783 MHz $S_{\rm{p}}$ values to be above the 2783 MHz 3.05$\sigma$ detection threshold will be detected at 2783 MHz.  Low $(S/N)_{\rm{FB}}$ sources without a 2783 MHz detection may have typical spectral indices ($-1 < \alpha < -$0.5), but they are missed at 2783 MHz because they fall below the 2783 MHz 3.05$\sigma$ detection threshold, leaving them with two sub-band detections and therefore less constrained spectral indices than those with higher $(S/N)_{\rm{FB}}$.  Therefore, the low $(S/N)_{\rm{FB}}$ source sub-sample contains a relatively low fraction of sources with typical spectral indices and a relatively high fraction of flat-spectrum sources ($\alpha_R > -$0.5). As a result, the $\alpha_R$ distribution for sources at low $(S/N)_{\rm{FB}}$ is systematically skewed toward flatter spectral indices.  This does not imply that the spectral indices at low $(S/N)_{\rm{FB}}$ are incorrect, but that they indicate an incomplete sampling of all possible spectral indices of faint sources in the XXL-S radio source catalogue.

\begin{table}
\caption{Median radio spectral indices ($\alpha_R$) for given full-band peak flux density ($S_{\rm{p,FB}}$) bins.}
\begin{tabular}{c c}
$S_{\rm{p,FB}}$ bin (mJy) & $\alpha_R$\\
\hline
\hline
0.150 < $S_{\rm{p,FB}}$ < 0.281 & -0.363\\
0.281 < $S_{\rm{p,FB}}$ < 0.527 & -0.455\\
0.527 < $S_{\rm{p,FB}}$ < 0.987 & -0.530\\
0.987 < $S_{\rm{p,FB}}$ < 1.849 & -0.548\\
1.849 < $S_{\rm{p,FB}}$ < 3.465 & -0.549\\
3.465 < $S_{\rm{p,FB}}$ < 6.493 & -0.541\\
6.493 < $S_{\rm{p,FB}}$ < 12.167 & -0.541\\
12.167 < $S_{\rm{p,FB}}$ < 22.799 & -0.507\\
22.799 < $S_{\rm{p,FB}}$ < 42.721 & -0.559\\
42.721 < $S_{\rm{p,FB}}$ < 80.050 & -0.622\\
80.050 < $S_{\rm{p,FB}}$ < 150.000 & -0.756\\
\hline
\end{tabular}
\label{tab:sb_alpha_vs_SpFB}
\end{table}

\begin{table}
\caption{Peak rms noise values for each ATCA sub-band.}
\begin{tabular}{c c}
Sub-band (MHz) & Peak rms noise (mJy beam$^{-1}$)\\
\hline
\hline
1417 & 0.066\\
2100 & 0.057\\
2783 & 0.076\\
\hline
\end{tabular}
\label{tab:sb_peak_rms}
\end{table}

\begin{figure}
        \includegraphics[width=\columnwidth]{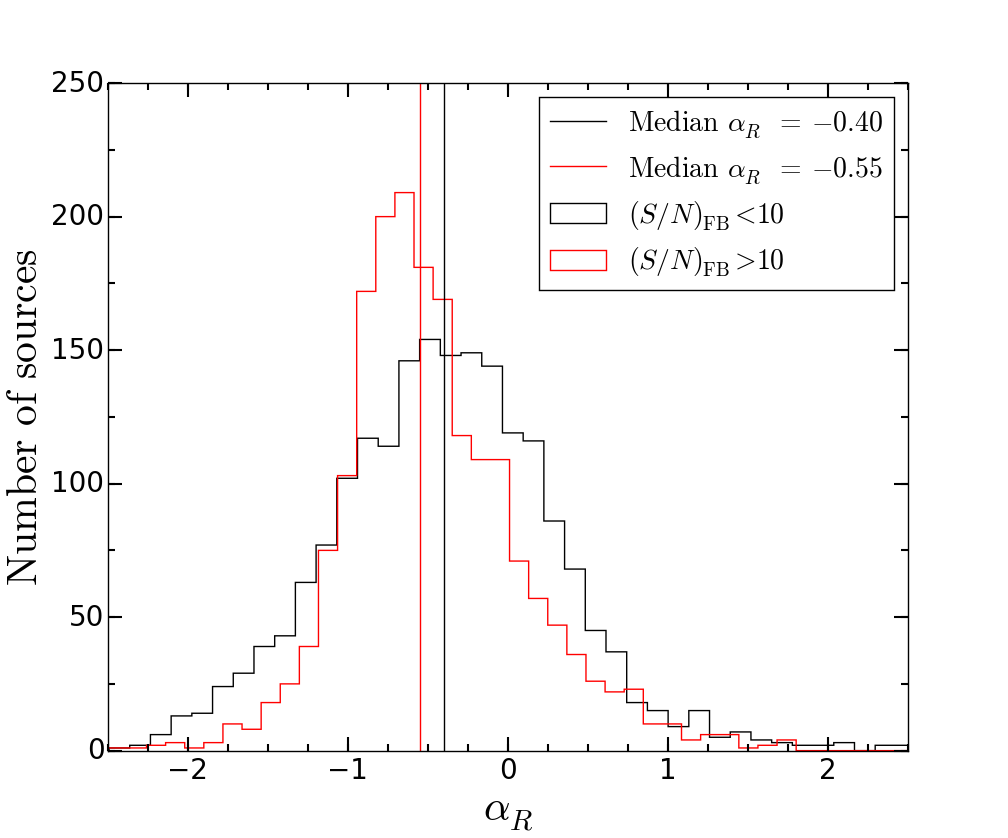}
    \caption{Distribution of ATCA in-band radio spectral indices for XXL-S radio sources with at least two 3.05$\sigma$ radio sub-band detections, separated into $(S/N)_{\rm{FB}} < 10$ (black histogram) and $(S/N)_{\rm{FB}} > 10$ (red histogram) sub-samples.  The median spectral index for $(S/N)_{\rm{FB}} < 10$ is $\alpha_R = -0.40$, whereas the median for $(S/N)_{\rm{FB}} > 10$ is $\alpha_R = -0.55$.}
    \label{fig:alpha_dist_det_SNR_all_plot}
\end{figure}

Figure \ref{fig:alpha_dist_det_SNR_all_plot} shows the distribution of ATCA in-band radio spectral indices for the 3827 XXL-S full-band radio sources with at least two sub-band detections, separated into those with $(S/N)_{\rm{FB}} < 10$ and those with $(S/N)_{\rm{FB}} > 10$. The median spectral index for sources with $(S/N)_{\rm{FB}} < 10$ is $\alpha_R = -0.40$, and the median for  those with $(S/N)_{\rm{FB}} > 10$ is $\alpha_R = -0.55$. Due to the systematic flattening of spectral indices at low $(S/N)_{\rm{FB}}$, if a source had $(S/N)_{\rm{FB}} < 10$ or $S_{\rm{p,FB}} < 0.527$ mJy and did not have an in-band spectral index measurement, its spectral index was assigned to be the median of all sources with $(S/N)_{\rm{FB}} > 10$, ($\alpha_R = -0.55$).  If a source had $(S/N)_{\rm{FB}} > 10$ and $S_{\rm{p,FB}} > 0.527$ mJy, and did not have a spectral index measurement available the value of the spectral index was determined by the median $\alpha_R$ value in its corresponding $S_{\rm{p,FB}}$ bin shown in Figure \ref{fig:alpha_vs_S_all_plot}.  For sources with $S_{\rm{p,FB}} > 150$ mJy and no $\alpha_R$ value, a value of $\alpha_R = -0.75$ was assumed.  All sources with a spectral index calculation available were assigned their corresponding $\alpha_R$ value.  Of the 3827 sources with at least two sub-band detections, 2974 were cross-matched to optical counterparts.
 
\label{lastpage}
\end{document}